%% file: MastersThesis.tex
\documentclass[12pt,twoside,leftblank]{mitthesis}

\usepackage{lgrind}
\usepackage{cmap}
\usepackage[T1]{fontenc}
\usepackage{physics}
\usepackage{epstopdf}

\usepackage{textcomp}
\usepackage{amsmath, amsthm, amssymb}
\usepackage{acronym}

\usepackage{courier}
\usepackage{bbold}

\usepackage{mathtools}
\usepackage{cite}
\usepackage{url}
\usepackage{float}
\usepackage{pdfpages}

\usepackage{graphicx}
\usepackage{amssymb}
\usepackage{wrapfig}
\usepackage{xspace}

\usepackage{subcaption}
\usepackage{multicol}
\usepackage{siunitx}

\usepackage{hyperref}

\usepackage{xspace}

\usepackage[dvips]{epsfig} 
\usepackage{tocloft}
\usepackage{xcolor}
\definecolor{darkblue}{RGB}{22,73,99}

\newcommand{\listofdefinitionsname}{List of definitions}
\newlistof[section]{defns}{dfn}{\listofdefinitionsname}

\usepackage{datatool}

\begin{document}

\include{cover}

\include{signature}

\pagestyle{plain}
\include{contents}

\include{DocumentOverview}

\include{Sources}

\include{Interactions}

\include{DetectionMethods}

\include{CosmicWatch}

\include{ExampleMeasurements}
\include{Conclusion}

\appendix

\bibliographystyle{ieeetr}
\bibliography{Master} 

\end{document}

%% file: cover.tex
\title{\textsc{\Huge{ \linebreak \color{black} The Physics Behind the CosmicWatch Desktop Muon Detectors}}}

\author{\LARGE{Spencer Nicholas Gaelan Axani \\ \hspace{1mm}}}
\prevdegrees{Power Engineering Technologies Dip, Southern Alberta Institute of Technology~(2008)\\
B.Sc. (Hons), University of Alberta (2014)}
\department{Department of Physics}

 \degree{Master of Science in Physics}

\degreemonth{February}
\degreeyear{2019}
\thesisdate{February 26$^{\mathrm{th}}$, 2019}


\supervisor{Janet M. Conrad}{Professor of Physics, MIT}
\supervisor{Scott A. Hughes}{Professor of Physics, MIT}


\chairman{Nergis Mavalvala}{Associate Department Head of Physics, MIT}

\maketitle

\title{\textsc{\Huge{ \linebreak The Physics Behind the CosmicWatch Desktop Muon Detectors }}}
\author{\LARGE{Spencer Nicholas Gaelan Axani}}



\cleardoublepage
\pagestyle{empty}
\setcounter{savepage}{\thepage}
\begin{abstractpage}
\input{abstract}

\end{abstractpage}


\cleardoublepage

\section*{Acknowledgements}

The CosmicWatch project owes a lot to Dr. Katarzyna Frankiewicz. For both of us, this began as hobby and desire to contribute to the community, and eventually ballooned into a popular and well-developed program thanks in large part to her work. Prof. Janet M. Conrad's support, input, and encouragement throughout the development of this project was equally important for the success of this project. I am very thankful for the opportunity to work on this project while engaged in my PhD.

This work was supported by grant NSF-PHY-1505858 (PI: Prof. Janet Conrad), MIT seed funding, and funds from the National Science Centre, Poland (2015/17/N/ST2/04064). I would like to thank SensL (currently On Semiconductor) and Fermilab, for donations that made the development of this project possible, as well as P. Fisher at MIT and IceCube collaborators at WIPAC, for their support in developing this as a high school and undergraduate project; J.~Moon, D.~Torretta, and J.~Zalesak for their aid in making the measurements at Fermilab;  B. Jones for the idea of developing this project for the IceCube detector; and those who taught Phys 063 at Swarthmore College for the inspiration. I would also like to thank Prof. Scott A. Hughes from MIT for helping in the development of this document. Finally, I would also like to thank the hundreds of students, professors, and enthusiasts who have contact me with their questions related to the detectors and the encouragement that they often provide.

\begin{figure}[h]
\begin{center}
\includegraphics[width=0.75\columnwidth]{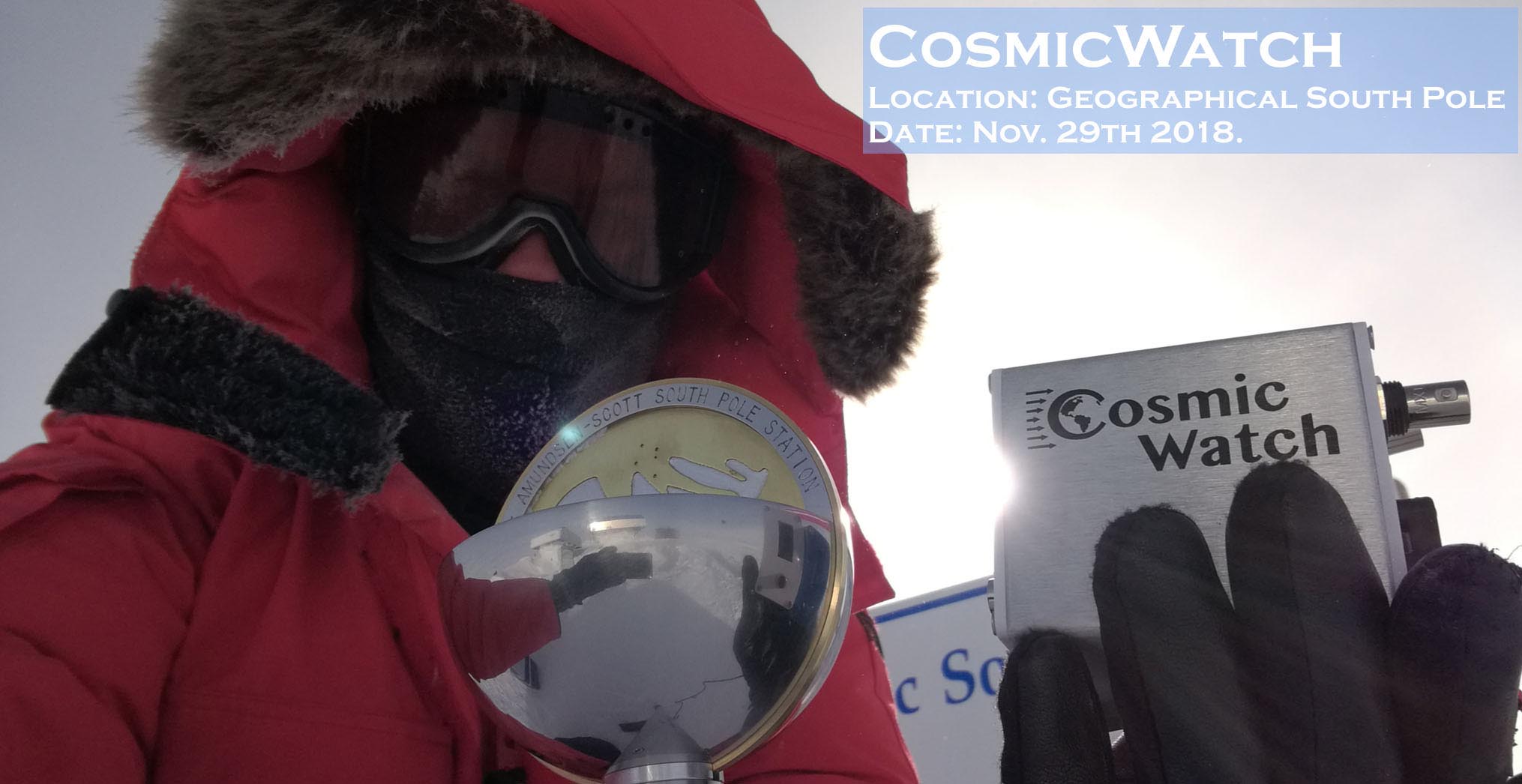}
 \end{center}
\end{figure}


%% file: abstract.tex
The CosmicWatch Desktop Muon Detector is a Massachusetts Institute of Technology (MIT) and Polish National Centre for Nuclear Research (NCBJ) based undergraduate-level physics project that incorporates various aspects of electronics-shop technical development. The detector was designed to be low-power and extremely portable, which opens up a wide range of physics for students to explore. This document describes the physics behind the Desktop Muon Detectors and explores possible measurements that can be made with the detectors. In particular, we explore various physical phenomena associated with the geomagnetic field, atmospheric conditions, cosmic ray shower composition, attenuation of particles in matter, radioactivity, and statistical properties of Poisson processes.

%% file: signature.tex
%
%
%

\begin{titlepage}
\begin{large}

\includegraphics[width=0.4\columnwidth]{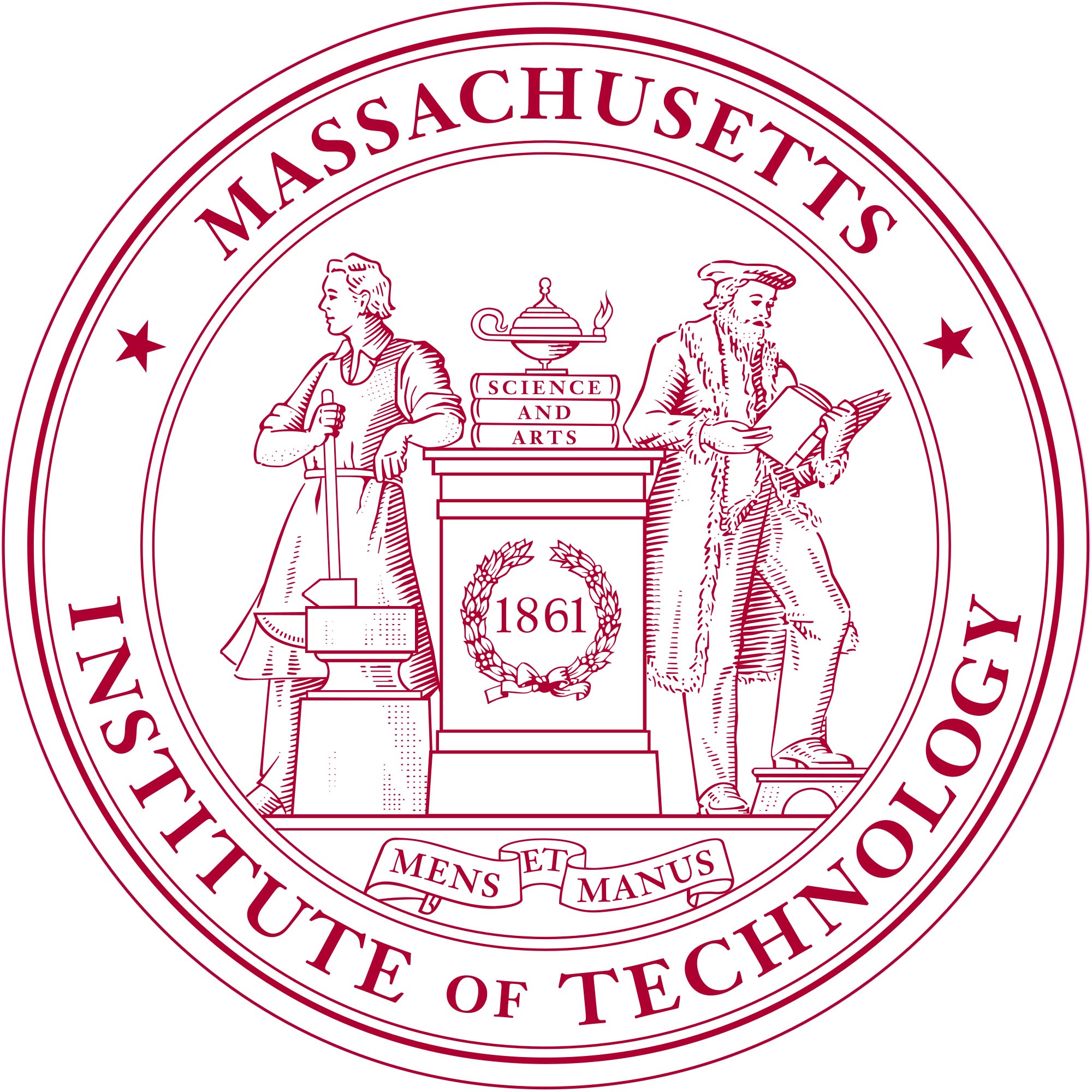} 
\linebreak 
\linebreak 
\linebreak 
\linebreak 
This master's thesis has been examined by a Committee of the Department
of Physics as follows:

\signature{Professor Janet M. Conrad}{Thesis Supervisor \\
    Professor of Physics, MIT}
    
\signature{Professor Scott A. Hughes}{Thesis Co-Supervisor \\
   Professor of Physics, MIT}

\end{large}
\end{titlepage}

%% file: contents.tex
\tableofcontents
\newpage
\listoffigures
\newpage
\listoftables

%% file: DocumentOverview.tex
\chapter{\textsc{Document Overview}}\label{sec::overview}

This document is concerned with describing the physics associated with the CosmicWatch Desktop Muon Detectors. It will begin by outlining the forms of ionizing radiation that can trigger the detector. Then, we describe what effect these forms of ionizing radiation have on the detector and why they are important for our consideration. Reading these chapters will be useful for those interested in developing their physics knowledge or teaching particle/astro physics in the classroom. We then move to the information specific to the Desktop Muon Detectors, by giving an overview of the technology used and the physics behind them.

The latter half of the document is dedicated to illustrating various physical phenomena described in the earlier chapters using measurements made with the detectors. Along the way, we'll briefly mention some fascinating, and perhaps enlightening aspects of physics, which hopefully encourages novices to dive a bit deeper than what we go into here. As new ideas come in, and new data is analysed, we hope this chapter will be continuously updated.

All supplementary material can be found in the GitHub repository located here:

\url{https://github.com/spenceraxani/CosmicWatch-Desktop-Muon-Detector-v2}

There are several documents that have been particularly useful when putting this together. Experience physicists will undoubtedly be familiar with the Particle Data Group's (PDG) summary of cosmic rays~\cite{pdg_cosmic_rays} and energy loss in matter~\cite{pdg_energy_loss}, as it is a fantastic reference. A non-physicist may find Prof. Bruno Rossi's \textit{cosmic rays}\cite{rossi1964cosmic}, a great entry level book that gives an early account of the initial investigation into cosmic-ray physics. It's a great read, although it obviously does not contain information on our modern understanding, as it was published in 1964. A more modern perspective that particularly focuses on the the history can be found in Prof. Michael W. Friedlander's book entitled \textit{A Thin Cosmic Rain}\cite{friedlander2002thin}. The most thorough and comprehensive overview of cosmic-ray physics can be found in Dr. Peter K. F. Grieder's textbook \textit{cosmic rays at Earth}\cite{grieder2001cosmic}; we will be referencing it often. When discussing higher energy cosmic-ray physics, Prof. Thomas K. Gaisser's book entitled \textit{cosmic rays and Particle Physics}\cite{gaisser2016cosmic} is extremely useful. Although more heavily focused on the neutrino side of cosmic rays and particle physics, Prof. Masataka Fukagita's two textbooks \textit{Physics of Neutrinos} and \textit{Physics of Neutrinos and Applications to Astrophysics}\cite{fukugita2013physics}, was found to often have interesting bits of information that we could not find in the other readings. The overall best description that on energy loss in matter comes from \textit{Techniques for Nuclear and Particle Physics}\cite{leo2012techniques} by William R. Leo, which is  a great resource for experimental particle physics. We would also recommend Prof. Claus Grupen's textbook entitled \textit{Particle Detectors}\cite{grupen2008particle}, which is a great reference for describing detection methods in particle physics.

%% file: Sources.tex
\chapter{\textsc{Sources of ionizing radiation}}\label{sec::sources}

We'll begin by discussing the various sources of ionizing radiation that are capable of triggering the Desktop Muon Detectors. There are a few forms of ionizing radiation that are particularly important due to their abundances and their energies. In passing, we will also discuss why we are not sensitive to certain other forms of ionizing radiation, however a description of them is often important for completeness.

\begin{figure}[t]
\begin{center}
\includegraphics[width=0.8\columnwidth]{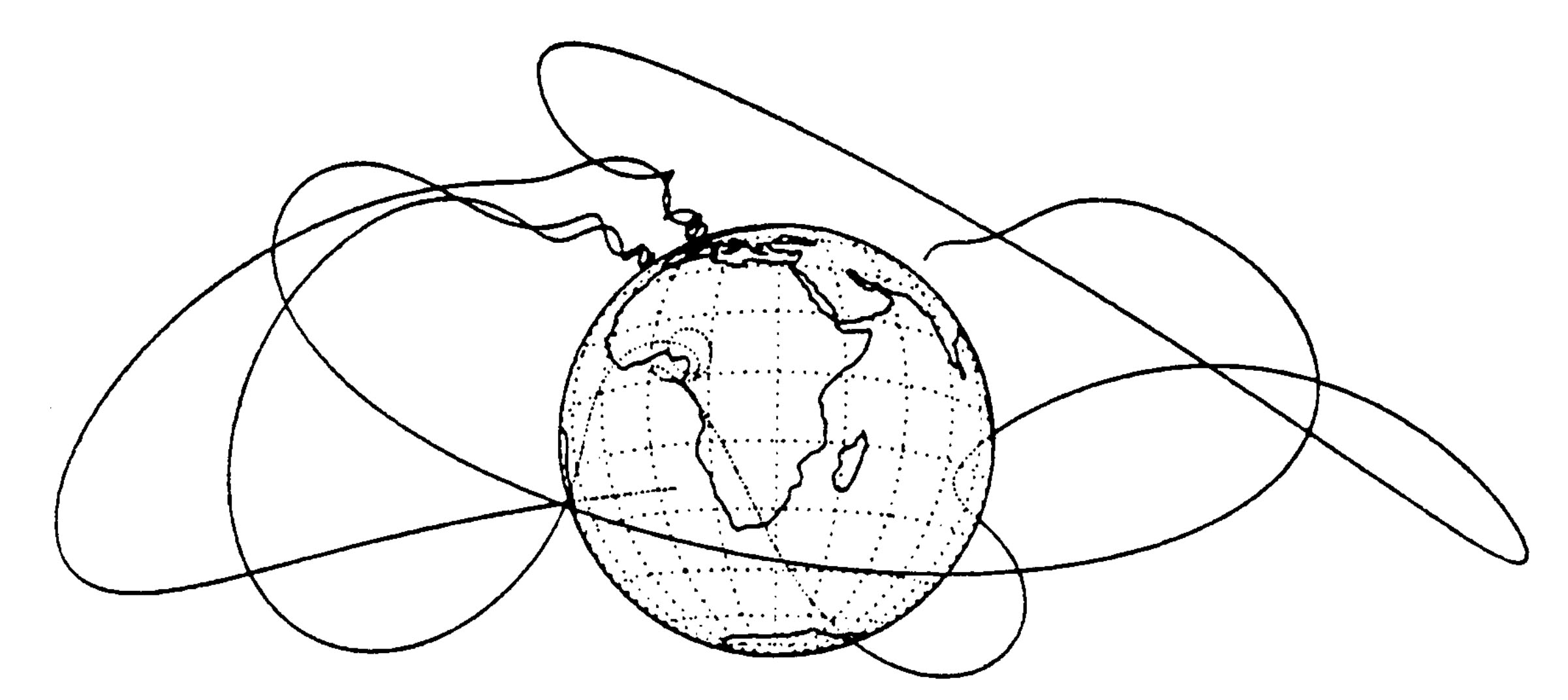}
\caption{Simulated trajectories of low energy cosmic-rays interacting with the geomagnetic field. From Ref.~\cite{friedlander2002thin}}
  \label{fig::LE_cosmic_rays}
 \end{center}
\end{figure}

\section{Cosmic radiation}\label{sec::cosmic_rays}
The Earth is continuously being bombarded by a flux of particles called \textit{cosmic-rays}. Approximately 74\% by mass of this flux comes from ionized hydrogen (free protons), 18\% from helium nuclei (two protons and two neutrons), and the remainder is trace amounts of heavier elements~\cite{pdg_cosmic_rays}. 
The majority of the cosmic-ray flux observed at Earth is relativistic, in that the individual nuclei have kinetic energies greater than their rest mass (E$_k$/m$c^2 > 1$). The lower energy cosmic-rays (GeV-scale) are greatly influenced by the solar wind as well as the geomagnetic field (see for example Fig.~\ref{fig::LE_cosmic_rays}), which limits the flux interacting with the Earth. The high energy flux extends up to 10$^{11}$~GeV at which point the cosmic-ray loses energy from interactions with the cosmic microwave background. This is often referred to as the GZK cutoff\footnote{The highest energy cosmic-ray observed was measured to have approximately 3$\times$10$^{20}$\SI{}{\electronvolt}\cite{bird1993evidence} (48 joules), equivalent to a brick falling on your toe\cite{taubes1993pattern}, all contained in a single proton, and later named the \textit{Oh-My-God Particle}.} (GZK stands for the initials of the three principle authors of the first theoretical prediction: K. Greisen~\cite{greisen1966end},  G. T. Zatsepin, and V. A. Kuzmin~\cite{zatsepin1966upper})). The energy of the cosmic-rays falls off rapidly with energy: below 10$^6$~GeV, the flux falls off as E$^{-2.7}$, and above this it steepens to approximately E$^{-3.1}$\cite{gaisser2016cosmic}. For perspective, the number of \SI{1}{\giga\electronvolt} cosmic-ray protons is 8.1 orders of magnitude higher than that at 1000~GeV (i.e. 2.7$\times$3), or 16.2 orders of magnitude higher\footnote{The steep fall of in energy of the cosmic-ray flux is why we require large detectors to measure the rare high energy events.} than that at 10$^6$~GeV (i.e. 2.7$\times$6).

\begin{figure}[h]
\begin{center}
\includegraphics[width=1.0\columnwidth]{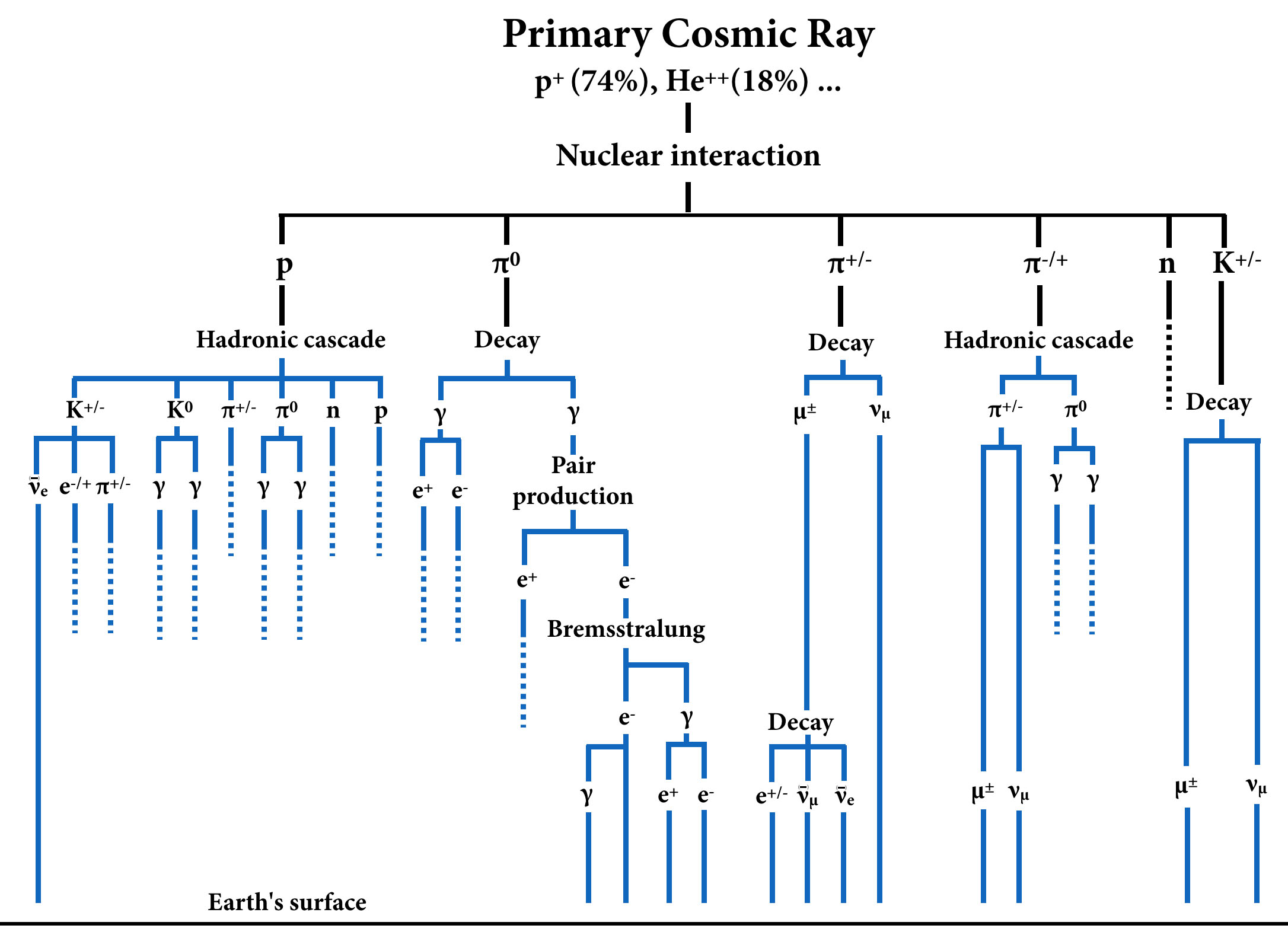}
\caption{An schematic representation of the various decay and interactions chains that result from the interaction of a cosmic-ray in Earth's atmosphere. This is a modified figure from Ref.~\cite{rossi1964cosmic}.}
  \label{fig::cosmic_rays}
 \end{center}
\end{figure}

When a primary cosmic-ray collides with a nucleus in the upper atmosphere (typically with the nucleus of an oxygen or nitrogen molecule), the energies can be large enough to break apart both or either of the primary particle or the target nucleus through a nuclear interaction.
Much of the energy of the collision goes into producing short lived particles known as \textit{mesons}\footnote{Mesons are particles that, unlike the proton and neutron, contain only two quarks: one quark and one anti-quark. The lightest meson is the pion, then the kaon. There exists many other combinations of quark/anti quark pairs, however due to their higher masses, they are not preferentially produced and will not be discussed here. More information on this side of particle physics can be found in Ref.~\cite{halzen1984quarks,griffiths2008introduction}},
 the most common being the $\pi$-meson or pion ($\pi^+$,$\pi^-$,$\pi^0$) and then the K-meson or kaon ($K^+$,$K^-$,$K^0$).  
The charged pions ($\pi\pm$) decay within approximately ten billionths of a second~\cite{pdg} producing a same charge muons and a neutrinos (the charged kaons, $K\pm$, are a bit more complicated, but they also preferentially decay this ways as well, or to pions). The neutral mesons ($\pi^0$, $K^0$), decay approximately one billion times faster (10$^{-17}$s) than the charged mesons, preferentially to gamma rays. Unlike the neutral mesons, the charged mesons are able to travel far enough before decaying to interact with another molecule in the atmosphere. This interaction in turn can be in the form of another nuclear interaction (since mesons are made of quarks, they experience the strong force responsible for the nuclear interaction), much like the original cosmic-ray interaction. The interaction may then produce even more mesons, contributing to the shower of particles induced by the primary interaction. The primary cosmic-rays do not penetrate directly to Earth's surface due to the shielding provided by the atmosphere, however, a small flux of nuclear fragments (such as protons and neutrons) from these interactions can occasionally cascade down and make it to the surface. An illustrative diagram of a cosmic-ray interaction is shown in Fig.~\ref{fig::cosmic_rays}. The first interaction of vertical cosmic-rays takes place at an altitude approximately between 15 and 20~km. cosmic-rays entering at an angle, will interact at higher altitudes due to their path passing through more atmosphere\cite{fukugita2013physics}. 

The high energy photon from the decay of the neutral mesons quickly materialize into an electron-positron pair, also referred to as \textit{pair production}. These electron-positron pairs then radiate high energy photons, which can again materialize into another electron-positron pair. This electromagnetic cascade process continues, dividing the original energy of the photons between the numerous electrons, positrons, and lower energy photons at the end of the cascade. Photons with energies less than \SI{1.022}{\MeV} cannot further pair produce and their interactions will be dominated by Compton scattering and photoelectric absorption. At lower altitudes, there isn't a fresh supply of high energy neutral mesons due to the rapid decrease of nuclear interactions at lower altitudes.

The cosmic-ray muons ($\mu^\pm$) originate from the decay of the charged mesons. A charged pion will decay to a same-sign muon (and muon-neutrino) with a branching fraction of 99.98\%, whereas a charged kaon decays to a muon (and muon-neutrino) 63.5\% of the time~\cite{tanabashi2018aps}. The neutrinos are not electrically charged and only interact through the weak force, therefore they can be fully ignored in this discussion.

\begin{equation}\label{eq::1}
\begin{split}
&\pi^{\pm} \rightarrow \mu^{\pm}+\nu_\mu (\bar{\nu}_\mu )~...~(99.98\%) \\
&K^{\pm} \rightarrow \mu^{\pm}+\nu_\mu (\bar{\nu}_\mu )~...~(63.5\%)
\end{split}
\end{equation}

Approximately 80-90\% of the muon flux in the energy range of interest (GeV to TeV-scale) comes from the decay of pions, and the remainder from kaons\cite{gaisser2002flux}.  
The muons are particularly penetrating, that is, they essentially only lose energy due to ionization as they travel through the atmosphere and other matter and can make it through a large amount of material. This is contrasted with \textit{baryons} (particle comprised of quarks), which also interact through the strong force.
This makes muons the most numerous charged particle showering down on the Earth's surface. 
Muons have a mass of 105.65~MeV and are also unstable particles with a half-life of 2.2~$\times10^{-6}$s. They decay to an electron and two neutrinos. 
\begin{equation}\label{eq::2}
\mu^{\pm} \rightarrow  e^{\pm} +\nu_e(\bar{\nu}_e) +\bar{\nu_\mu}(\nu_\mu)~...~(100.0\%),
\end{equation}
however, again we can ignore the neutrinos. A cosmic-ray muon with an energy greater than 2.4~GeV will be sufficiently relativistic that its half-life, as seen by an observer on the Earth, will be dilated enough that it has a significant chance of reaching the Earth's surface before decaying. In other words, the muon decay length becomes greater than 15~km -- the approximate altitude of the cosmic-ray muon production. Further, a typical muon will lose approximately 2~GeV of energy due to ionization as it passes through the atmosphere on its way to the ground. Combining these two facts with the fact that the geomagnetic field and interstellar solar winds drive back the GeV-scale cosmic-rays, as well as the steeply falling cosmic-ray energy spectrum, we can expect the average muon energy at Earth's surface to be greater than a few GeV. We typically quote the mean cosmic-ray muon energy at Earth's surface to be approximately 4~GeV\cite{pdg_cosmic_rays}. The muons that do not survive the journey to the Earth's surface decay. The resulting electrons (or positrons in the case of a $\mu^+$), also referred to as \textit{Michelle electrons}, contribute to the low energy electromagnetic component from cosmic-ray showers at sea level.

Let us now compare the number of particles showering down on us at sea level. We'll limit ourselves to thinking about particles that are coming from one steradian about the zenith, this can be thought of as circular disk around the vertical (zenith) part of the sky whose area represents approximately 1/6$^{th}$ of the total visible sky, or equivalently, a half-angle of 32$^\circ$ from the zenith. From this direction, we can describe the number of particles passing through a 1$\times$1~cm$^{2}$ horizontal surface per minute (cm$^{-2}$~ min$^{-1}$~sr$^{-1}$) by following the measurements outlined in Ref.~\cite{pdg_cosmic_rays}. We expect approximately 
0.4~cm$^{-2}$ min$^{-1}$sr$^{-1}$  from $\mu^{\pm}$ with energies greater than 1~GeV; 0.2 e$^\pm$ cm$^{-2}$ min$^{-1}$sr$^{-1}$ above 10~MeV, but the flux falls off fast with energy, becoming negligible above 1~GeV;  and 0.0054 cm$^{-2}$ min$^{-1}$sr$^-1$ from protons above 1~GeV; and a charged meson flux above 1~GeV two orders of magnitude lower than that of the proton flux.  This means that the protons and charged mesons are insignificant at sea level, however, there is a significant muonic and low-energy electromagnetic component. In Ref.~\cite{grieder2001cosmic}, the flux is divided into a hard component (essentially fully muons), which can penetrate 15~cm of lead, and soft component (approximately 60-65\% muons and the remainder is electrons, positrons, and photons), which cannot. As we'll soon see, there are a variety of physical phenomena that influence fluxes, but the relative contributions listed here represent a useful approximation.

As we increase in altitude, the relative contribution from the ionizing radiation fluxes change. In particular, we see a larger contribution from both the protons and electromagnetic component, whereas the charged mesons are still sub-dominant. Once we pass the primary interaction region where the primary cosmic-rays are most likely to interact (typically around 15-20~km), the secondary particles produced by the initial nuclear interaction die off and we see a decrease in the ionization radiation flux. The shape of the curve describing the ionizing radiation flux as a function of altitude is called the Pfotzer curve, and where the ionizing particle production reaches a maximum is termed the Regener-Pfotzer maximum~\cite{regener1935vertical}. 
 
\subsection{Flux variations due to solar system properties}\label{sec::variations_space}
There are several properties associated with the interstellar medium that modulate the cosmic-ray flux, and in particular the observable cosmic-ray muon flux. These properties are primarily associated with behaviour of the Earth and Sun's magnetic field. 

\textbf{The latitude effect}: Roughly speaking, the Earth has a magnetic field that behaves similarly to a magnetic dipole orientated from north to south.  The magnetic field points parallel to the surface of the Earth near the equator, and perpendicular to the surface near the poles. Particles traveling towards the Earth will be less deflected (F=q$\vec{v}\times\vec{B}$) near the poles compared to the equator. Low energy charged particles passing through the magnetic field may even become trapped in what's known as the \textit{Van Allen radiation belt}. This presents a low energy cutoff, where the magnetic field is able to deflect protons below approximately 10~GeV near the equator (corresponding to a rigidity of 10~GV) and near 1~GeV at higher latitudes~\cite{allkofer1968muon,allkofer1972sea}.

\textbf{The East-West asymmetry}: The cosmic-ray muon flux is larger looking towards the west compared to the east due to the Earth's magnetic field. This is an effect produced by primary cosmic-ray particles being predominately positively charged. The positively charged muons curve towards the east, meaning that the intensity from the west is stronger. This effect is more evident in the upper atmosphere\cite{grieder2001cosmic}, and obviously a larger effect at the geomagnetic equator than at the poles. 

\textbf{Magnetic anomalies}: There are local geomagnetic field variations, which causes a change in the cosmic-ray intensity. The most prominent being the \textit{South Atlantic Anomaly (SAA)}~\cite{augusto2008muon}, which extends from the east coast Brazil to the west coast of southern Africa (-50.0 to 0.0 geographic latitude and from -90.0 to 40.0 longitude). This is the region where Earth's inner Van Allen radiation belt extends closest to the planet's surface and provides the smallest amount of protection from cosmic-rays. In fact, the increased level of ionizing radiation when passing through the SAA is responsible for radiation damage to electronics on-board Low-Earth Orbit (LEO) satellites\footnote{The International Space Station (ISS) passes intermittently through the SAA~\cite{website:nasa} and has dedicated instrumentation for measuring the increased radiation dose to astronauts~\cite{dachev2006observations}. Astronauts from NASA missions as early as Apollo 11~\cite{pinsky1974light} have also reported seeing flashes of light while being in orbit~\cite{fuglesang2006phosphenes}. These flashes of light are attributed to high-energy particles in the space radiation environment, however many details on the origin are still unknown~\cite{rinaldi2009phosphenes}}.

\textbf{Solar modulation}: The observed cosmic-ray flux at the top of the Earth's atmosphere depends partially on solar activity, which manifests itself as an 11-year cycle\footnote{There is also a 22-year cycle since the solar magnetic dipole flips polarity at every solar maximum, which occurs every 22 years~\cite{grieder2001cosmic}}. Solar winds can drive back low energy cosmic-rays entering the solar sphere and the modulation effect decreases with an increase in energy. According to Ref.~\cite{fukugita2013physics}, the 1~GeV cosmic-ray proton flux is twice as small during maximum solar activity compared to minimum solar activity; similarly there is a 10\% reduction in the 10~GeV cosmic-ray protons during the solar maximum.

\textbf{Solar Flares}: Solar-flares can eject protons with energies up to several GeV, the upper end of which is able to produce muons through nuclear interactions. These events are rare transients, and since the energy is low, it primarily has an effect on the low energy muon flux~\cite{augusto2011muon}.

\subsection{Flux variations due to atmospheric properties}\label{sec::variations_atm}
Similar to the previous subsection, there exists terrestrial phenomena that also modulate the cosmic-ray muon flux.

\textbf{The Cosine Squared Law}: At greater angles from the vertical, cosmic-ray muons must travel through a much larger distance, and therefore amount of matter, to reach a ground-based observer. A cosmic-ray muon traveling vertically downwards may only travel through 15~km of atmosphere, whereas one traveling in the horizontal direction must pass through approximately 500~km of atmosphere. The larger path length means that the muon will lose more total energy due to ionization in the atmosphere and also have a higher probability of decaying before reaching the ground. As a function of zenith angle, the cosmic-ray muon intensity is expected to follow a cosine squared dependence~\cite{pdg_cosmic_rays}. 

\textbf{The atmospheric attenuation}: Recall that the nuclear interactions between the primary cosmic-ray and atmospheric nucleus happen in the upper atmosphere. Therefore, particles reaching sea level must have had sufficient energy to penetrate the remainder of the atmosphere. An increase in atmospheric density (perhaps due to atmospheric pressure changes) will cause secondary particles to lose more energy as they propagate to the Earth's surface. Due to this, the muon rate turns out to be anti-correlated with the pressure (i.e. if the atmospheric pressure increases, the cosmic-ray muon rate decreases). The density of the atmosphere changes with the season and therefore exhibits a time-dependence.  From other measurements, this is expected to be a percent level effect~\cite{savic2015pressure}. 

\textbf{The positive temperature effect}: To produce a muon, we require a charged meson to decay. However, recall that the charged mesons are typically relativistic and have lifetimes on the order of nanoseconds\footnote{For example, a 5-GeV $\pi^\pm$ produced at 15~km will travel approximately 300~m before decaying. This distance is small compared to the interaction path length of approximately 13~km, which means that most charged pions will decay rather than interact. However, at approximately 115~GeV, the pion has an equal probability to interact or decay in the atmosphere.}. This gives the charged mesons sufficient time to potentially interact with another nucleus in the atmosphere rather than decay. As the temperature increases, the atmosphere expands and there are fewer particles to interact with, thus increasing the probability of decaying rather than interacting~\cite{barrett1952interpretation}. 

Rather than correlating this with the ground based pressure (as in the paragraph above), it is more commonly correlated with atmospheric temperature -- taking into account the temperature profile of the atmosphere. This effect is larger at higher energies and therefore is typically measured in laboratories located deep underground where the low energy cosmic-rays have less of an influence~\cite{an2018seasonal,bouchta1999seasonal,ambrosio1997seasonal}. 

\textbf{The negative temperature effect}: As the temperature of the atmosphere increases, the atmosphere expands, moving the muon production region further out. This means that the muon path length increases, which gives them a higher probability of decay prior to making it to the ground. During the winter when the atmosphere is colder, shallower and more dense, cosmic-ray interactions happen closer to the Earth's surface. The charged mesons quickly begin to lose energy and have a less likely chance of decaying into muons.

\section{Radioactive backgrounds}\label{sec::backgrounds}
The previous section described the ionizing radiation that we expect from showers of particles raining down from the upper atmosphere, and the expected phenomena that can modulate this flux. This section will describe ionizing radiation that originates on the surface of Earth and can also influence our measurements; we'll refer to these as the \textit{radioactive backgrounds}. Radioactive backgrounds are sub-divided into primarily three main processes called alpha, beta, and gamma radiation. Radioactivity is a quantum mechanical effect, which is non-deterministic, that is, we cannot predict when a particle will decay, rather we can only assign a probability to it. The energy scale of these processes are relatively low (MeV-scale) compared to the energies associated with the cosmic-rays (GeV and above), but their natural abundance on the surface of the Earth is sufficient that these are typically the dominant source of triggers in the Desktop Muon Detector. 

Alpha decay is the result of an unstable nucleus ejecting a helium nucleus (a bound state of two protons and two neutrons), $(Z, A) \rightarrow (Z- 2,A - 4)+\alpha$. This is a quantum mechanical effect, where a helium state (helium is a very tightly bound state) forms in the nucleus, then quantum tunnels through the nuclear potential barrier, exiting the nucleus. The emitted alpha particle is mono-energetic, and since the helium nucleus has a charge of +2e and mass of approximately 4~GeV (therefore, it moves slow and has a large charge), it will lose energy rapidly in matter.  A 5-MeV alpha particle will have a range of  3.5~cm in air before losing all of its energy, or equivalently, 23 micrometers in aluminum\footnote{The high energy loss rate that alpha radiation makes it useful for cancer therapies. An alpha particle will deposit all of its kinetic energy into a very local space (order micrometers in human tissue), which is capable of destroying cancerous cells.}~\cite{benenson2006handbook}. 

Beta radiation is described as the decay of a neutron to a proton\footnote{More fundamentally, during neutron decays, a down-quark in the neutron converts to a up-quark, emitting a virtual W-Boson $u \rightarrow d + W \rightarrow d + e^{-}+\bar{\nu}_e$. On the macroscopic level, this appears as the transmutation of an atom converting to another atom with an extra proton and one fewer neutron: $A(Z,N)\rightarrow A'(Z+1,N-1) + e^{-}+\bar{\nu}_e$. }: $n \rightarrow p + e^-+\bar{\nu}_e$. The proton remains in the nucleus, while the electron and electron-neutrino are ejected. Since this is a 3-body decay, the electron is not mono-energetic. It is emitted with a continuous energy spectrum whose maximum energy is approximately at the total energy available for the decay (the Q-value). Beta decays typically have energies that can range from tens of keV to a few MeV. 

Gamma radiation is simply a high-energy photon, emitted during the de-excitation of an atomic nucleus. When the nucleus is in an unstable state (for example, maybe the nucleus absorbed a neutron or was left in an excited state after a beta decay), it will de-excite into a lower energy configuration releasing a photon. This is analogous to the de-excitation of an atomic electron, emitting a characteristic mono-energetic photon. Since the energy levels in the nucleus are quantized, gamma ray are also mono-energetic (with a small spread due to nuclear motion). These energy scales are in the 100~keV to MeV-range.

%% file: Interactions.tex
\chapter{\textsc{Particle interactions with matter}}\label{sec::matter}
In order to detect a particle, it must undergo an interaction with the material of a detector. This section will go through the main interactions that transfer energy from the particle to the absorbing material in sufficient detail to understand the data from the Desktop Muon Detector. We'll begin with the muons, protons, pions/kaons, and generically other high-energy heavy charged particles, then move to high-energy electrons (and positron), and finally onto the interactions associated with the radioactive backgrounds. 

\section{High energy heavy charged particles}\label{sec::heavy}
The description below will be useful when thinking about any charged particles with a mass much greater than that of the electron (m $\gg$ m$_e$). This happens to be all charged particles except for the electron and positron (e.g. the muon, the next lightest charge particle, is 206 times more massive than the electron). Unlike the trajectories of heavy charged particles, the trajectories of electrons are not straight lines in a target and special consideration is required. The description below represents an approximation to the underlying processes responsible for energy loss in matter, since the breadth of this subject is far too large to cover in a single document. More information can be found in Ref.~\cite{pdg_energy_loss, grupen2008particle, green2000physics,li2013nu}.

\begin{figure}[t]
\begin{center}
\includegraphics[width=0.80\columnwidth]{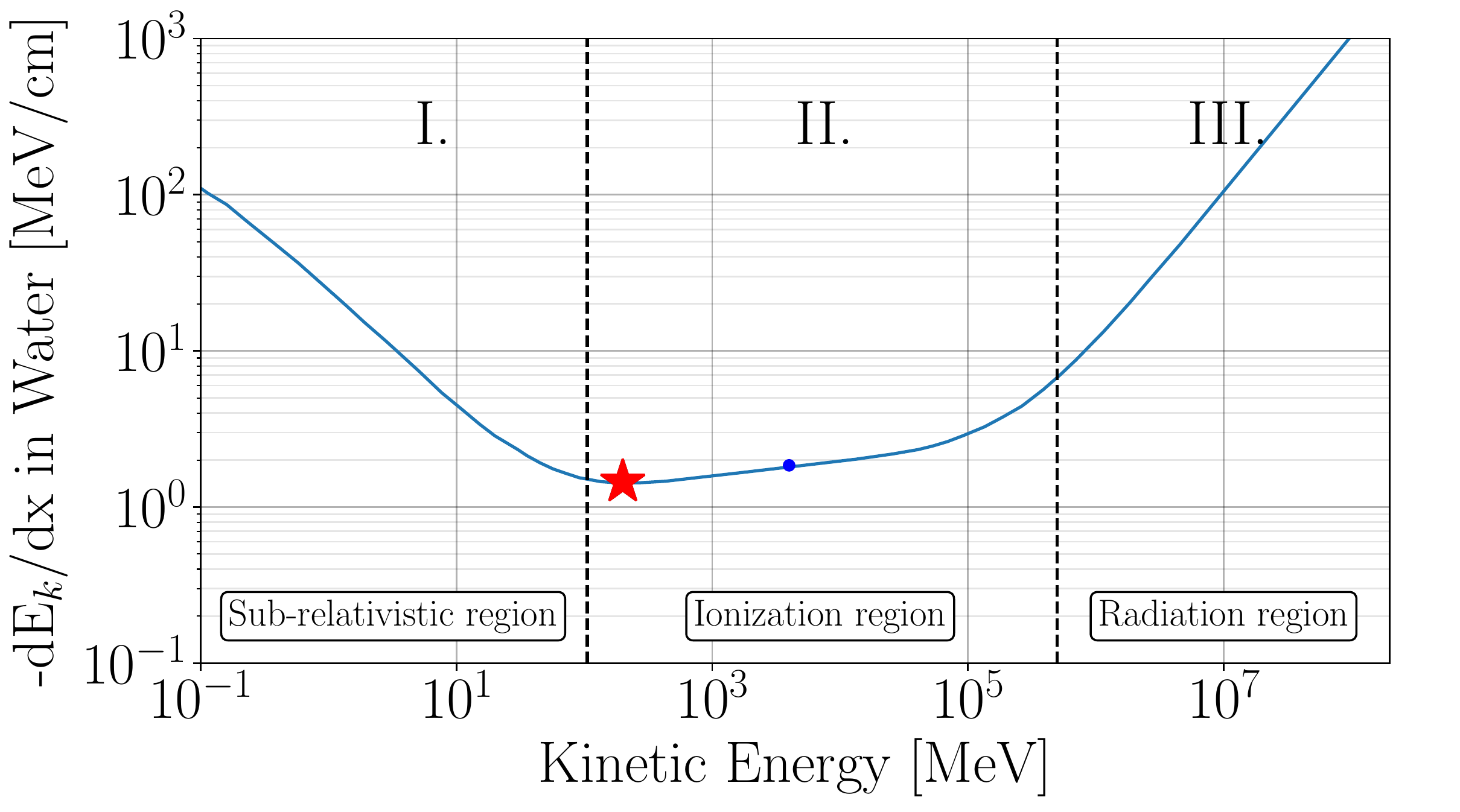}
\caption{The kinetic energy loss per centimeter traveled, of a muon traveling through H$_2$O ($\rho = 1.0 g/cm^3$). Modified from Ref.~\cite{pdg}. The blue dot represents the mean energy of the cosmic-ray muons at sea level (4~GeV), and the red star represents where the particle is minimally ionizing.}
  \label{fig::dedx}
 \end{center}
\end{figure}

The energy loss rate (often called the stopping power), -dE/dx, is a measure of how much energy is loss per unit distance traveled. It is often expressed in units of MeV cm$^2$/g (referred to as \textit{mass stopping power}), where one simply has to multiply by the density (in terms of g/cm$^3$) of the absorbed to get the energy lost per cm. We've taken care of this by expressing the energy loss rate in terms of water ($\rho = 1.0 ~g/cm^3$), which coincidentally has approximately the same density of plastic scintillator. As a particle travels through matter, it can be broken up into three energy ranges that are mass dependent: 

\begin{itemize}
\item I. The sub-relativistic region: (E$_k < $ mc$^2$)
\item II. The ionization region: (E$_k >~$mc$^2$ and  E$~<~$400~GeV~m$^2$/m$_\mu^2$)
\item III. The radiation region (E$~>~$400~GeV~m$^2$/m$_\mu^2$).
\end{itemize}

These three regions are shown in Fig.~\ref{fig::dedx} for a muon traveling in water, however the energy loss can be scaled to another material by simply multiplying by the density (in g/cm$^3$) of the material (for example, the stopping power in lead would be scaled by a factor of 11.34). The energy loss can be scaled to another particle by multiplying by the charge of the particle squared. For example, the energy loss of an alpha particle will be scaled by a factor of 4. This plot is often represented in terms of momentum, but we've scaled it in terms of the kinetic energy of the incident particle to simplify the description. The vertical-dashed lines separate the three regions and can be scaled to another charged particle using the equalities above. For example, a proton will enter the sub-relativistic region at approximately 1~GeV. 

In the \textbf{sub-relativistic region} (Muon: E$_k\approx$ 100~MeV; Proton: E$_k\approx$ 1~GeV), as the particle loses energy, the amount of energy loss per unit distance travels increases. Essentially, this means that once a particle enters this region, it comes quickly to a stop. This phenomena is also known as the \textit{Bragg peak}~\cite{leo2012techniques}. 

In the \textbf{high energy radiation region} (Muon: E$_k\approx$ 400~GeV; Proton: E$_k\approx$ 27~TeV), the energy loss is associated with \textit{bremsstrahlung}, pair production, and nuclear interactions, and scales linearly with energy. The radiation term begins dominating at approximately 400~GeV for muons, however, recall that the cosmic-ray flux falls off fast with energy. The muons that are in this regime represent a small percentage of the flux and lose energy fast.

Finally, the \textbf{ionization region} is where the majority of the cosmic-ray muon flux lies. Recall that the mean muon energy at sea level is 4~GeV (indicated as a blue marker in Fig.~\ref{fig::dedx}). The energy loss in this region is due to ionization (breaking electromagnetic bonds) and excitation (raise the electron to a higher-lying shell within the absorber atom) of the incident particle and described by the \textit{Bethe Bloch formula} (great descriptions of the formula can be found in Refs.~\cite{leo2012techniques,grupen2008particle,pdg}). This region is particularly interesting since the energy loss rate is nearly constant (it actually increases logarithmically), with an average energy loss rate of 2.2~MeV/cm in a density 1.0~g/cm$^3$ material, over many orders of magnitude. The minimum here, indicated by a red star in Fig.~\ref{fig::dedx}, is where the muon is said to be a minimum ionizing particle (MIP) and represents the energy at which the muon is most penetrating. The function is so flat in the region of this minimum (up to approximately 400~GeV), that any particle whose energy is near this red star is often called a MIP.  It is also coincidental that the majority of the cosmic-ray flux falls into this region. This means that if we want to approximate the penetrating depth of a typical cosmic-ray muon, we can simply divide the energy by 2.2~MeV/cm and multiply by the density of the absorber. As an example, a 10~GeV muon will penetrate through approximately 17~m of concrete ($\rho =$ 2.7~g/cm$^3$). What about the other heavy charged particles discussed in this document: the protons, pions, and kaons? These will also lose energy through ionization, but since they are composed of quarks, they can also interact via the strong force. The strong force is responsible for the nuclear collisions that can greatly impact the particle and trajectory. This is what makes the muons so unique -- they do not interact via the strong nuclear force and they are heavy, which allows them to penetrate through materials with minimal loses due to collisions with the electrons in the absorbers and with minimal deflection on their trajectory\footnote{Since high energy muons are able to penetrate very large distances through material, many experiments are buried kilometers underground to shield against them. For example, the neutrino detector Super-Kamiokande, is buried underneath a 1-km mountain in Japan, in order to reduce the muon flux by a factor of 10$^5$ so that they are not swamped when looking for the rare, less energetic interactions from neutrino. For similar reasons, the IceCube neutrino detector is buried under 1.4~km of ice in the Antarctic glacier at the South Pole.}.

\section{High energy electrons/positrons and photons}\label{sec::HE_electrons}
\begin{wrapfigure}{R}{0.4\textwidth}
\centering
\includegraphics[width=.4\textwidth]{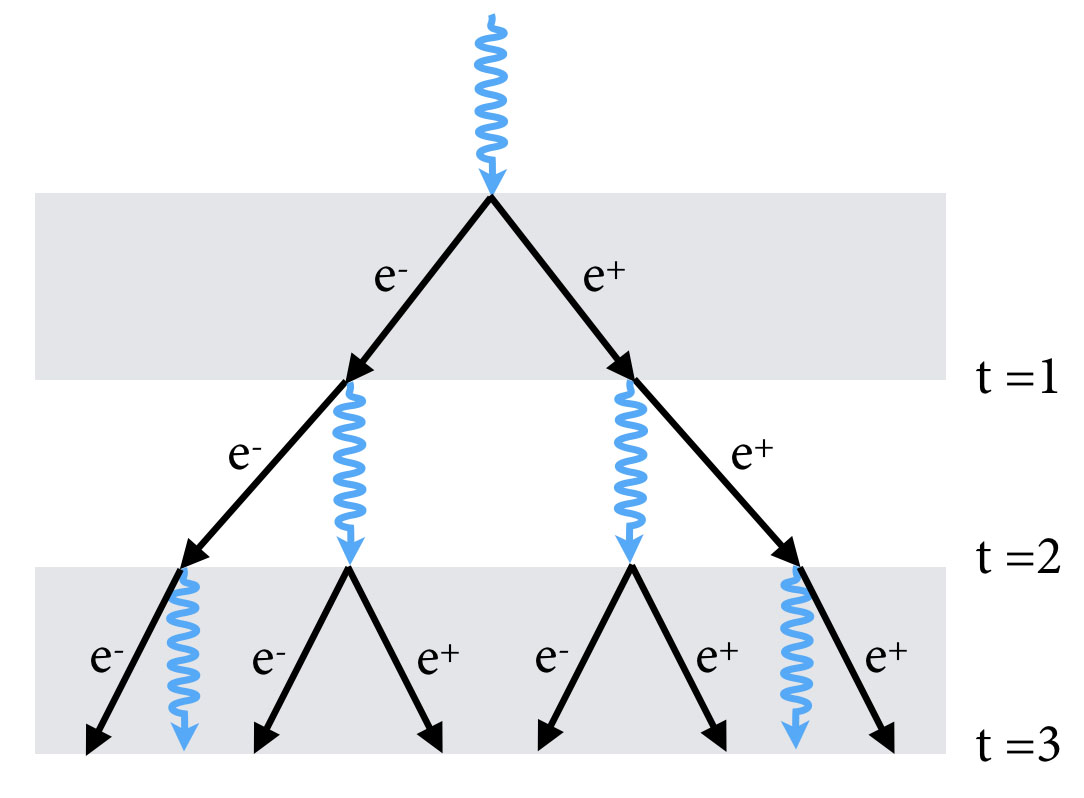}
\caption{The Heitler Model for electromagnetic cascades.}
\label{fig::pair}
\end{wrapfigure}
As described in Section~\ref{sec::cosmic_rays}, there is a non-negligible source of electrons/positrons with energies below 1~GeV showering down onto the Earth's surface. This section will describe the energy loss in terms of high energy electrons, but the description is valid for positrons as well. It turns out that for an electron above a few tens of MeV, the energy loss will be dominated by radiation losses, predominantly \textit{bremsstrahlung radiation} ("braking radiation" in German). Bremsstrahlung radiation is the emission of photons produced by a particle accelerating and decelerating as it passes near the electric field of the nucleus of the material. 

A bremsstrahlung photon with sufficient energy can pair produce an electron and positron, which will subsequently radiate other bremsstrahlung photons, thereby creating a cascade of electrons, positrons, and photons. This process dominates the energy loss of the electron until the energy drops below a few tens of MeV (typically referred to as the critical energy). The radiation energy loss rate actually scales with energy, so in the region above the critical energy, dE/dx is proportional to energy. This means that a 20~GeV electron will start off by losing 1000 times more energy per centimeter traveled then a 20~MeV electron in the form of bremsstrahlung radiation. This causes the high-energy particles to lose their energy rapidly as a function of distance. 

A \textit{radiation length}, $X_0$, is defined is the average thickness of a material that reduces the mean energy of the electron or positron by a factor 1/e (Euler's number = 2.71828) due to bremsstrahlung radiation. This means that an electron will lose a factor of e$^{-t}$ energy after traveling through $t$ radiation lengths. For example, after four radiation lengths, a 1~GeV electron will end up with approximately 20~MeV. A simplified model known as the \textit{Heitler Model for electromagnetic cascades}, approximates this saying that one electron-positron pair will be created per radiation length, each of which will receive half of the energy of the photon that produced them. After t radiation lengths, the cascade will contain 2$^t$ particles (electron, positrons, and photons) each with an average energy of approximately E = E$_0$/2$^t$~\cite{martin2006nuclear}. We've illustrated this in Fig.~\ref{fig::pair}. 

A useful list of radiation lengths for various materials can be found in Table~\ref{table::radiation}. This table includes the information regarding pertinent materials used in the Desktop Muon Detectors or during some measurements below.  

\begin{table}[h]
\footnotesize
\begin{center}
\begin{tabular}{|| l l  l l ||}
\hline
\textbf{Material}   & Density [g/cm$^3$] &  Radiation length [cm] &  Critical Energy [MeV] \\ [0.5ex]
\hline
\hline
Water (H$_2$O)		       & 1.00 	 & 36.1 & 92\\
\hline
Lead (Pb)		   &11.35	 & 0.56 & 9.51\\
\hline
Concrete			   & 2.5 & 10.7&\\
\hline
Air at STP$^*$  &1.2931	& 30420& 102\\
\hline
Scintillator (Polystyrene)			&1.032 & 42.4& 109 \\
\hline
Aluminium (Al)			& 2.70 & 8.9   & 51.0 \\
\hline
\end{tabular}
\caption{A table of materials that are mentioned in this document and their corresponding radiation length. $^*$STP indicates that the air is at the standard temperature and pressure: 20$^\circ$C and 101.325~kPa. The data was collected from Ref.~\cite{pdg_cosmic_rays,leo2012techniques}.}
\label{table::radiation}
\end{center}
\end{table}

\section{Low energy electrons/positrons}\label{sec::LE_electrons}
Electrons and positrons are the lightest charged particles. The qualitative behavior for electron scattering is different from the high-energy particles in two ways. First, the energy loss by electrons fluctuates much more than for heavy particles. For example, the maximum transferable kinetic energy of an electron from a 4~GeV electron is the full 4~GeV (since they have the same mass), whereas a muon with the same energy has a maximum transferable energy of approximately 1~GeV. Further, because of its small mass, electrons are particularly susceptible to large angle deflections by scattering off a nucleus. This probability is so high, in fact, that multiply scattered electrons may be turned around in direction altogether, defined as \textit{backscattering}.  The backscattering probability is higher at lower energies, and if backscattered, the electrons do not deposit all their energy in the absorbing medium. A 1~MeV electron has approximately a 10\% chance of backscattering off of a thick slab of aluminium, and a 50\% chance of backscattering off a slab of gold~\cite{tabata1971empirical}.

The previous section described electrons and positrons with energy above the critical energy of a material (typically tens of MeV), where their energy loss is completely dominated by bremsstrahlung radiation. At lower energies, the electrons-positrons can in-elastically interact through Coulomb collisions with atomic electrons to lose energy~\cite{berger1982stopping}. This leads to ionization and excitation like the heavier particles. At lower energies still, MeV-scale, the electrons (positrons) also exhibit \textit{M{\o}ller} (\textit{Bhabha}) scattering.

\section{Low energy gamma rays}\label{sec::LE_gammas}
\begin{figure}[!h]
\begin{center}
\includegraphics[width=0.9\columnwidth]{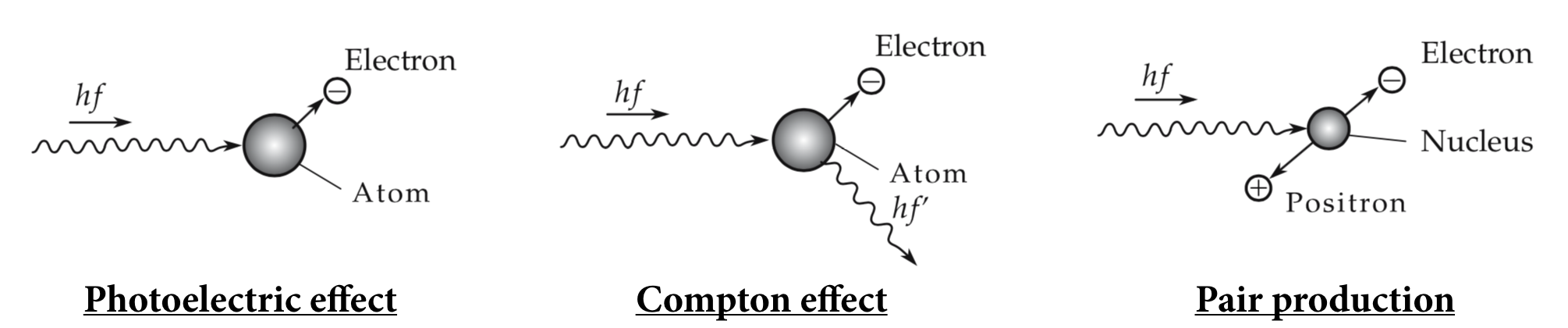}
\caption{Modified from Ref.~\cite{benenson2006handbook}}
  \label{fig::gamma_proc}
 \end{center}
\end{figure}

Gamma-rays interact slightly differently from the charged particles due to their lack of electric charge. The three main interactions of gamma rays (and X-rays) are shown in Fig.~\ref{fig::gamma_proc}.

\begin{wrapfigure}{R}{0.5\textwidth}
\centering
\includegraphics[width=0.5\textwidth]{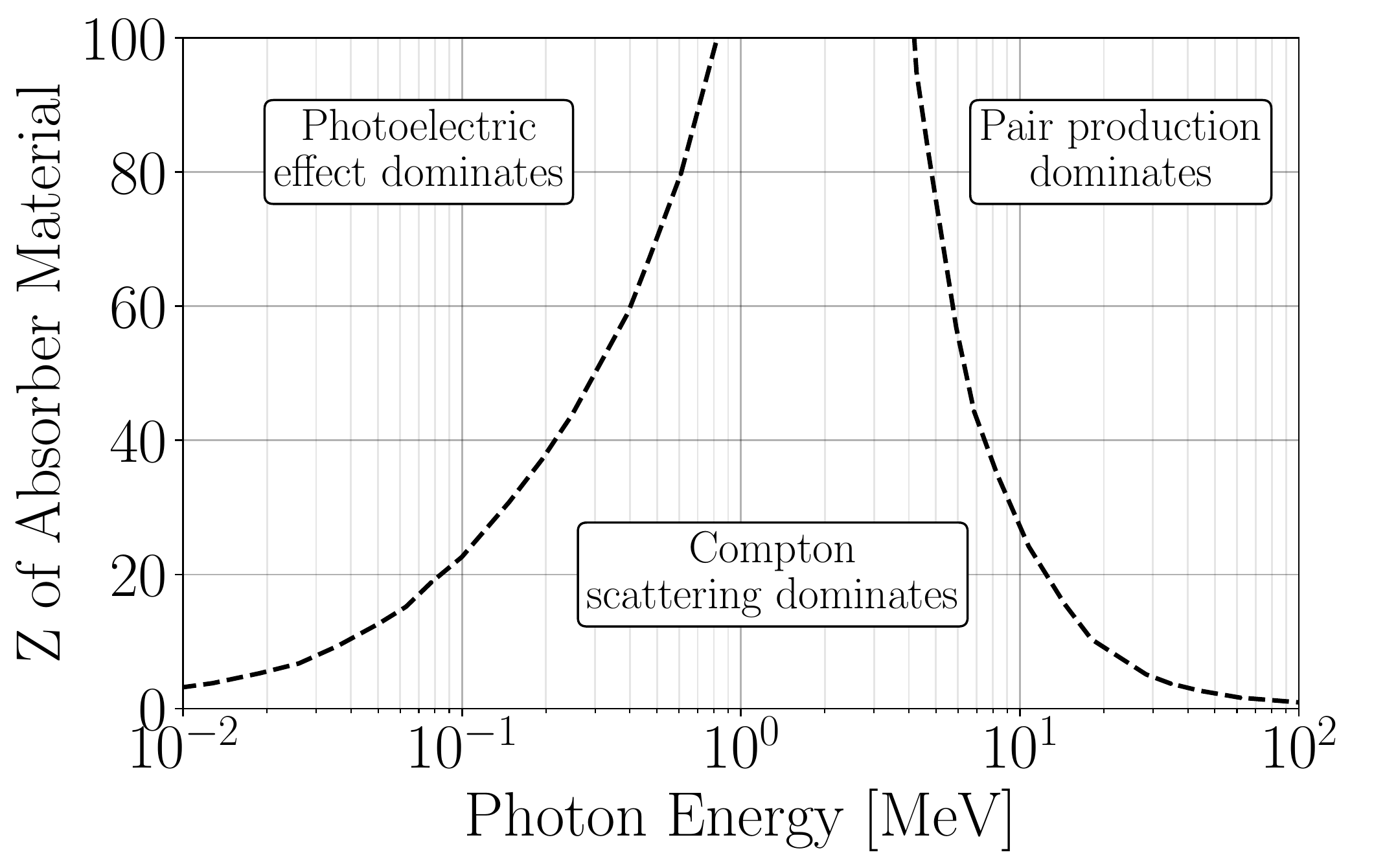}
\caption{Dominant region for the three gamma ray interactions as a function of the photon energy and the target charge number Z. Modified from Ref.~\cite{grupen2008particle}}
\label{fig::interactions}
\end{wrapfigure}

An atomic electron can fully absorb the energy of a gamma ray\footnote{In order to conserve momentum, the photoelectric effect cannot occur on a free electron, it requires a nucleus to absorb part of the recoil.}, resulting in an electron with the energy of the initial gamma ray (MeV-scale) minus the binding energy of the atomic electron (eV-scale). This process is known as the \textit{photoelectric effect}. As shown in Fig.~\ref{fig::interactions}, the photoelectric effect dominates for low-energy gamma-rays with a moderate-to-high density absorber. 

\textit{Compton scattering} results in the partial transfer of energy from the incident gamma-ray to an atomic electron, resulting in the electron being bumped into a higher energy level or being ionized. The gamma-ray may change direction, exiting the material that contained the electron, or perhaps it scatters again off another electron. Compton scattering dominates at low energy for very low-Z materials, and the probability of scattering will be proportional to the electron density and therefore proportional to the proton number of the material.

At energies above 1.022~MeV, electron-positron pair production plays a role. Pair production follows the same description as that found in the description on high-energy electrons (Sec.~\ref{sec::HE_electrons}). The only difference being that the chain begins with a photon rather than an electron. 

Beyond this, there are second order effects, such as \textit{Rayleigh scattering} where the photon wavelength is large enough that it coherently scatters off of the entire atom, and photonuclear interactions at higher energies that break up the nucleus~\cite{pdg}.

\section{Neutrons}\label{sec::neutrons}
Like the photon, the neutron is not electrically charged, therefore it is not subject to Coulomb interactions with the electrons and nuclei. Rather, it interacts through the strong force with nuclei. Due to the short range nature of the strong force, these interactions are comparatively rare (the neutron needs to get close to the nucleus in order to interact). Several interactions that may occur are~\cite{leo2012techniques}:
\begin{enumerate}
\item Elastic scattering off a nucleus
\item Inelastic scattering off nucleus that leaves the nucleus in an excited stated that may decay emitting a gamma ray. To excite the nucleus the neutron must transfer MeV-scale energies.
\item Neutron capture. At low energies, the neutron might be captured by the nucleus, emitting a gamma ray.
\item Fission
\item Hadronic shower, particularly at high energies ($>$~100~MeV).
\end{enumerate} 

High energy neutrons produced in the primary cosmic-ray interaction, will often collide with another nucleus creating a similar interaction as to the primary cosmic-ray.

%% file: DetectionMethods.tex
\chapter{\textsc{Detection methods}}\label{sec::detection}

\section{Single photon detections: photomultipliers}\label{sec::sipm}
One of the most common instruments used by particle physicists are photomultipliers. Photomultipliers are devices capable of producing a measurable electrical signal from the interaction of a single photon. Photon detection enables us to extract information  related to the incident particle by measuring the photon emission of a particle as it loses energy in a material. The most prolific single photon sensing device is a photomultiplier tubes (PMT), which offers a large photo-sensitive area of coverage at modest cost. They are however bulky and require high-voltage. Other common technologies, such as the Avalanche Photodiode (APD) and P-type and N-type semiconductor photodiode (PIN photodiode), both have their benefits and drawbacks. Recent improvements in the manufacturing of silicon chips have made it possible to make a new type of photon detector called a \textit{silicon photomultiplier}, or SiPMs (often abbreviated as SPM). SiPMs have many advantages over PMTs, such as being able to operate at low voltages, being insensitive to magnetic fields, being robust, and having a compact form factor. They are also single photon detectors and have a peak responsivity near the peak emission from typical scintillator. This modern technology is what we use in the Desktop Muon Detectors.

\begin{figure}[h]
\begin{center}
\includegraphics[width=.9\columnwidth]{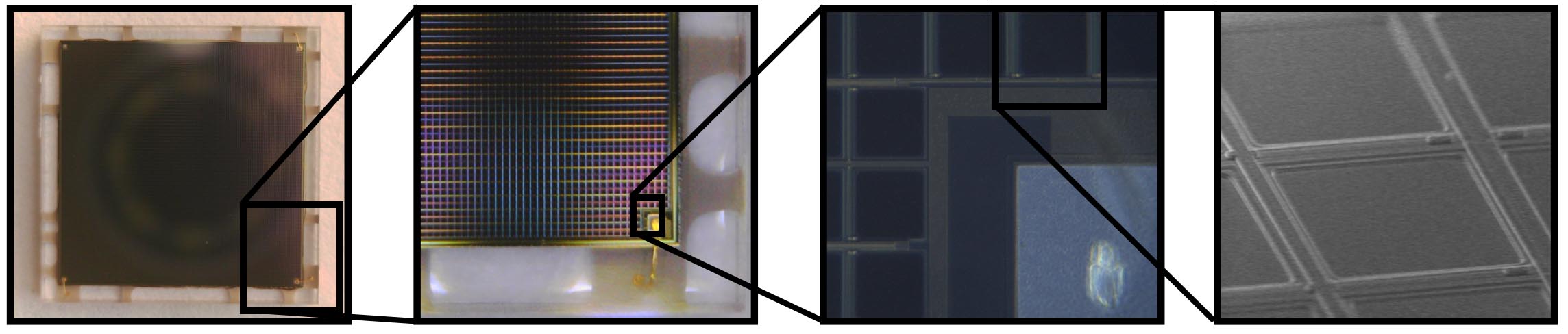}
\caption{An image of a On Semiconductor MicroFC 60035 C-Series SiPM~\cite{sipms}. The SiPM has a length and width of  7.00$\pm$0.05~mm and a thickness of 0.65$\pm$0.05 ~mm. Each one of the tiled squares represents a single microcell, each of which operate independently in Geiger mode. }
  \label{fig::sipm}
 \end{center}
\end{figure}

SiPMs are constructed out of densely arranged \textit{microcells} (see Fig.~\ref{fig::sipm}), each of which is a separate P-type and N-type semiconductor junction (P-N junction). When a P-N junction is first formed, the free electrons from the N-type semiconductor diffuse towards the P-type semiconductor and annihilate (similarly, the holes from the P-type diffuse into the N-type). Upon annihilation, the boundary region between the P and N-type semiconductors becomes an insulator known as the \textit{depletion region}. When a photon travels through the depletion region, and deposits sufficient energy to a bound electron, the electron can be transported to the conduction band, thereby producing an electron-hole pair. If a potential difference is applied between the P-N junction, the electron will gain energy and at a certain point collide with other electrons also transporting them into the conduction band. When the potential difference is sufficiently high ($>$~5$\times$10$^5$V/cm), an electron avalanche (or cascade) can occurs (a Geiger discharge) in which a single electron turns into a current of order millions of electrons. Once the flow of electrons is initiated, the silicon becomes conductive at which point a quenching resistor lowers the potential difference across the PN junction sufficiently to stop the electron cascade. In this way, each microcell acts as a photon triggered switch, which lets a small amount of current to briefly flow if it was struck by a photon.  The sum of the total current flow is proportional to how many microcells were triggered and therefore is proportional to the incident photon-flux (when the number of triggering photon $ll$ number of microcells).

The Desktop Muon Detector employs the On Semiconductor MicroFC 60035 C-Series~\cite{sipms}. These are most sensitive in the 450~nm range~\cite{website:sensl}, which is a deep blue to purple. If a photon wavelength is too large ($>$1000~nm), the absorption length in the silicon is also too large and the necessary size of the SiPM would be too bulky. If the photon wavelength is too short, it will not penetrate into the sensitive region region of the SiPM, which is required for the detection. 

The applied potential difference is termed "bias voltage," and dictates both how large of a region is able to produce the avalanche (the depletion region), and the amount of energy gained by the electron-hole pair. The "breakdown voltage" defines the voltage at which the voltage gradient in the depletion region is large enough to create a Geiger discharge. This is typically between 24.2 and 24.7~V for the C-series SiPMs. If the bias voltage is increased beyond the breakdown voltage, the microcells will still operate in Geiger mode, however the electron cascade in the P-N junction will carry more energy, thus increasing the charge output (or gain) linearly. The difference in the bias voltage to the breakdown voltage is termed \textit{over-voltage}. An over-voltage between 1.0 and 5.0~V is recommended. The Desktop Muon Detectors operate at an over-voltage of 5.0~~V, which corresponds to a gain of roughly 5$\times$10$^6$~\cite{sipms}.

Thermal fluctuations can produce electron-hole pairs, which mimic single photon events. For our SiPMs, this occurs at rate of approximately 100~kHz per mm$^2$, or at several MHz for the full SiPM. This can be undesirable for many applications that rely on distinguishing between small numbers of photons. The breakdown voltage required to initiate the electron cascade is temperature dependent; a lower temperature has a lower breakdown voltage.

\section{Scintillators}\label{sec::scintillators}

Scintillators are simply a material that absorbs energy (through Coulomb interactions) and re-emits that energy in the form of electromagnetic radiation (scintillation light). Scintillators can come in different forms: for example, the scintillator could be grown as a crystal (referred to as an inorganic scintillator) with an added dopant, or the scintillator could have a fluorescing material embedded in a plastic (such as polystyrene or acrylic) or  mixed into a liquid (like toluene or mineral oil) -- these would be examples of organic scintillators. Inorganic scintillators are typically more expensive, however they can also have a higher density and emit more photons per unit energy deposited. This makes them more useful for calorimetry. Organic scintillators on the other hand are typically cheaper since the florescent material is suspended in a common, often low density material like a plastic, which enables them to be easily manufactured.

\begin{wrapfigure}{L}{0.3\textwidth}
\centering
\includegraphics[width=0.3\textwidth]{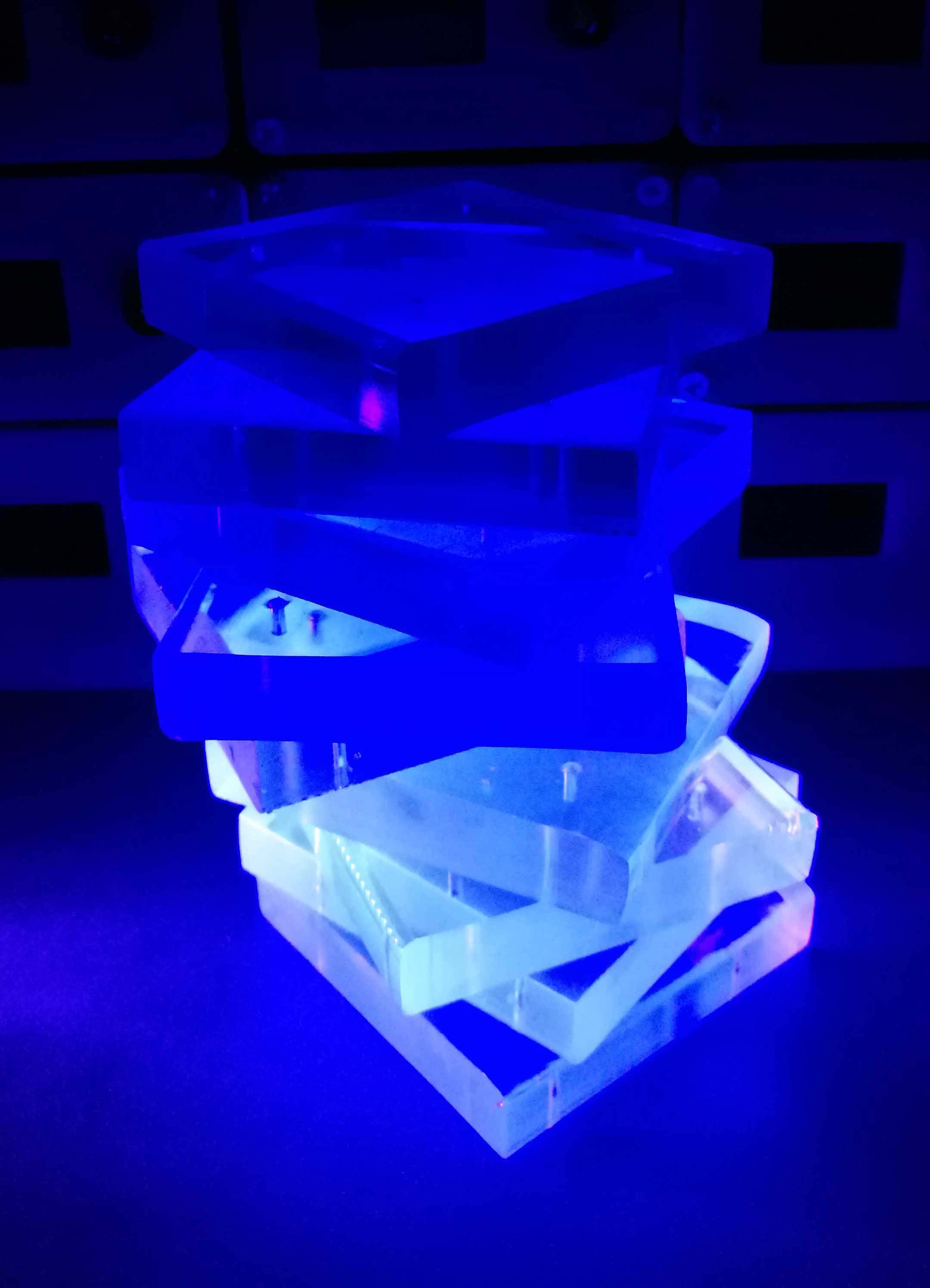}
\caption{A UV flash light illuminating scintillator.}
\label{fig::uv}
\end{wrapfigure}

Scintillators are particularly useful since they emit light proportional to the energy deposited in the material, therefore, one will often see a metric of the quality of the scintillator expressing how many photons are emitted per absorbed MeV of energy (often called the scintillator efficiency). A common organic scinitllator may have an efficiency of 10,000 photons/MeV. Another important quantity associated with scintillators photon emission profile of the scintillator, which determines what photon wavelengths are emitted after de-excitation. Scintillators must also be transparent to the scintillation light so that it can propagate to the photon detector. Plastic scintillators may have attenuation lengths on the order of 0.3 meters to 3 meters~\cite{pla2005extruded, eljen}, whereas liquid scintillators like Linear Alkyl Benzene (LAB) can have attenuation lengths of up to 25 meters~\cite{yang2017light}. Scintillators also have a very fast response and recovery (excitation and de-excitation of the foreclosing molecules), which happens on the order of nanoseconds for organic scintillators and hundreds of nano-seconds for inorganic scintillators.

The CosmicWatch Desktop Muon Detector was designed using an organic plastic scintillator, which consists of a polystyrene base (essentially just an inexpensive transparent plastic) mixed with a primary dopant of 1\% by-weight of POP (2,5-diphenyloxazole) and 0.03\% secondary dopant POPOP (1,4-bis[2-(5-phenyloxazolyl)]benzene)~\cite{beznosko2004fnal}\footnote{Interestingly, PPO was one of the earliest compounds to be investigated as a scintilator solute by Hayes et. al (see Ref.~\cite{horrocks2012applications})from 1953-1958 and is still one of the most widely used}. This plastic scintillator does not emit below 400~nm, and has a maximum emission around 420~nm (deep-purple light). This particular scintillator was developed by FermiLab for the MiNOS~\cite{ambats1998minos}/MINER$\nu$A~\cite{aliaga2014design} experiments. We'll focus on a description of organic scintillators below, but a good description of inorganic scintillators can be found in Ref.~\cite{horrocks2012applications}.

The plastic scintillator requires three components: 
\begin{itemize}
\item A transparent (in the visible spectrum) base that is used to suspend the florescent material. This can be some sort of plastic (like polystyrene) or transparent liquid like mineral oil.
\item A primary fluorescing agent that is excited by the energy transfer from the incident charged particle. The de-excitation of the primary fluorescent material releases ultra-violet light. Ultra-violet light will not travel far in the base (order mm), before being absorbed. 
\item A secondary fluorescent agent that absorbs the UV light and converts it to the visible spectrum. The visible light then travels through the scintillator, internally reflecting off the walls until it is absorbed. Some of the visible light will hopefully strike the photon sensor that is coupled to the scintillator.
\end{itemize}

\begin{figure}[t]
\begin{center}
\includegraphics[width=.8\columnwidth]{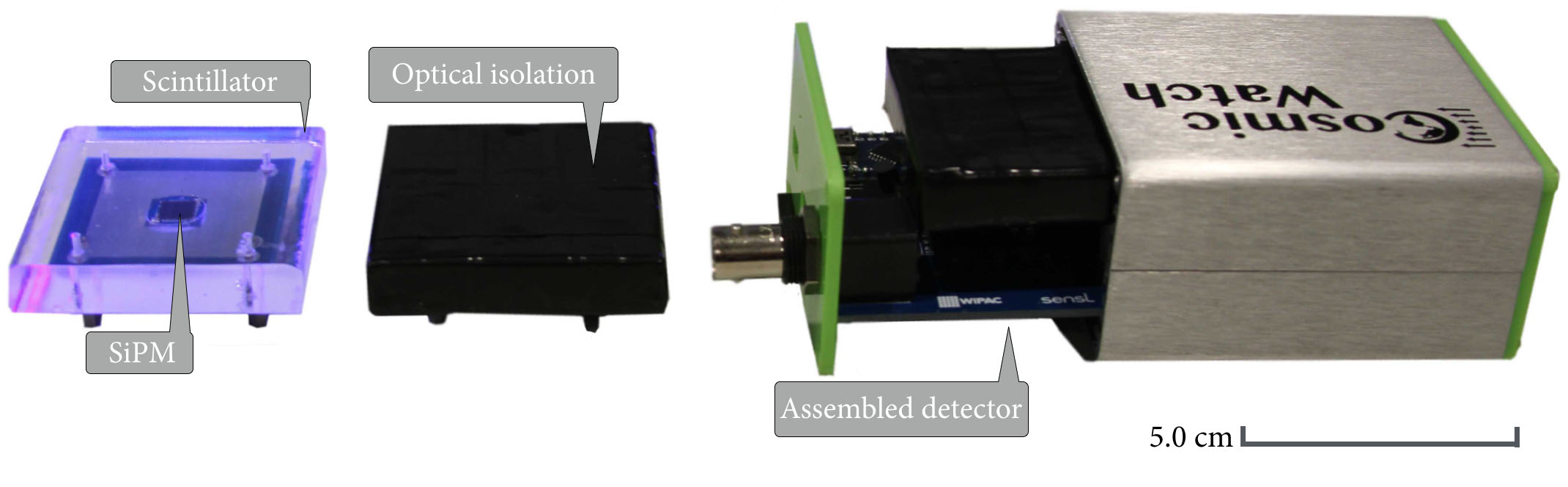}
\caption{The components of the Desktop Muon Detector. Modified from Ref.~\cite{axani2018cosmicwatch}.}
  \label{fig::det}
 \end{center}
\end{figure}

A simple illustration of the UV conversion is shown in Fig.~\ref{fig::uv}. Here, we have illuminated several pieces of scintillator with a UV flash light. The UV is absorbed by the secondary fluorescent agent and re-emitted as a deep blue/purple light. Polystyrene-based scintillator has a density of roughly 1.032 g/cm$^3$ (similar to water) and a refractive index at standard atmosphere pressure of n~=~1.581~\cite{grupen2008particle}.

The Desktop Muon Detector couples a SiPM (see Sec.~\ref{sec::sipm}) to a slab of scintillator via optical gel, which reduces the probability of a photon being reflected at the interface by matching the index of refraction from the scintillator to the housing of the SiPM, eliminating the n~=~1.0 air gap. We also wrap the remaining surface in aluminum foil to reflect photons that escape the scintillator. It is then wrapped in 2-3 layers of black electrical tape to make the whole thing light-tight. The assembly of a detector is shown in Fig.~\ref{fig::det}.

%% file: CosmicWatch.tex
\chapter{\textsc{The CosmicWatch program}}\label{sec::program}
Many excellent programs exist to provide physics outreach experiences to students, and they each tend to offer very different opportunities and unique ideas. Recently, there has been quite a bit of interest in both the US (such as the Distributed Electronic Cosmic-Ray Observatory, DECO~\cite{DECO}) and Poland-based (the Cosmic-Ray Extremely Distributed Observatory, CREDO~\cite{CREDO}) cosmic-ray programs that employs the CCD on a users smart phones to detect the interactions with cosmic-ray muons. We have also spoken with many people who work on the QuarkNet program~\cite{bardeen2018quarknet}, which brings elements particle physics to high school teachers using lab-grade muon detectors. 

Our approach is slightly different than those mentioned above. The Desktop Muon Detector was designed to be as portable and self-sufficient as possible, while giving the individual student the ability to record and analyze their own data. The full detector is open source, and we encouraged our students/teachers to go through the full build process themselves: acquiring materials, populating the circuit boards, troubleshooting the electrons, and finally testing the actual detector. This gives our students experience in the machine and electrical shop and introduces them to circuit design. Once the detector is built and functioning properly, students can either use our tools for analyzing the data, or they build their own programs for extracting physics from the data. The data is then used as an educational tool to illustrate various particle/astro physics phenomena. 

\section{The Desktop Muon Detector}\label{sec::DMD}

\begin{figure} [t]
\begin{center}
\includegraphics[width=1.\columnwidth]{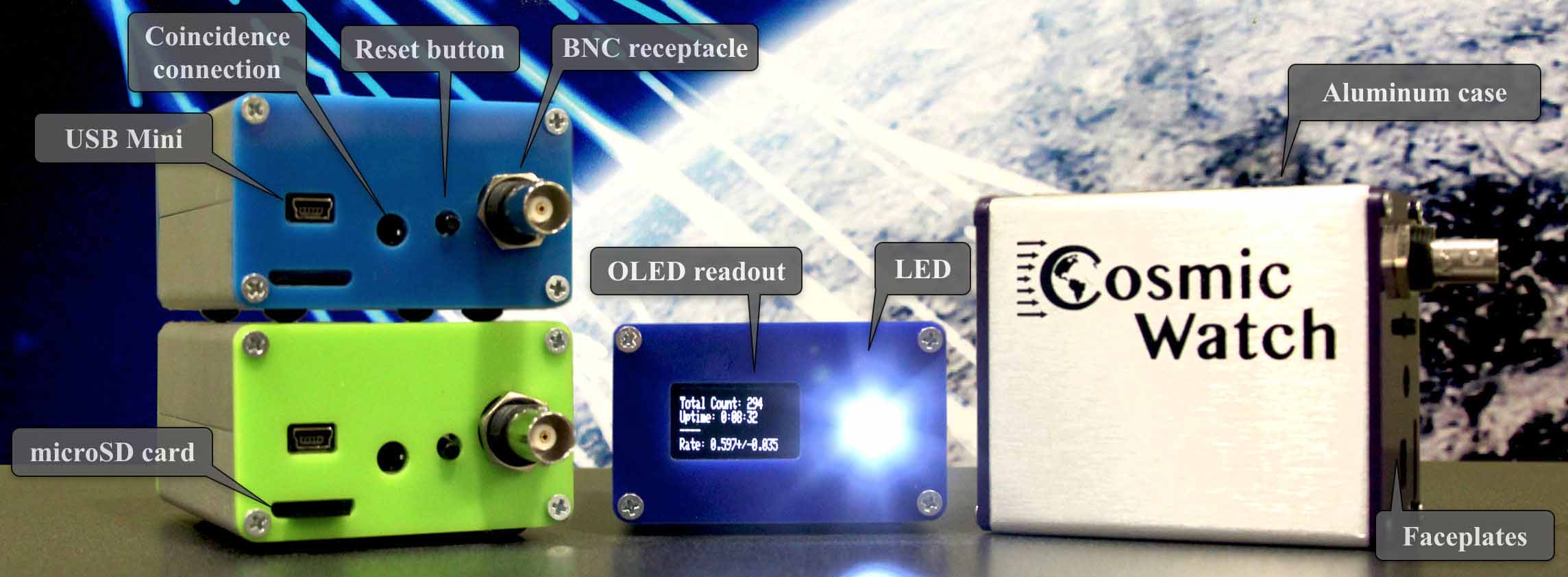}
  \caption{An arrray of Desktop Muon Detectors. From Ref.~\cite{axani2018cosmicwatch}.}
    \label{fig::array}
 \end{center}
\end{figure}

Particle detectors in general aim to identify each particle, measure the energy and momentum, as well as localize the particle with sufficient resolution in space and time. A single detector is never optimal for all of these requirements. The CosmicWatch Desktop Muon Detector aims to be a general purpose detector that can collect granular information on these requirements such that it has access to a plethora of physics. 

As described in Ref.~\cite{axani2018cosmicwatch}, the CosmicWatch Desktop Muon Detector consists of a 5$\times$5$\times$1~cm$^3$ slab of extruded plastic scintillator instrumented with a silicon photomultiplier (SiPM). When a charged particle passes through the scintillator and deposits energy, some of that energy is re-emitted isotropically along the particle track in the form of photons. Photons incident on the photosensitive area of the SiPM can induce a Geiger discharge in the SiPM microcells. When the microcells discharge, they produce a measurable current. A single photon can trigger a single microcell (neglecting second-order effects), whereas multiple photons may trigger multiple cells. The produced current is sent through a custom designed printed circuit board (PCB), which amplifies and shapes the signal such that it is suitable to be measured by an inexpensive micro-controller (in our case, an Arduino Nano~\cite{website:arduino}). The Arduino Nano measures the event time stamp and peak value on the analog-to-digital converter (ADC). The measured peak ADC value can then be converted back into a SiPM peak voltage, which is proportional to the number of photons incident on the SiPM. If the measured ADC value is above a software defined threshold, the micro-controller records the event data either to a microSD card or directly to a computer through a USB connection. The CosmicWatch website also provides an online application to plot data in real-time or to record data. For each event, the detector records the event number, event time, average measured 10-bit ADC value, calculated SiPM peak voltage (which is proportional to the SiPM pulse charge), total dead-time, and temperature. When plugged directly into a computer, the data can be recorded through a python-based program that also includes the computer time and date of each event.

An array of complete detectors is shown in Fig.~\ref{fig::array}. The front of the detector includes a built-in 0.96" OLED screen, which displays the count number since the last reset, total uptime, count rate (compensated for the detector dead-time), and an indication bar whose length is proportional to the calculated SiPM peak voltage of the last triggered event.  Multiple detectors can also be linked together using a 3.5~mm male-to-male audio cable to make a \textit{coincidence measurement} (see Sec.~\ref{sec::coincident}). The detector was measured to draw 0.27~W and can be powered through any USB connection (provided it supplies greater than 4.5~V). This includes the USB port on a computer, USB power bank, or power outlet USB. The total mass of the detector, including the aluminum case, is 178~g and the outer dimensions of the detector are 66.4~mm$\times$101.6~mm$\times$39.9~mm (including the protruding BNC receptacle and LED holder). Excluding the aluminum enclosure and end-plates, the detector has a mass of 76~g. 

The backside of the detector includes a port for inserting and removing a microSD card; a USB-mini port for powering the detector, uploading new code to the Arduino Nano, or recording data directly through a USB port; a 3.5~mm female audio connection port used for connecting multiple detectors together to make a coincidence measurement; a reset button used for resetting the detector or assigning the detector as either the \textit{master} or \textit{coincident} detector in coincidence mode (this will be described in Sec.~\ref{sec::coincidence}); and a BNC connection, which is connected directly to the SiPM output, and can be used for injecting signals into the PCBs for testing. 

The supplementary material pertaining to this project can be found in the GitHub repository located in Ref.~\cite{github} and will be referred to as needed throughout. Included in the supplementary material is a document (instructions.pdf) that describes the detailed process of building, testing, and troubleshooting the detector. 

\section{Building and operating the detector}\label{sec::building}
The detectors were designed to be as openly available as possible, and therefore we attempt to use commonly available components all of which have been chosen to reduce the cost of the detector. Occasionally, we find that the manufacturers of the components are out of stock, so it is important for students to understand the provided information on the products we use. The purchasing list (purchasing\_list.pdf) found in the supplementary material includes a detailed description of each components and where they can be purchased. Instructions regarding purchasing the material and going step-by-step through a complete build can be found in the instructions.pdf document, is located in the supplementary material.

We've also produced a set of YouTube~\cite{youtube} videos, in which we build a detector from scratch, perform a simple measurement, then illustrate how to troubleshoot the detector. These can be found here:

Introduction:\\
\url{https://www.youtube.com/watch?v=e4IXzNiNxgU}

Populating the Main PCB:\\
\url{https://www.youtube.com/watch?v=zdVC8El_Xt8}

Populating the SiPM PCB:\\
\url{https://www.youtube.com/watch?v=yFgin5wlw4I}

Troubleshooting:\\
\url{https://www.youtube.com/watch?v=Ck24HGrjBfY}

Beyond this, there are several pieces of software required to operate the detectors. They are used for programming the Arduino and a python based program that we use for recording data or plotting data in real-time on the website.

Regarding the Arduino, the supplementary material contains code used to program the Arduino Nano (in the /Arduino directory). The code requires the user to download the Arduino IDE, and install several libraries. A list of the libraries are found at the top of all the Arduino code and within the installation instructions document. A description of the three pieces of code follows:
\begin{enumerate}
\item \textbf{Naming.ino}: This code is used to permanently name the detector (permanent up until this code is uploaded again). When this is uploaded to the Arduino, the variable, det\_name, is written to the EEPROM memory of the detector. It is recommended that every detector is given a unique name. 
\item \textbf{OLED.ino}: This code can be uploaded if the user would like information to print out on the OLED screen. 
\item \textbf{SDCard.ino}: This code sets the detector up to record the data directly to the removable microSD card. If you are running this code, the OLED screen will not be updated.
\end{enumerate}

Ideally, both the OLED.ino and SDcard.ino code would be contained in a single script, however, due to the limited SRAM and flash storage on the Arduino, only one of the codes can be run at a time. 

Depending on the operating system that the student is using there may be special drivers needed in order to get the Arduino Nano to communicate with their computer. In particular, when using a Mac OS, the CH340g driver needs to be installed. This can be found in Ref.~\cite{website:ch340g}. It's best to try and simply get the Arduino Nano working on your computer prior to working with the code above. Arduino provides an example code called Blink.ino, which simply blinks the LED connected to Pin 13 at 1~Hz. Once this is working, it's simple to transition to our software.

The python code is used primarily for recording data directly to a computer, or to connect to the website in order to plot the data in real-time. Recording the data directly to the computer has the added benefit that the code will automatically add an extra time stamp to each event from the computer. This gives an accurate time stamp to within approximately 1~ms. The python code, import\_data.py, can be found in the Recording\_Data directory in the supplementary material. In order to use the Arduino code, the user should have python 2.7 installed, and will have to install two libraries:
\begin{enumerate}
\item \textbf{Tornado}: This library is used to communicate with the CosmicWatch server. On a Unix-based machine, it can be installed through terminal using "sudo pip install tornado" or through another package manager.
\item \textbf{Pyserial}: This is used to use python to read the data that is flowing into the USB port. It can also be install through terminal using "sudo pip install pyserial." 
\end{enumerate}
Once the libraries are installed, the import\_data.py code can be run locally with python 2.7. 

\section{Recording data}

Data can be collected from the Desktop Muon Detector in many ways:

\begin{enumerate}
\item \textbf{Through a microSD card}. To save data to a microSD card, the SDcard.ino Arduino code must be uploaded to the detector and a microSD card (128~Mb is sufficient for all applications in this documents) inserted into the port on the back of the detector. Each time that the detector is reset or powered on, a new file is created with a file name that count sequentially upwards from the previous file, and an "M" or "S" indicating if the detector was in \textit{master} or \textit{coincident} mode respectively. 
\item \textbf{Directly to a computer through a mini-USB cable}. To record the data directly to the computer, either the SDcard.ino or the OLED.ino code can be uploaded to the detector. When the detector is plugged into a computer USB port, and the import\_data.py is run using python 2.7, the user is prompted to supply the path and name of the file to where the data is to be stored. It will then begin recording the data in real time to the output file. This method also includes a computer time and date stamp.
\item \textbf{Through the website}. Either the OLED.ino or the SDcard.ino code can be uploaded to the Arduino. When running the import\_data.py code with python, there is an option to connect to the website (option: "Connect to server: www.cosmicwatch.lns.mit.edu"). After selecting this option, it will prompt the user to go to the website, where data from the detector will be plotted in real-time. At any point, the data can be saved in the form of a .txt file with the "save" button. 
\item \textbf{Through the Arduino IDE}. When a detector is plugged into the computer, and the Arduino IDE is opened, the data can be seen accumulating in real-time in the serial monitor. The data can be copied and pasted into a text editor for later analysis. 
\end{enumerate}

\begin{figure}[h]
\begin{center}
\includegraphics[width=.9\columnwidth]{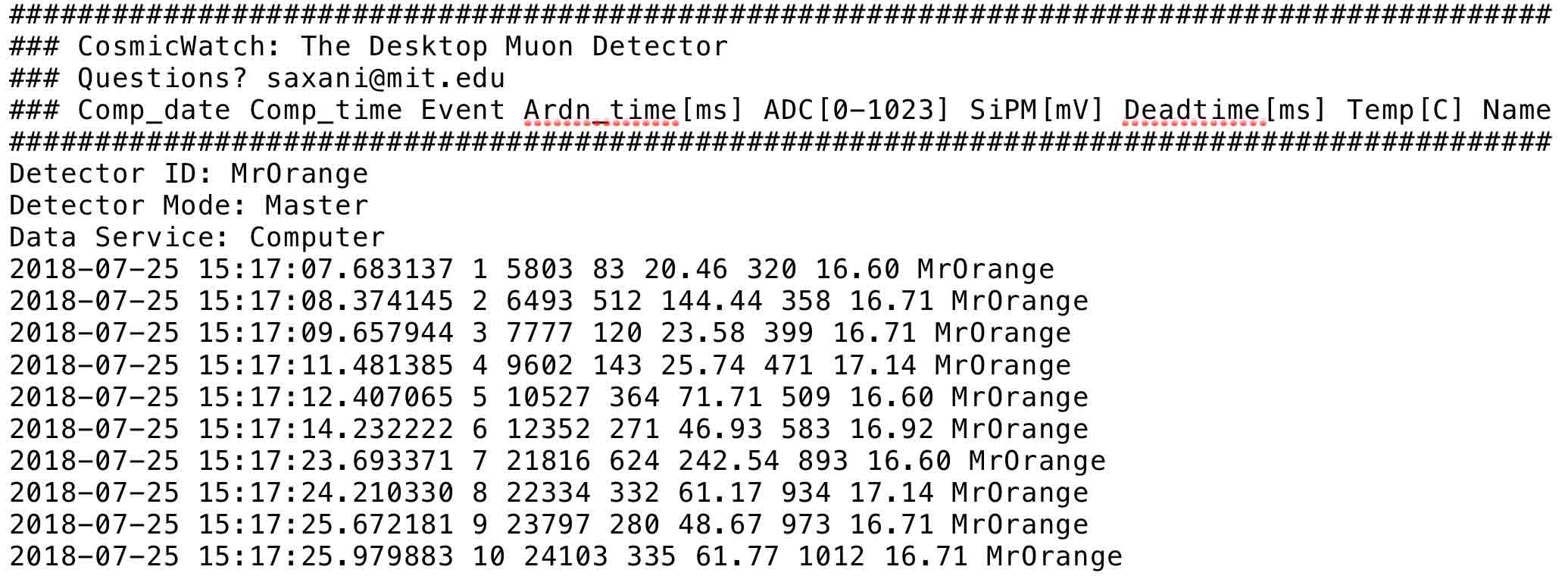}
\caption{Example data format. The header for the file is the first 8 lines and 10 example events follow. This data was recorded using the detector named "MrOrange," in \textit{master} configuration, recorded directly to the computer through the import$\_$data.py code. The definitions of the columns are listed in the header, as well as a more descriptive description in the text below.}
  \label{fig::example}
 \end{center}
\end{figure}

When data is saved directly through the import\_data.py script to the computer (either through method 2 or 3 above), the computer puts a time stamp with an precision and accuracy of +/-1~ms, that includes the computer date, computer time, and the detector name. An example of the data from the computer is shown in Fig.~\ref{fig::example}.

The data shown in  Fig.~\ref{fig::example} is formatted into space delimited columns, each of which is labelled in the header, and defined as:
\begin{itemize}
\item \textbf{Comp$\_$date}: The date given by the computer. Requires data saved using method 2 or 3 from above.
\item \textbf{Comp$\_$time}: The time stamp given by the computer. Requires data saved using method 2 or 3 from above.
\item \textbf{Event}: The event number of the detector.
\item \textbf{Ardn$\_$time [ms]}: The total elapsed Arduino time, measured in milliseconds. This is only accurate to roughly $\pm$1~min per day, and precise to the nearest millisecond.
\item \textbf{ADC [0-1023]}: The ADC measurement for the event. The Arduino Nano has a 10-bit ADC, meaning the values reported are from 0-1023 (2$^{10}$ values). The ADC is referenced between ground and 3.3~V using the internal Arduino voltage regulator.
\item \textbf{SiPM [mV]}: The calculated SiPM peak voltage. This a number calculated from the measured ADC value. It represents a number roughly proportional to the number of photons that triggered the SiPM. The conversion between the ADC value and the SiPM peak voltage is described in the calibration section of the instructions.pdf document.
\item \textbf{Dead-time [ms]}: The total dead-time of the detector. This must be subtracted for any rate measurement from the elapsed time to determine the detector livetime. 
\item \textbf{Temp [$^\circ$C]}: The measured temperature of the detector via the on-board TMP36 temperature sensor. Measured in degrees Celsius.
\item \textbf{Name}: The detector name. The name is set using the naming.ino program. Requires data saved using method 2 or 3 from above.
\end{itemize}

When recording data to the microSD card using the SDCard.ino Arduino code, the \textbf{Comp$\_$date}, \textbf{Comp$\_$time}, and \textbf{Name} fields are omitted.

\subsection{Setting detectors in master and coincidence mode}\label{sec::coincidence}
Many measurements in this document rely on the detectors operating in \textit{coincidence mode} (e.g. see Sec.~\ref{sec::coincident}). This mode allows us to improve the purity of the comic ray muon sample by rejecting events that likely came from interactions from radioactive backgrounds. Coincidence mode requires two or more detectors connected together using a 3.5~mm male-to-male audio cable (I'll often refer to this as the \textit{coincidence cable}). Once connected together, only one of the detectors requires power while the others are powered through the tip connection of the 3.5~mm cable. To set the detectors into the coincident configuration, simply reset the detectors while they are connected with the coincidence cable. The first detector to be reset becomes the \textit{master} and the subsequent detectors, when reset within 10~ms to 2000~ms of the \textit{master}, will become \textit{coincident} detectors.

The master detector triggers on all events, which create a pulse larger than the software defined threshold in the Arduino code. The coincident detector will only record events in which the master detector also triggered.  These event are likely to be due to a cosmic-ray muon, since the backgrounds and accidental coincidence are unlikely to trigger both detector simultaneously. A summary of why the purity of the cosmic-ray muon signal increases when in coincidence mode is below.

\begin{itemize}
\item Alpha particles will not penetrate a single detector (either the aluminum enclosure or even the black electrical tape) and therefore cannot trigger both the master and coincident detector at the same time.
\item Beta particles can be significantly attenuated by the aluminum case, and have a significant chance of scattering, thus losing energy. It's unlikely that the beta particle will be able to deposit sufficient energy within the scintillator of the master, exit, then depositing sufficient energy in the coincidence detector. 
\item Gamma rays can penetrate the aluminum enclosure and plastic scintillator, however they have a significant chance of Compton scattering, which will change the direction. If a gamma ray does interact with both detectors, this means that it likely Compton scattered off the scintillator slabs, lost sufficient energy to trigger the detector, then Compton scattered or photoelectrically absorbed in the second detector, also depositing sufficient energy to trigger the detector. This process is unlikely, and represents a small part of the coincidence signal (an estimate of this rate is found in Sec.~\ref{sec::sk}).
\item Accidental coincidences from random events are unlikely. This will be elaborated on in Sec.~\ref{sec::measurements}.
\item A typical minimum ionizing muon passing through the slab of scintillator will typically deposit more than 2~MeV of energy in the scintillator, without being deflected. If the muon passes through both scintillators, it will likely trigger both detectors simultaneously. 
\end{itemize}

Now that we understand how to build a detector, record data, and set up the detector into coincidence mode, we can begin making measurements. The remainder of this document will be dedicated to describing measurements we have previously made performed.

%% file: ExampleMeasurements.tex
\chapter{\textsc{Example measurements}}\label{sec::measurements}
This Chapter will describe a set of measurements made using the CosmicWatch Desktop Muon Detectors (version 2)~\cite{axani2018cosmicwatch,github}. This will be useful for those intending to use the detectors for educational purposes and students who are trying to make their own measurements. The individual measurements below attempt to provide compelling evidence for the physical processes describe in the first half of this document. Some of these measurements may be difficult for students to perform, therefore we also provide the majority of the data and the scripts used to produce these plots in the supplementary material. Reproducing these plots will require iPython to be installed with a few extra libraries that are left up to the user.

When performing a measurement using the Desktop Muon Detectors, there are a few important things to keep in mind.

1. The Arduino Nano is a relatively slow device. It is based off of an old processor called an ATmega328P. Although this makes it cheap to purchase, it also means that there are subtleties that we should be aware of. A critical aspect of any rate measurement performed with the Desktop Muon Detector is that every command (such as adding two and two, or printing the detector information to the serial port) takes time to perform.  Accounting for the time that each command takes is important since the detector is unable to trigger on an event if the detector is busy. The term associated with the time we are unable to make a measurement is \textit{dead-time}. The dead-time is calculated in the Arduino code by measuring the time each command takes in microseconds. This must be subtracted from the up-time in order to accurately calculate the time that the detector was able to make a measurement; the result is called \textit{live-time}. Dead-time is a common feature in all particle physics detectors, however for us it is particularly important due to the limited speed of the Arduino.

\begin{figure}[t]
\begin{center}
\includegraphics[width=1.0\columnwidth]{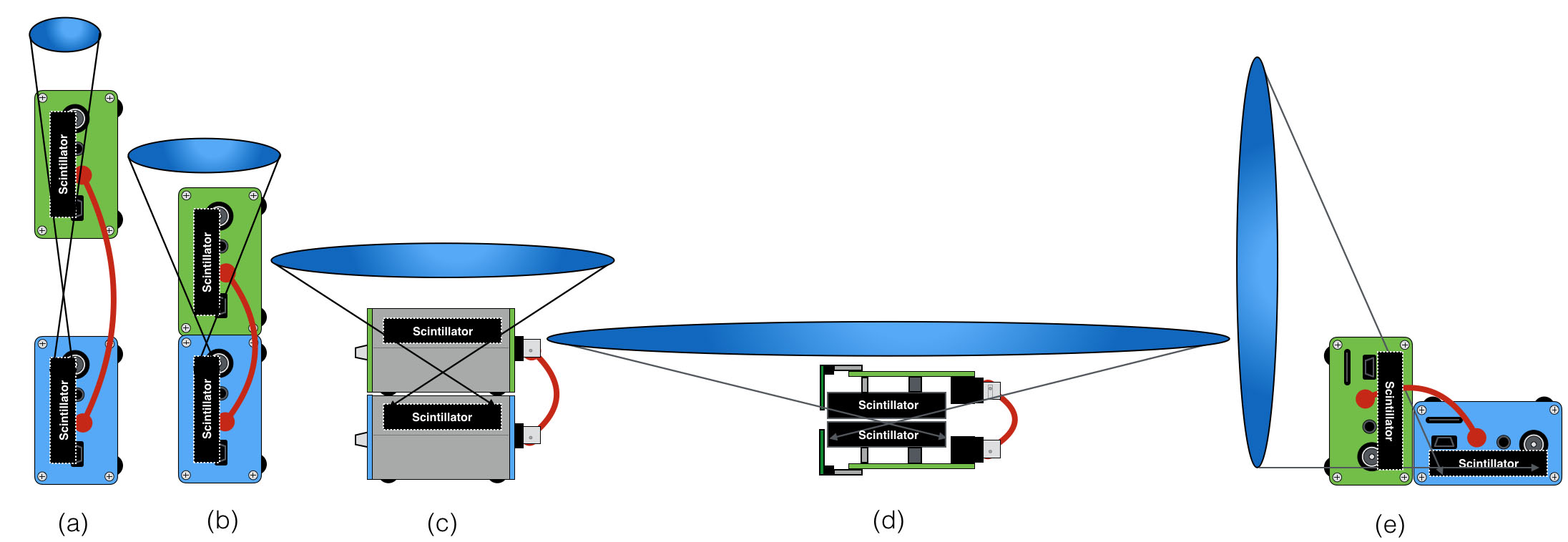}
\caption{This figure shows several configurations for setting up the detectors in coincidence mode. Moving from left to right the solid angle subtended between detectors increases (illustrated as the blue ovals above the detectors) as well as the coincidence rate as measured by the coincident detector.} 
  \label{fig::opening_angle}
 \end{center}
\end{figure}

2. The orientation of the master and coincident detector when making a measurement will have different characteristics. When making a coincidence measurement, it is important to think about what is being measured, and how the orientation will effect the measurement. For example, if we are interested in the angular spectrum of cosmic-ray muons, we want to only accept muons coming from a small solid angle. Fig.~\ref{fig::opening_angle} shows several possible configurations. The left side of Fig.~\ref{fig::opening_angle}, labelled as (a), shows two detectors spaced a few centimetres apart and connected with a coincidence cable. We see that only the muons that travel downwards through the blue area are able to trigger both detectors. Fig.~\ref{fig::opening_angle} (d) shows a configuration in which both detectors will trigger from down-going muons over a much larger solid angle.

3. Two independent events could have triggered both the master and coincident detector accidentally if they happened to arrive within the coincidence time window. This becomes much more likely at higher count rates. If we assume that the individual count rates are determined by Poisson statistics and that the two detectors are independent, then the probability of observing \textit{n} events when the mean number of events is expected to be $\mu$, is given by the Poisson distribution:
\begin{equation}
f(\textit{n},\mu) = \frac{\mu^\textit{n}  e^{-\mu}}{\textit{n}!}
\end{equation}\label{eq::poisson}

Following Ref.~\cite{grupen2008particle}, let the count rate on the two detectors be N$_1$ and N$_2$, and the coincidence window (the time window in which an event will be labelled as a coincident event) be $\tau$. The average count rate for both detectors at sea level is approximately 1~Hz, and given a coincident time window of roughly 30$\mu$s, we expect on average 0.000030 events in the time window (N$_{1,2}\times \tau$). Therefore, the probability of  observing zero event in the time window is $f(0,30\times10^{-6} )$ and the probability of one detector observing an event in the coincident time window is 1 - $f(0,30\times10^{-6})$. Since the second detector could have also triggered first, the rate of accidental triggers is: R = 2N$_1$N$_2\tau$. This corresponds to an accidental coincident rate for two detectors of 1 every 15,000 events, or roughly six events per day using two detectors, at sea level.

More generally, assuming that there are \textit{q} detectors, each with the same baseline count rate N, then the number of \textit{p}-fold coincidence out of \textit{q} detectors is:
\begin{equation}
R_p(q) = {q \choose p} pN^p \tau^{p-1}
\end{equation}\label{eq::accidental}

\begin{table}[!ht]
\footnotesize
\begin{center}
 \begin{tabular}{||c c c c c c||}
 \hline
\textbf{Number of detectors} & \multicolumn{4}{c}{ \textbf{p-fold coincidence}} &\\ [0.5ex]
q &  p = 2 & p = 3 & p = 4 & p = 5 & p = 6 \\ [0.5ex]
\hline
\hline
 2 & 0.0006   & & & &  \\
\hline
3 & 0.0018   &  2.7$\times$10$^{-7}$  & & &   \\
\hline 
4 & 0.0036 &  1.1 $\times$10$^{-6}$  & 1.1 $\times$10$^{-10}$  & &   \\
\hline
5 & 0.006  & 2.7$\times$10$^{-6}$  &  5.4$\times$10$^{-10}$  & 4.1$\times$10$^{-14}$ & \\
\hline
6 & 0.009  & 5.4$\times$10$^{-6}$  & 1.6 $\times$10$^{-9}$  & 2.4$\times$10$^{-13}$ &  1.5$\times$10$^{-17}$  \\
\hline
\end{tabular}
\caption{A table of values for Eq.~\ref{eq::accidental}. Given q detectors, the rate of p-fold coincidences is shown assuming a time window of $\tau=30\mu$s and a count rate of N=1~Hz on all detectors.}
\label{table::change}
\end{center}
\end{table}

\section{A coincidence measurement}\label{sec::coincident}
A coincidence measurement refers to a measurement in which we have extracted only the events which have triggered multiple detectors within some time window.  The time window when using the 3.5~mm coincidence cable is approximately 30$\mu$s. The implementation of this isn't obvious when examining the Arduino code. It works by having the master detector immediately send Pin 6 on the Arduino HIGH, each time it registers an event. The detector in coincidence mode has its Pin 6 set-up as an input. The 3.5~mm coincidence cable connects Pin 6 on both detectors together through the ring terminal. When the coincidence detector registers an event, it first waits approximately 30$\mu$s before checking to see if the Pin 6 is HIGH; if the Pin is HIGH, it counts the event, otherwise the event is dropped. The time window was chosen to accommodate the slow ADC sampling time. A single ADC sample takes approximately 10$\mu$s (we've increased the sampling speed using a \textit{pre-scaler} in the Arduino code), which means that the master detector may have triggered up to 10$\mu$s after the coincidence detector. Note that since the time window is so large, we cannot determine, which detector actually triggered first (a relativistic particle will travel 9~km is 30~$\mu$s), therefore it doesn't matter if the coincidence detector is on-top or below the master detector.

\begin{wrapfigure}{R}{0.6\textwidth}
\centering
\includegraphics[width=0.6\textwidth]{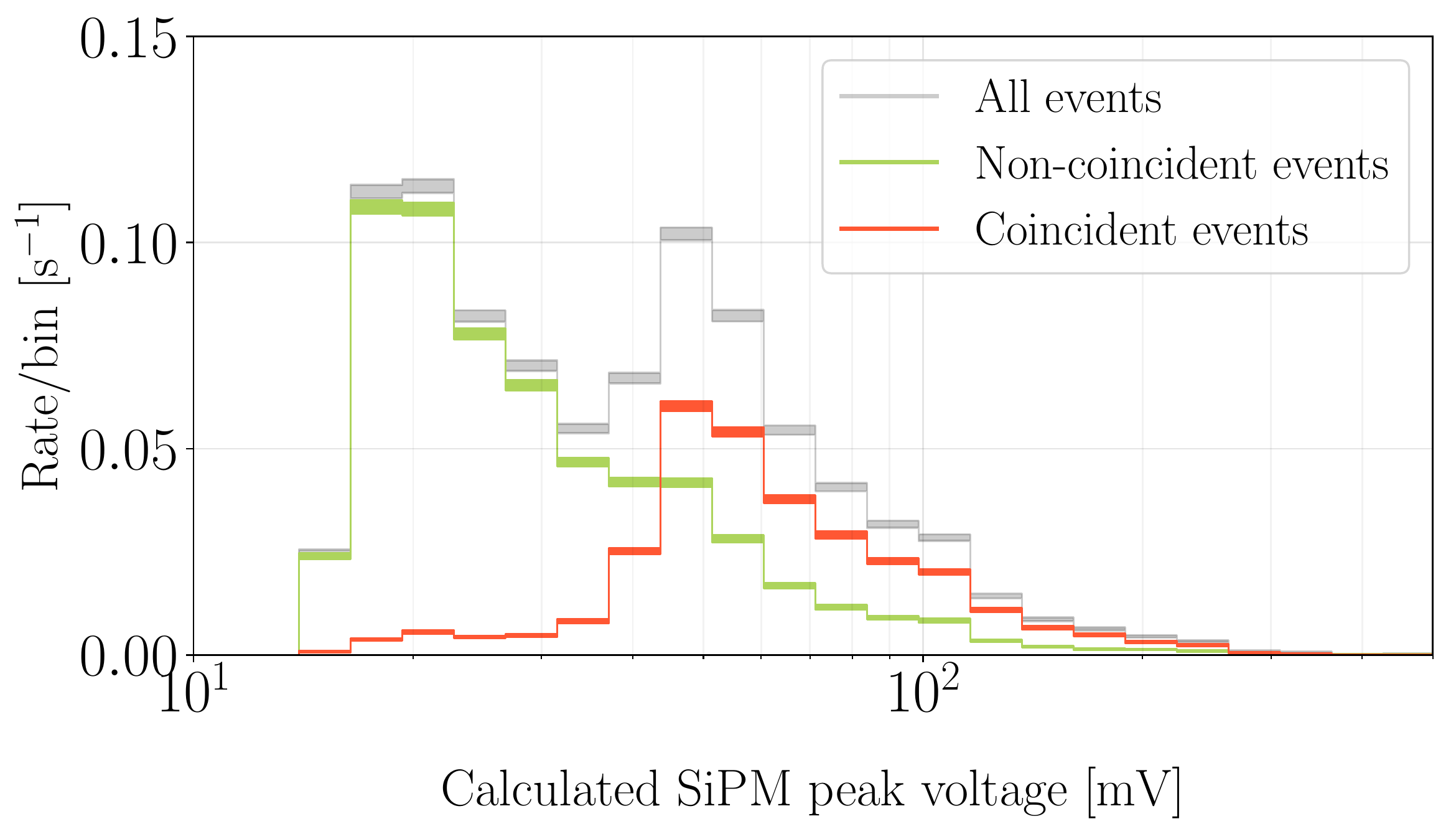}
\caption{A coincidence measurement showing the calculated SiPM pulse amplitude for the coincident events and non-coincident events. } 
\label{fig::coincident} 
\end{wrapfigure}

Typically, when a detector is placed in coincidence mode, only the coincident events are saved. For this measurement however, we slightly modified the Arduino code such that both the coincident and non-coincident events are saved, and we label each event to indicate what type it was. Two detectors were placed into configuration (d) of Fig.~\ref{fig::opening_angle}, and connected via a 3.5~mm coincidence cable. The calculated SiPM pulse amplitude for this data set is shown in Fig.~\ref{fig::coincident}. We see that the two data subsets (coincident versus non-coincident events) populate different regions. This is due to the discussion found in Sec.~\ref{sec::heavy} and ~\ref{sec::coincidence}

In areas of high radioactive backgrounds, the non-coincident events may dominate the spectrum even at 50~mV. The width of the peak shown in the coincident events at a SiPM peak voltage of 50~mV can vary depending on how well the detector was built. For example, if the coupling between the SiPM and scintillator was poor, the SiPM will observe fewer photons per event thus reducing the resolution of the detector.

\section{Measuring the cosmic-ray muon rate in an airplane at 33,000~ft \label{sec::airplane}}
A rate measurement was performed during a flight from the Boston International Logan Airport (BOS, latitude = 42.4$^\circ$) to the Chicago O'Hare Airport (ORD, latitude = 42.0$^\circ$) using a single detector. The data was recorded to a microSD card with and plugged into a 10,000~mAh USB power bank. The altitude of the airplane was collected from the flight records found at FlightAware.com~\cite{flight_data}. 

\begin{figure}[h]
\begin{center}
\includegraphics[width=1\columnwidth]{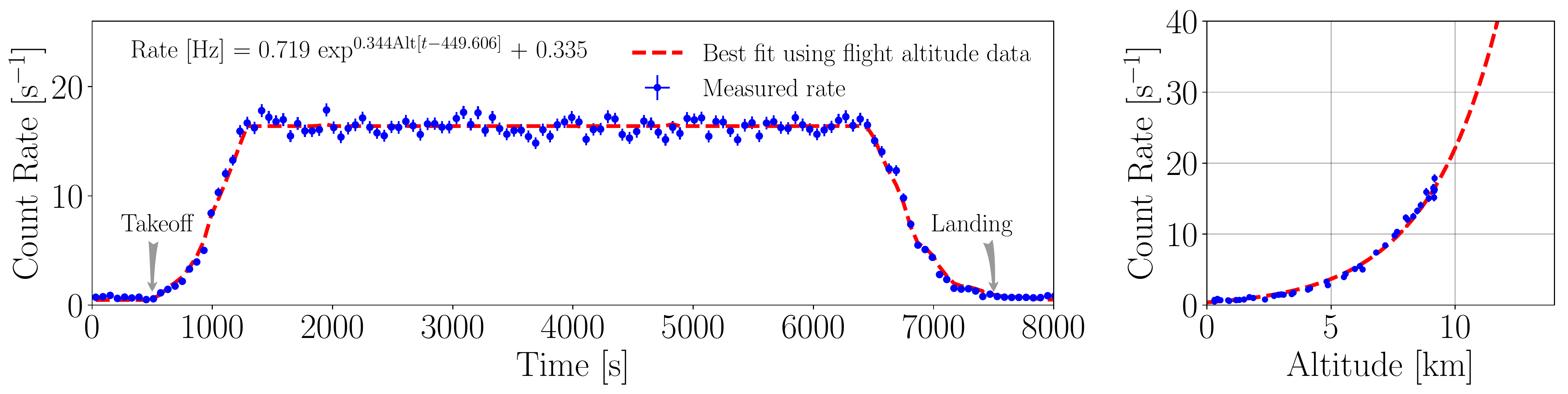}
\caption{(Left) The count rate measured during a flight from Boston to Chicago as a function of flight time. The dashed red line shows the actual amplitude of the airplane~\cite{flight_data} scaled by a fitted exponential shown at the top left of the plot. (Right) The measured count rate as a function of altitude. The dashed red line shows the fit.}
  \label{fig::flight}
 \end{center}
\end{figure}

Fig.~\ref{fig::flight}  (left) shows the trigger rate of the detector in blue as a function of time, binned into 60 second intervals. The error bars shown here are purely statistical. The altitude data of the airplane was linearly interpolated between points to estimate the altitude at any given minute. The interpolated altitude data was the fit to the detector data using a simple exponential plus an offset. Since we did not know the absolute take-off time (data was recorded to the microSD card), we allow the altitude time stamps to shift during the minimization. The best-fit equation is shown at the top left of this figure, where ALT[t] represents altitude measured in kilometres as a function of time. The best-fit is also plotted as a dashed red line.

Fig.~\ref{fig::flight}  (right) shows the measured trigger rate as a function of true altitude. Here, we show the exponential fit extended beyond the measured values. The count rate uncertainties were calculated by taking the square root of the sum all the events measured at a particular altitude. One thing to note is that this is data taken with a detector in master mode, which means it is also sensitive to the background radiation from the interior of the plane. 

The cosmic-ray muon flux is known to vary also as a function of latitude, which we have avoided by measuring the flux at a near constant latitude. We also expect the exponential fit to fail as we extend to higher altitudes due to the change in flux composition near the primary cosmic ray interaction region. 

\section{High altitude balloon measurement at 107,000~ft}\label{sec::hab}

\begin{figure}[h]
\begin{center}
\includegraphics[width=1.\columnwidth]{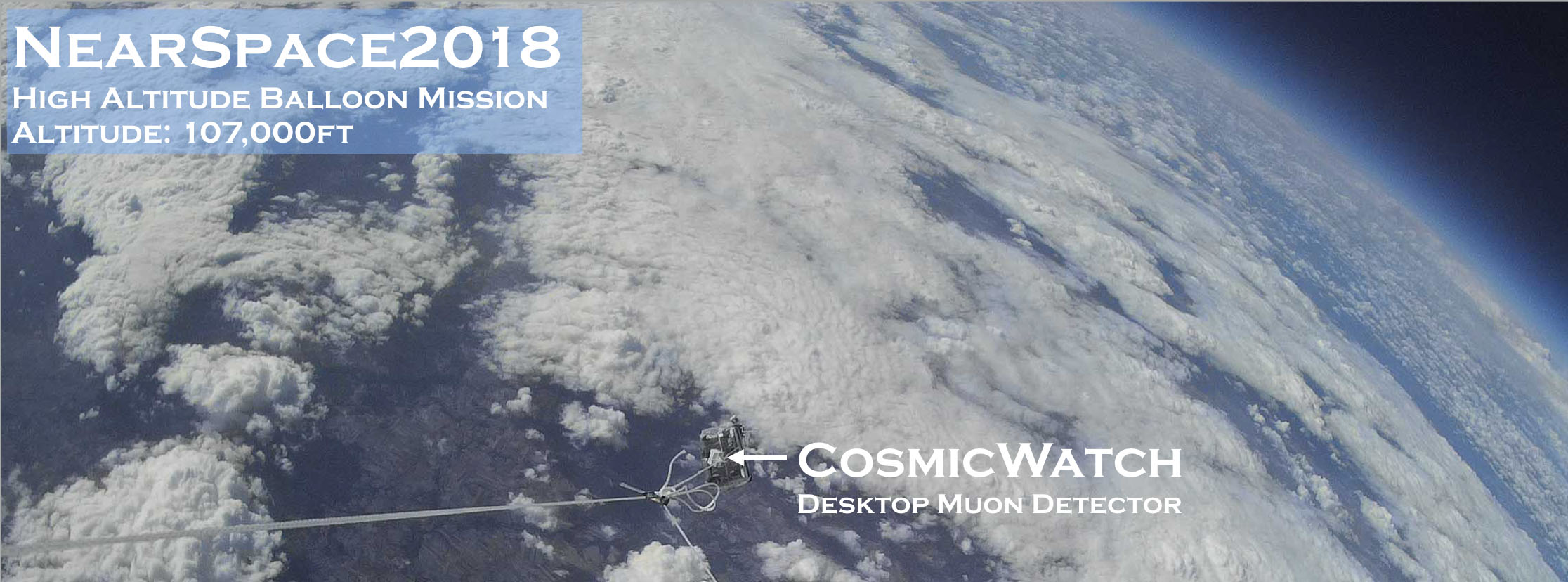}
\caption{An image from the high-altitude balloon flight at 107,000~ft. Photo from \textit{Daniel Kaczmara - DNF Systems}. }
  \label{fig::nearspace}
 \end{center}
\end{figure}

\begin{wrapfigure}{R}{0.45\textwidth}
\centering
\includegraphics[width=0.45\textwidth]{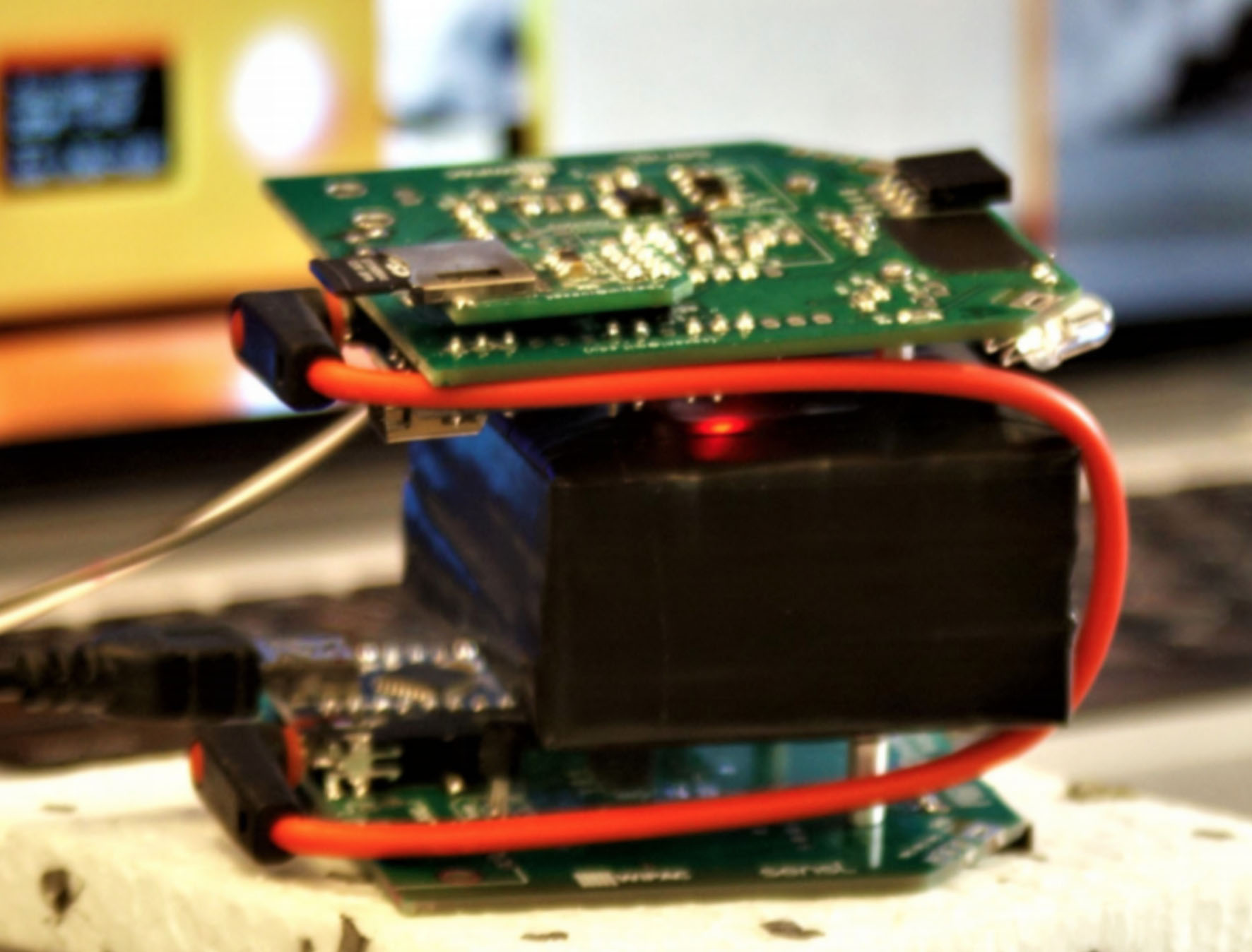}
\caption{The two detectors flown in the HAB flight. The scintillators were taped together in order to preserve the detector orientation relative to each other.}
\label{fig::hab_det} 
\end{wrapfigure}

As noted in Sec.~\ref{sec::cosmic_rays}, primary cosmic rays interacting in the upper atmosphere produce showers of particles, some of which decay to muons. Muons are typically produced near an altitude of 15~km. At higher altitudes, there is an increase in the contribution from other ionizing particles, primarily from electrons/positrons, and protons.

During the NearSpace2018 conference~\cite{nearspace} in Torun Poland, we participated in a high-altitude balloon (HAB) flight to measure the ionizing radiation flux as a function of altitude. Two detectors were used for the flight so that we could measure both the total rate on the master detector and the down-going rate on the coincident detector. The detectors were placed one-on-top of another (configuration (d) in Fig.~\ref{fig::opening_angle}) and taped together to ensure their orientation relative to each other remained the same throughout the flight. The BNC connectors and the OLED screens were removed from the PCB to reduce the weight. An 8" 3.5~mm audio cable was used to connect them into coincidence mode and the SDCard.ino code was uploaded to both detectors. An image of the two detectors is shown in Fig.~\ref{fig::hab_det}. Both detectors were powered by single cell lithium ion battery. 

The temperature during the accent was expected to reach -60$^\circ$C, therefore thermal protection was required both for the battery and in order to minimize the effect on the SiPM described in Sec.~\ref{sec::sipm}. A 10$\times$10$\times$10~cm$^3$ Styrofoam enclosure was constructed with a wall thickness of 1~cm, to house the components. This was sufficiently large that we could place the two detectors and a small heating element in the enclosure with two single-cell Lithium-ion batteries (one to power the detectors, and the other to power the heater). A micro-switch was connected to the battery and wired outside the enclosure so that we could initialize the detectors from outside the enclose, just prior to the flight.

\begin{figure}[h]
\begin{center}
\includegraphics[width=1\columnwidth]{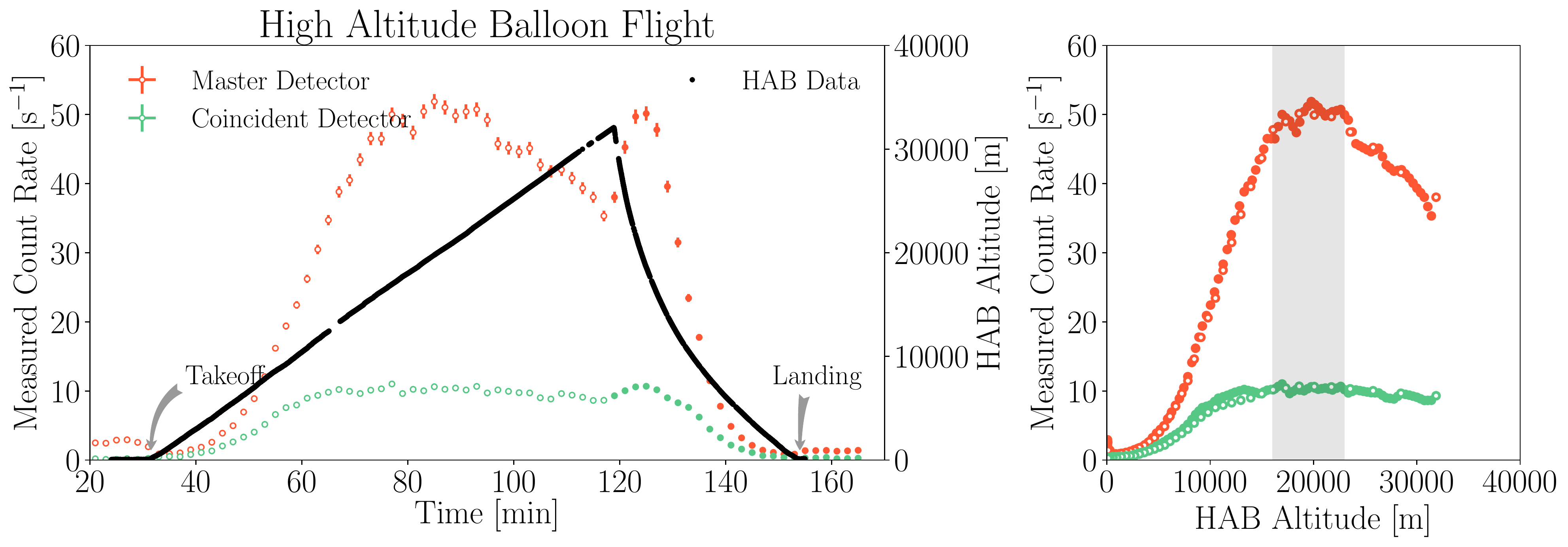}
\caption{(Left) The measured trigger rate for the master (orange) and coincident (green) detector during the high-altitude balloon flight. Take-off occurred 30 minutes after powering on the detectors. The black data points correspond to the altitude as measured through the on-board GPS. (Right) The GPS altitude as a function of trigger rate. The uncertainty in both these plots are statistical.}
  \label{fig::hab}
 \end{center}
\end{figure}

The HAB was launched on September 22$^{\mathrm{nd}}$, 2018 at 12:53~pm. DFN System recorded the balloon altitude and location using on-board GPS, they also mounted a camera to the balloon that looked down at the payloads. An image near the maximum altitude of the flight is shown in Fig.~\ref{fig::nearspace}. The master (orange) and coincident (green) detector count rate, binned in to 60 second intervals, is shown in Fig.~\ref{fig::hab}, along with the altitude data from the GPS (black). 

The shape of the measured spectrum in Fig.~\ref{fig::hab} corresponds to the Pfotzer curve. We find an initial maximum count rate (Regener-Pfotzer maximum) for the master detector at an altitude from approximately 16-23~km approximately 70-95 minutes into the measurement.  After the balloon popped (at minute 118), the detectors fell through the Regener-Pfotzer maximum. The decrease in the trigger rate after passing the maximum occurs due to the detectors ascending beyond the primary interaction region. 

The coincidence detector shows a flatter maximum at an altitude from approximately 12-25~km. The peak begins at lower altitudes since we are now preferentially triggering on vertically down-going particles. As described in Sec.~\ref{sec::variations_atm}, primary particles entering the Earth's atmosphere at larger angles from the zenith will interact at higher altitudes. In agreement with the data.

\section{Muon rate measurement while flying to the South Pole}\label{sec::sp_flight}

\begin{figure}[t]
\begin{center}
\includegraphics[width=1\columnwidth]{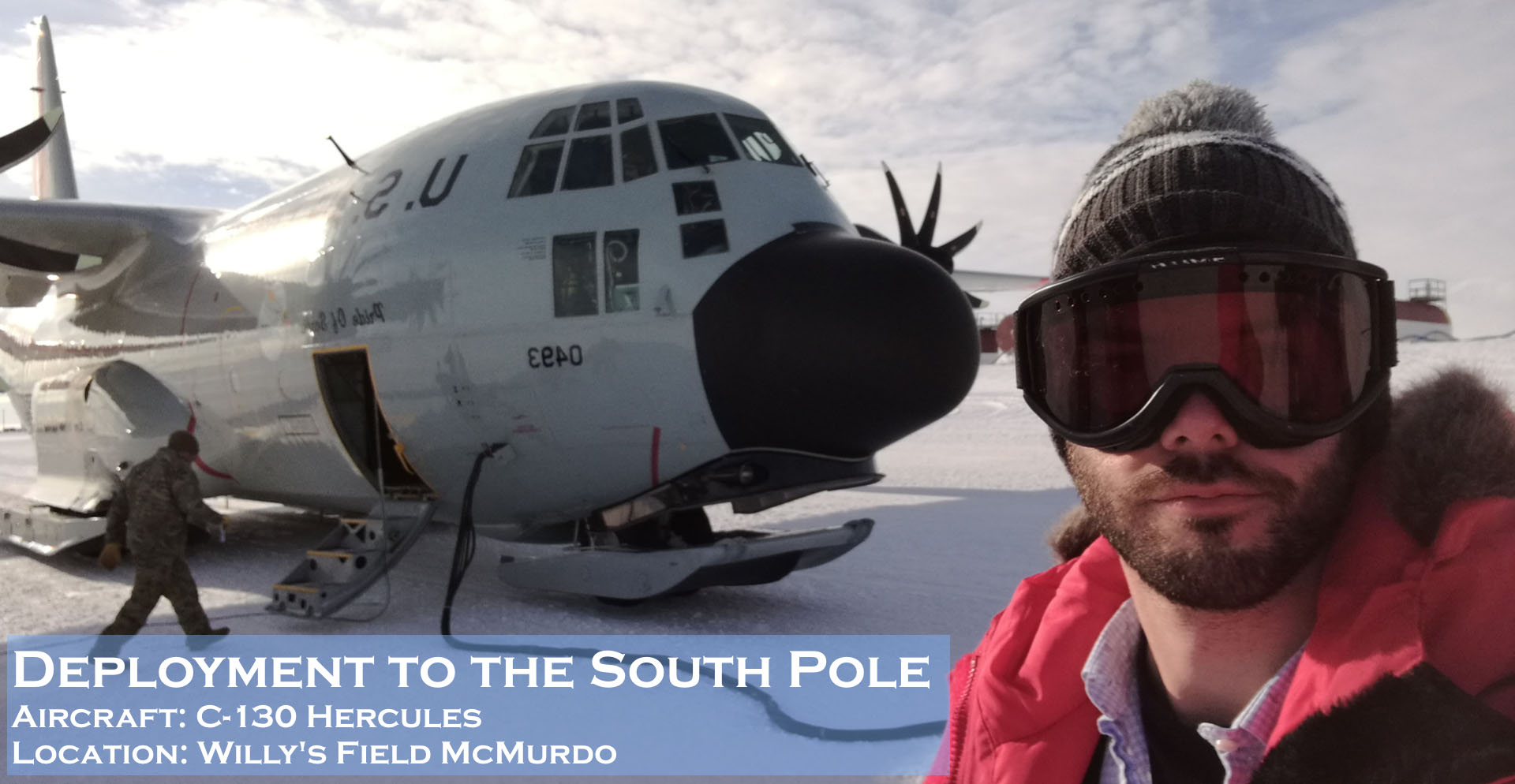}
\caption{Leaving McMurdo Station, Antarctica on a C-130 Hercules for the South Pole Station.}
  \label{fig:sp_flight}
 \end{center}
\end{figure}

In December 2018, I flew to the South Pole as part of a field team to perform maintenance and upgrades on the IceCube Neutrino Observatory~\cite{icecube}. During the flight I measured the ionizing radiation with two detectors orientated in configuration (d) of Fig.~\ref{fig::opening_angle}. Data was recorded to the microSD cards and powered through a single 30,000~mAh power bank. 

The first four flights (from Madison, WI USA to Christchurch, New Zealand) were operated by United Airlines and New Zealand Airlines. The altitude data was publicly available for these flights on FlightAware.com~\cite{flight_data}.  The flight leaving from Christchurch New Zealand, to McMurdo Antarctica, was on a C-17 military jet operated by the US Air Force. Similarly, we flew on a C-130 Hercules, the day after to the South Pole. Since these were military flights, the altitude of this flight was not available; however, several altitude measurements on the second flight were made using GPS. We landed several days later on the 2820~m thick South Pole glacier, approximately 0.5~km from the actual Geographical South Pole.

\begin{figure}[h]
\begin{center}

\includegraphics[width=1\columnwidth]{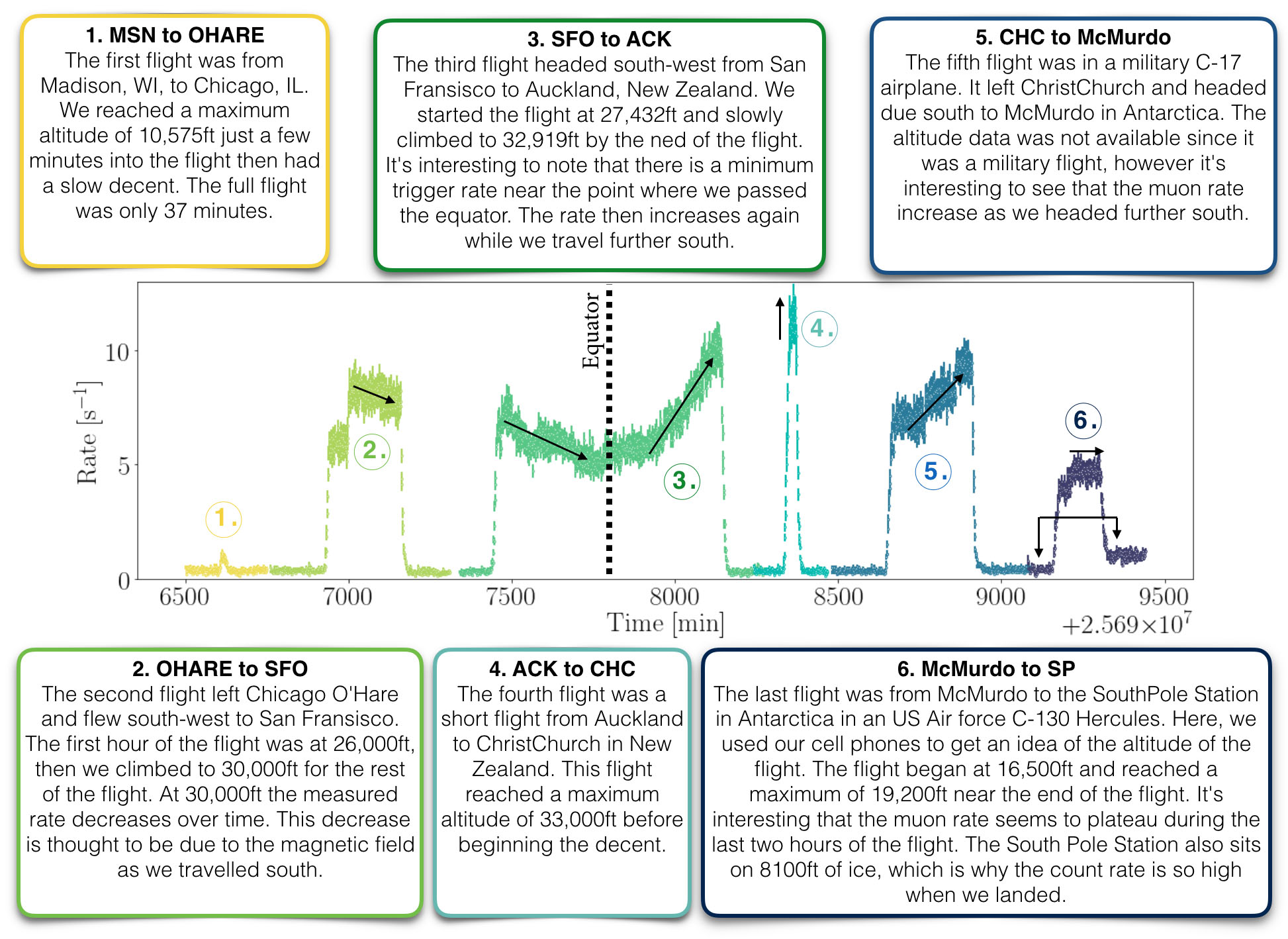}
\caption{}
  \label{fig::long_flight}
 \end{center}
\end{figure}

The full master and coincident detector data are shown in Fig.~\ref{fig::long_flight}, with descriptions of each flight in the text boxes. To give perspective for other measurements, the total data collected by the master detector was about 50~Mb, whereas the coincidence detector was approximately 15~Mb.

Fig.~\ref{fig::long_flight} illustrates a several very interesting properties. First, there is a trend towards lower count rates near the equator. This is due to the latitudinal variation in the cosmic-ray flux described in Sec.~\ref{sec::variations_atm}. This effect is most obvious in the flight from SFO to ACK (labelled as flight 3 in Fig.~\ref{fig::long_flight}), which travelled at nearly a constant altitude and a constant rate of latitude change from +32$^\circ$ to -32$^\circ$. We see that the rate is not symmetric. This is because the magnetic latitude is offset from the geographical latitude, which in turn is because the magnetic field is not symmetric about the equator. Second, it's interesting to see that when we landed at the South Pole, there is a noticeable change in trigger rate due to the combination of the elevation and change in Earth's magnetic field.

While flying through the equator at 35,000~ft, I also performed an East-West measurement using configuration (e) of Fig.~\ref{fig::opening_angle}. We measure a count rate coming from the east of 0.69~+/-~0.02~cps, while from the west 0.84~+/- 0.03~cps. This represents a 22.2$\pm$7.4\% increase in the westward direction. This is due to the east-west asymmetry described in Sec.~\ref{sec::variations_atm}.

\section{Latitude correction to the cosmic-ray muons}\label{sec::lat_correction}

\begin{figure}[h]
\begin{center}
\includegraphics[width=1\columnwidth]{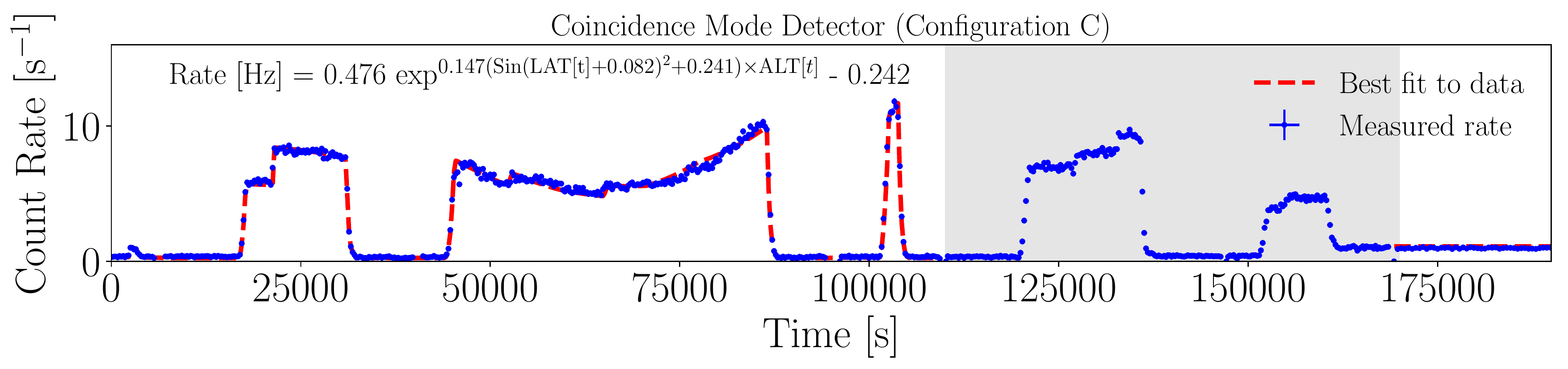}
\caption{The fitted data based on the altitude and latitude data from the first four flights. The fit also includes a measurement after landing at the South Pole, where the altitude and latitude were known (2700~m and -90$^\circ$). The two military flights were not included in the fit.}
  \label{fig::sp_fit}
 \end{center}
\end{figure}

We observed latitudinal variation in the cosmic-ray muon rate in the previous measurements. This was an expected effect due to the change in the Earth's magnetic field as a function of latitude. Here, we will empirically attempt to account for the variation in the latitude and altitude based on the previous measurement. We will assume that the change in the rate as a function latitude follows a sine-squared form; where the minimum occurs near the equator and the maximum occurs near the poles:

\begin{equation}\label{eq::sp_fit}
R[Hz] = N \mathrm{exp}^{\alpha (\mathrm{sin(LAT[t]} +\theta)^2+\beta ) \times \mathrm{ALT[t]}} - b
\end{equation}

\begin{wrapfigure}[11]{R}{0.45\textwidth}
\centering
\includegraphics[width=0.44\textwidth]{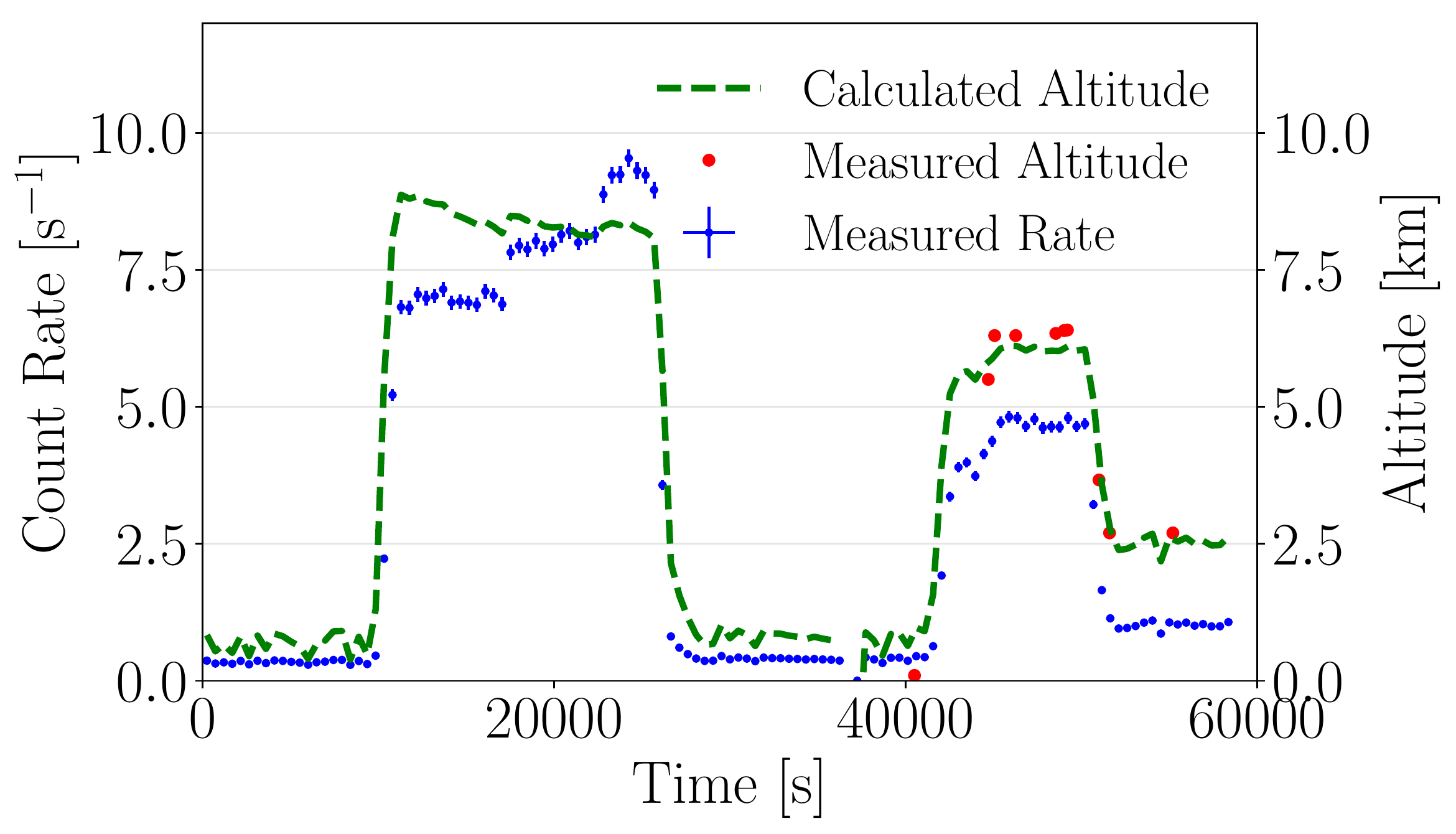}
\caption{The measured muon rate in blue along with the calculated flight altitude in green. The red markers show the few GPS altitude measurements during the flight.}\label{fig::flight_calc} 
\end{wrapfigure}

Here, $\theta$ represents a phase that offsets the latitude to account for the difference between the geographical latitude and the magnetic latitude, $\beta$ is a factor that permits an altitude effect at the magnetic equator (set the sine term to zero), $\alpha$ is a scale factor that dictates the strength of the latitudinal and altitude effect, $N$ is a normalization that sets the scale for the exponential component, and $b$ is an offset that accounts for a constant background radiation.

Fig.~\ref{fig::sp_fit} uses the function in Eq.~\ref{eq::sp_fit} to fit the data for the first four flights, plus the data after landing at the South Pole. The best fit values are shown at the top left of Fig.~\ref{fig::sp_fit}. Using this result, we can then invert  Eq.~\ref{eq::sp_fit} to calculate the altitude of the two military flights (assuming we flew at a constant velocity directly south).  The calculated altitude is shown in Fig.~\ref{fig::flight_calc} for the military flights in green, as well as the measured altitude from GPS in red.

\section{Cosmic-ray muon rate at various locations on Earth}\label{sec::locations_on_earth}
\begin{figure}[h]
\begin{center}
\includegraphics[width=0.9\columnwidth]{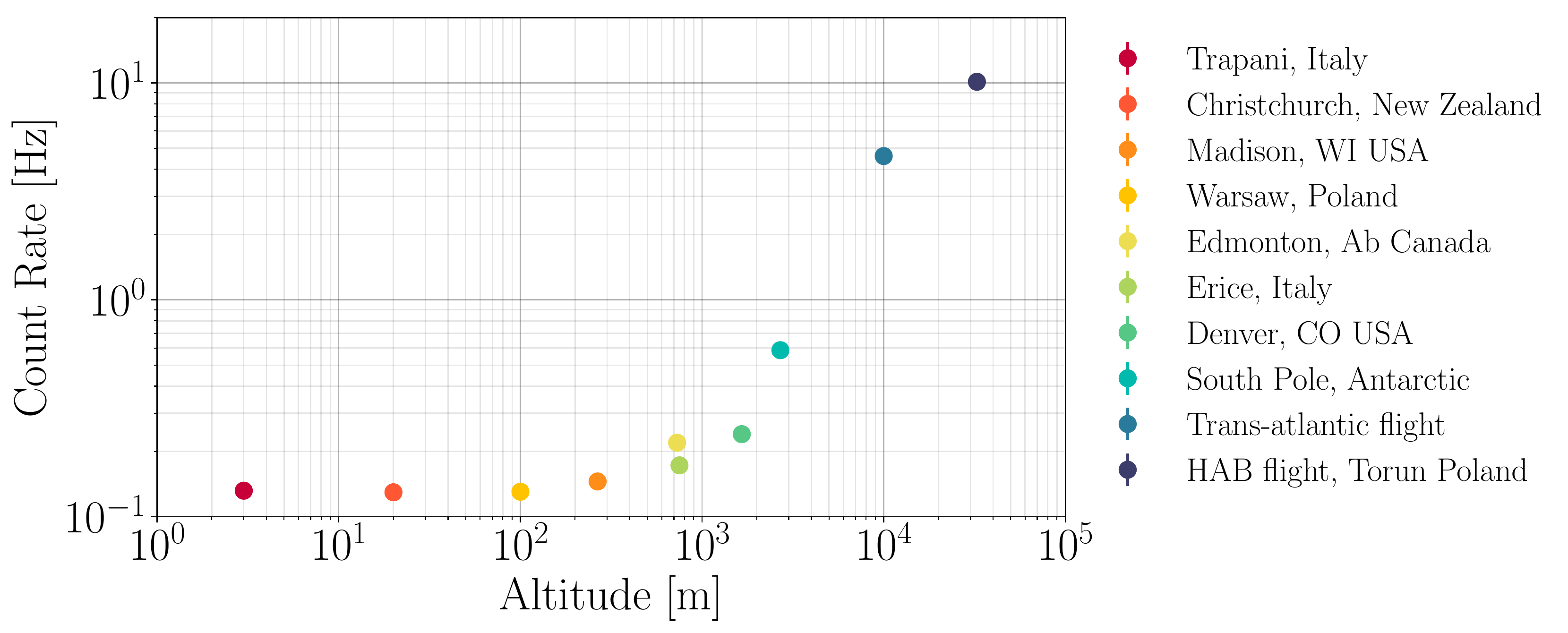}
\caption{The coincident detector count rate at various locations throughout the world. The trans-Atlantic flight rate was measured at 30,000ft, while the high-altitude balloon data was simply the count-rate at the Pfotzer maximum described in Section~\ref{sec::hab}. The statistical error bars of each point are smaller than the marker size.} 
  \label{fig::cities}
 \end{center}
\end{figure}

Acquiring data at different elevations above sea level can be challenging since many locations do not have immediate access to large changes in elevations. During 2018, we recorded the muon rate at various locations on the planet that we visited. Each measurement was performed in configuration (c) of Fig.~\ref{fig::opening_angle} using the coincidence connection, and all data was recorded to the microSD card. The altitude data of each location was found through Google and not corrected for changes due to geological features or the altitude of the measuring location (future measurements will use GPS to give a more accurate altitude). The exceptions in this dataset are the cases of the trans-Atlantic flight and the HAB measurement, whose altitude data was found using FlightAware.com and GPS respectively. Fig~\ref{fig::cities} shows the measured count rate at various locations (the trans-Atlantic flight data was recorded at an altitude of 33,000~ft and high-altitude balloon measurement was taken from the Pfotzer maximum described in Sec.~\ref{sec::hab}).

As expected, at higher altitudes, we observe a statistically significant change in the measured count rate. An interesting exception is noted when comparing the measurement at Sherwood Park, Alberta Canada, versus Erice, Sicily Italy. They are both at approximately the same the same altitude and latitude, however there is a 27\% measured rate difference between the two. This is thought to be due to the electromagnetic component of the shower not being properly shielded in the measurement in Canada. The Erice measurement was performed in a large concrete tower with several concrete floors above the detectors, while the measurement in Canada was performed on the ground floor of a two story wooden house. The lack of shielding against the electromagnetic component of the atmospheric flux is thought to be the primary source of this discrepancy.

\section{Rate measurement 1~km underground at Super-Kamiokande \label{sec::sk}}
\begin{figure}[h]
\begin{center}
\includegraphics[width=1\columnwidth]{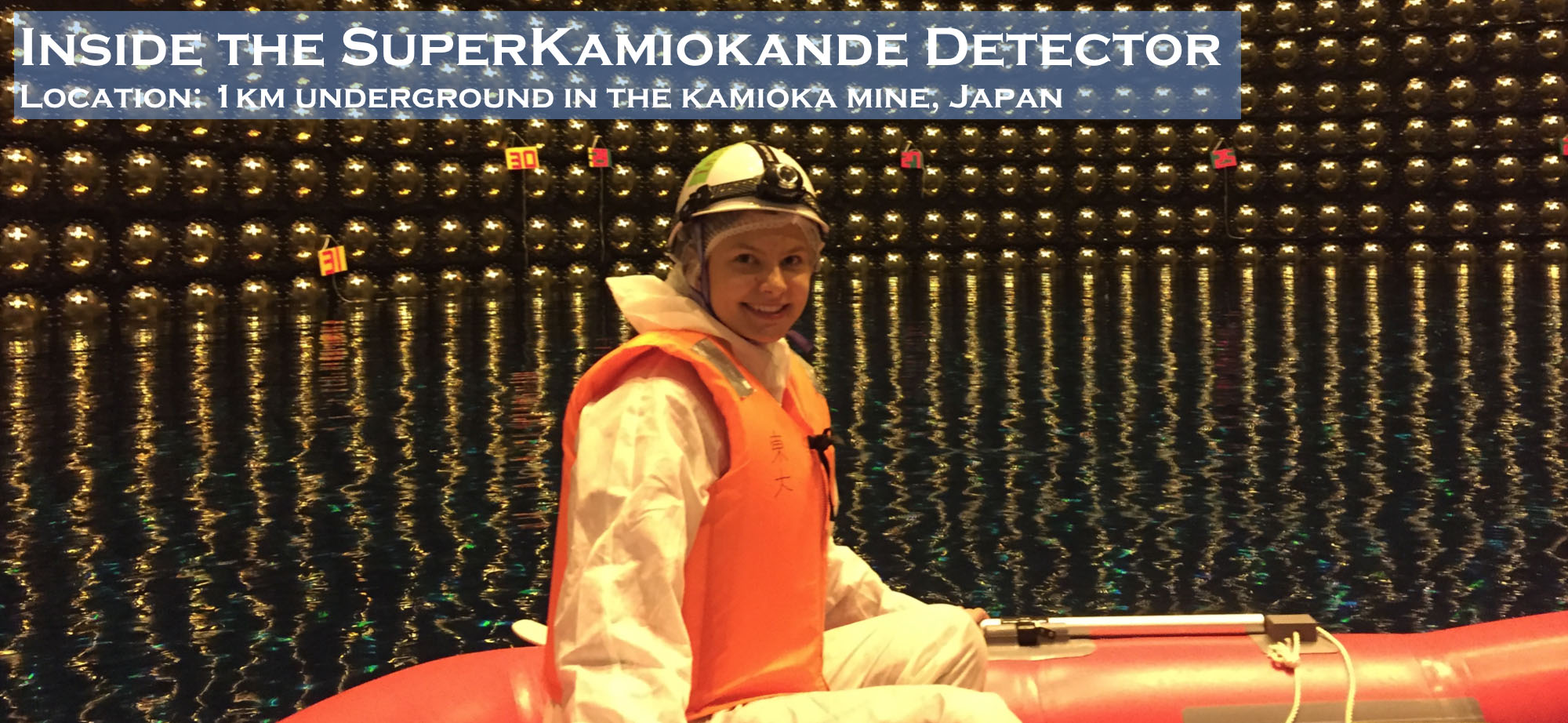}
\caption{Dr. Katarzyna Frankiewicz floating in the inner detector of Super-Kamiokande during the Gadolinium upgrade 2018.}
  \label{fig::kasia}
 \end{center}
\end{figure}

\begin{figure}[h]
\begin{center}
\includegraphics[width=0.7\columnwidth]{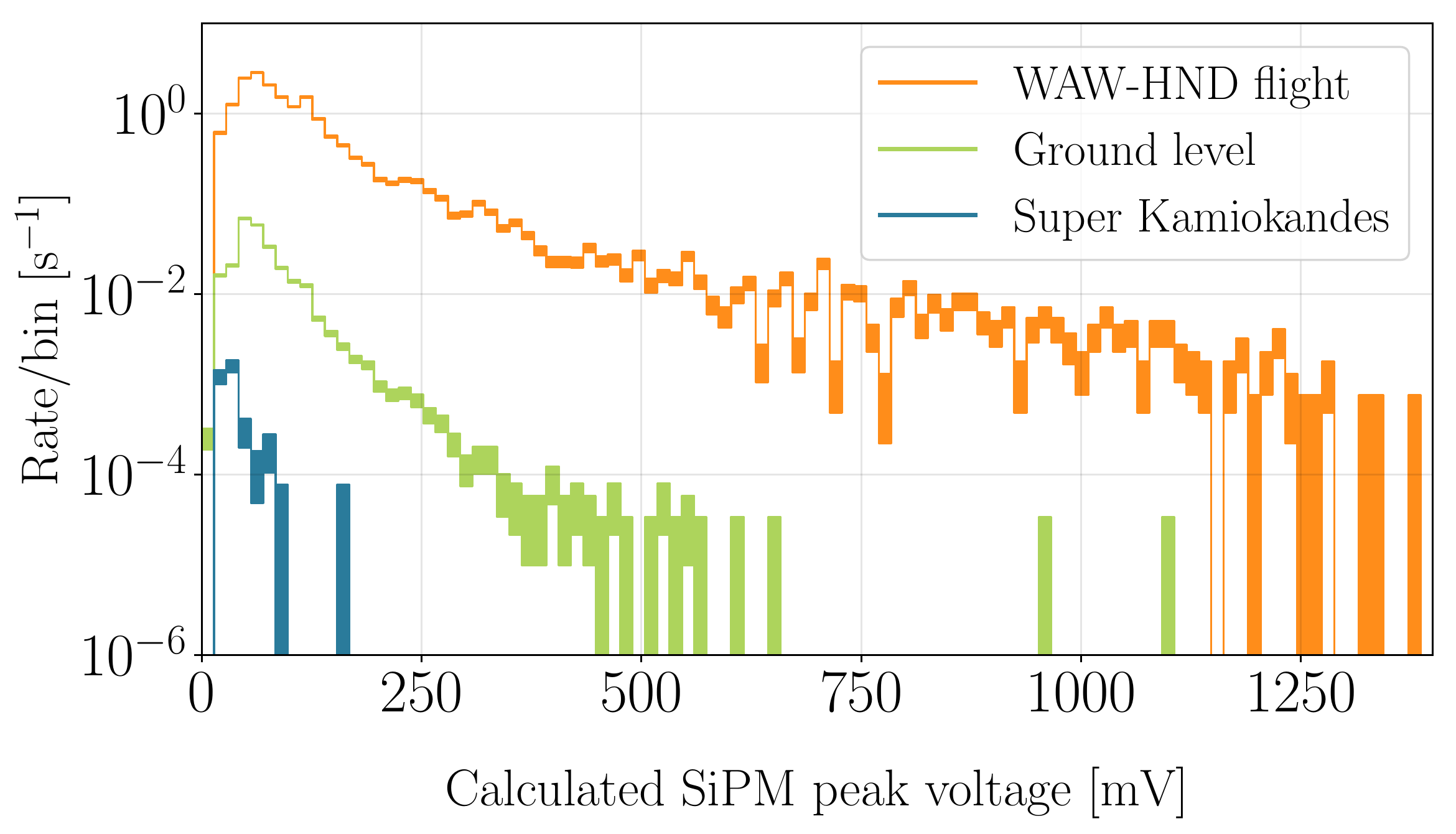}
\caption{The measured coincidence rate at three locations. Data from Ref.~\cite{axani2018cosmicwatch}.}
  \label{fig::sk}
 \end{center}
\end{figure}

Two Desktop Muon Detectors were brought to the Kamioka Observatory located 1~km underground in the Mozumi Mine, Japan.  This mine is home to several high profile experiments, perhaps most notably the 2015 Nobel prize winning particle physics experiment, Super-Kamiokande. Two detectors were placed in the Super-Kamiokande control room for 8 hours, and connected together via a 6-inch 3.5~mm audio cable in configuration (c) of Fig.~\ref{fig::opening_angle}. The data was recorded from the coincidence detector through the import\_data.py script directly to a laptop. Using the same detectors and set-up, a rate measurement was also performed outside the Kamioka mine in the observatory dormitory and in the airplane at 36,000~ft when travelling between Warsaw to Tokyo. Fig.~\ref{fig::sk} shows the trigger rate of the coincident detector for these three measurements, as a function of calculated SiPM peak voltage. 

The total number of measured coincident events measured inside the Super-Kamiokande control room was found to be 101. It was found that 96\% of these events were located below the 50~mV peak described in Sec.~\ref{sec::coincident}, indicating that these are likely not minimum ionizing cosmic-ray muons.

The average rock density in the mine was measured to be  2.7 g/cm$^3$, corresponding to approximately 2,700 m.w.e. (meter-water equivalent) of overburden\cite{fukugita2013physics}. Based on this, we expect the cosmic-ray muon rate to be attenuated by a factor of 10$^5$ compared to a ground level measurement. With this assumption, we only expect approximately 0.04 cosmic-ray muon events over the 8-hour measurement in the Super-Kamiokande control room.

The master detector count rate did not significantly change when it was brought into the mine, indicating that the radioactive background was still present in the control room. Given that the master detector count rate was 1~Hz, Eq.~\ref{eq::accidental} suggests that we should expect 1-2 accidental coincidence events over the 8-hour period (see Sec.~\ref{sec::measurements}). 

One unaccounted background (which was briefly mentioned in Sec.~\ref{sec::LE_electrons}) consists of events in which a gamma-ray from a radioactive decay Compton scattered off of the scintillator of the master detector, and was then absorbed or deposited sufficient energy in the coincident detector scintillator. This is thought to be the dominant source of triggers in this dataset, and a Monte Carlo simulation is currently being developed to investigate this. If these events originate from Compton scattering, we can estimate the rate for these types of events. Given the 101 events (and 1-2 of these are assumed to be accidental coincidences and cosmic-ray muons), the calculated accidental Compton scattering coincidence rate in configuration (c) from Fig.~\ref{fig::opening_angle} is found to be 0.0038 $\pm$ 0.0004. A second 8-hour run was performed using the same configuration and location, which found 92 events (with a similar SiPM peak voltage spectrum) corresponding to a count rate of 0.0035 $\pm$ 0.0004.

This result could be further investigated by repeating the measurement, however this time with a thin piece of lead between the scintillator. Lead, being a dense material, is likely to either absorb the gamma-ray or absorb some of the energy from the gamma-ray through Compton scattering. Both processes would reduce the probability of measuring the event with the coincident detector. Another potential source for these events is correlated noise. If the lead does not alter the coincident count rate, this is a potential source for this signal, however, thus far, we have not found any evidence of events due to noise.

\section{Cosmic-ray muon angular distribution}  \label{sec::angle}
This measurement illustrates the cosmic-ray muon angular dependence measured near sea level (it was performed in Madison WI, at 266~m above sea level) and was previously described in Ref.~\cite{axani2018cosmicwatch}. Here, two detectors were set to coincidence mode and placed side-by-side as in configuration (a) of Fig.~\ref{fig::opening_angle}, spaced 52~mm apart, inside their aluminium enclosures. The distance was chosen such that we gain sufficient statistics throughout a single day of measuring. If the detectors are placed too far apart, the count rate drops significantly and accidental coincidences can dominate the signal.

\begin{figure}[h]
    \centering
    \begin{minipage}{.6\textwidth}
        \centering
        \includegraphics[width=.9\linewidth]{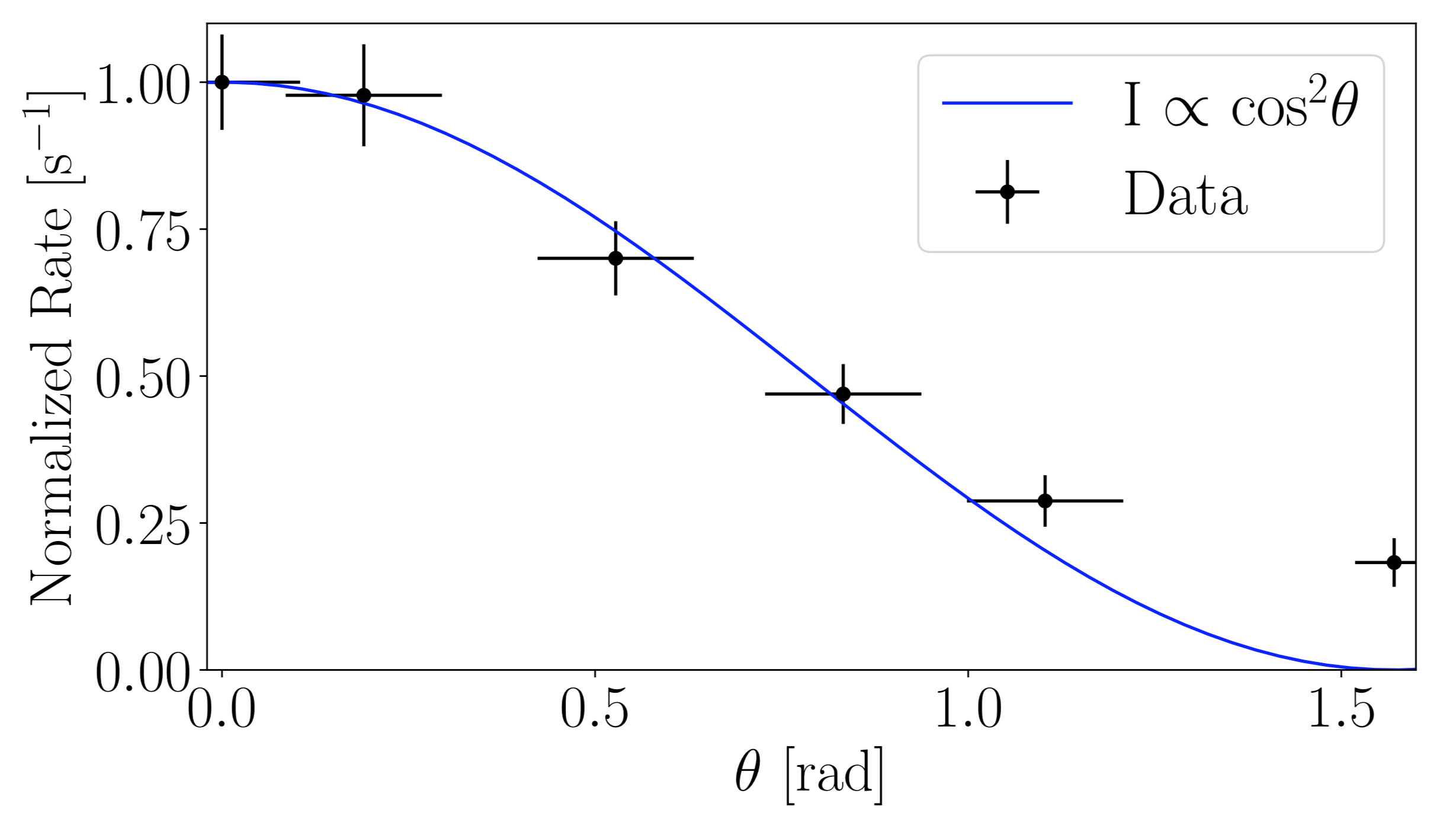}
        \subcaption{}
    \end{minipage}%
    \begin{minipage}{0.4\textwidth}
        \centering
        \includegraphics[width=0.9\linewidth]{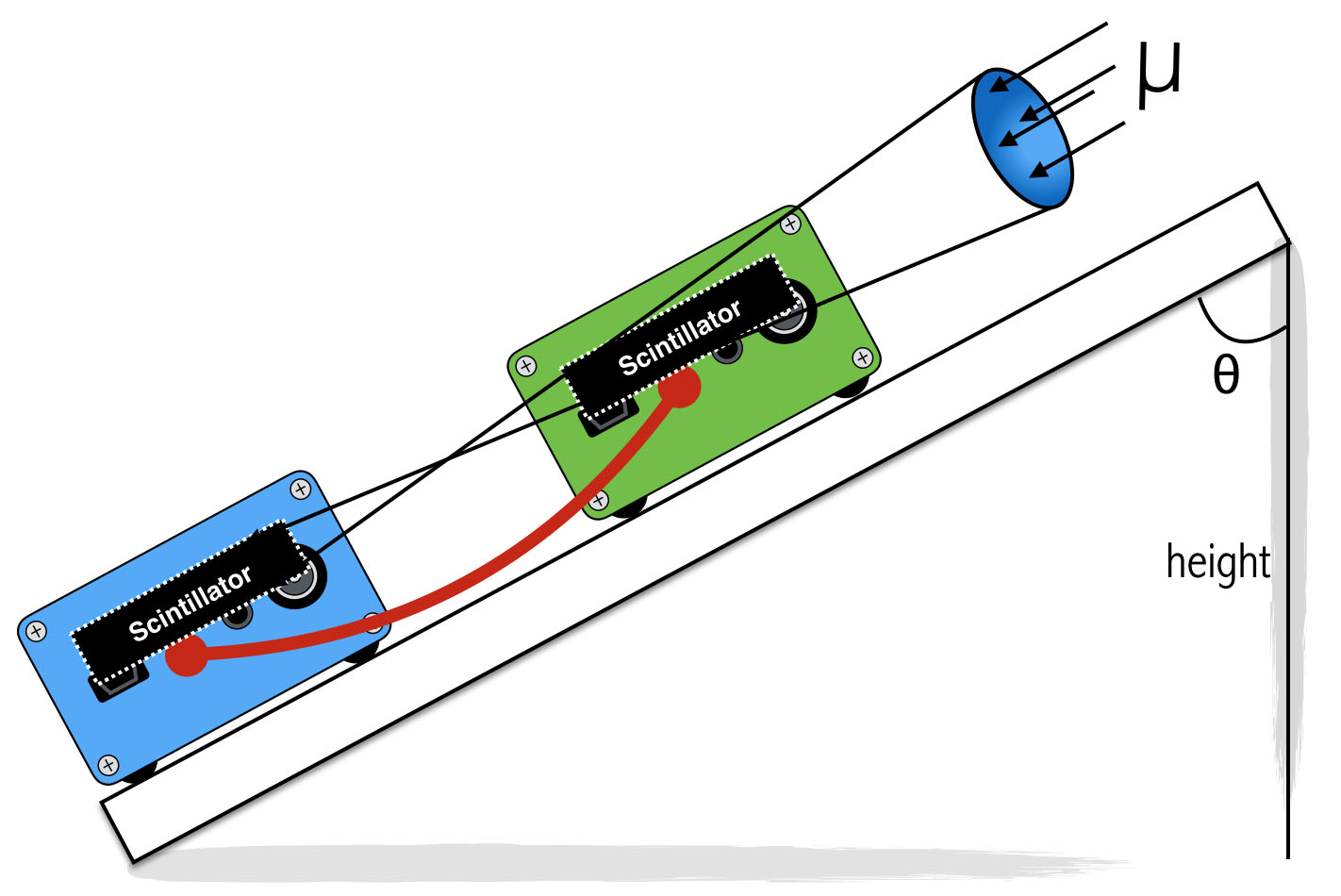}
        \subcaption{}
    \end{minipage}
       \caption{The measured cosmic-ray muon angular distribution measured by the coincident detector. The prediction by the PDG is shown in solid blue. From Ref.~\cite{axani2018cosmicwatch}.}
       \label{fig::angle}
\end{figure}

The angle of the detectors was determined by securing the detectors to a 100~cm long rectangular bar and then positioning the bar against a wall at a known height (see Fig.~\ref{fig::angle} (right)). It is important when making this measurement that the angle of the detectors are accurately measured. Fig.~\ref{fig::angle} (left) shows the measured relative rate as a function of zenith angle (with zero radians representing vertical). Each data point represents approximately 10 hours of data and the rate uncertainties are statistical. The horizontal uncertainties represent the calculated opening angle of the two detectors when spaced 52~mm apart. The measurement at $\theta = \pi/2$ is divided by 2, since at this angle it accepts cosmic-ray muons from both directions, whereas all the other angles only accept down going muons. 

As indicated in Sec.~\ref{sec::variations_atm}, the angular cosmic-ray muon dependence at sea level should follow a cosine squared dependence. The overall measured shape of the distribution is found to agree relatively well with the cosine squared prediction, however it is shown that the rate does not fall completely to zero in the horizontal configuration ($\theta = \pi/2$). This is currently under investigation, and could be related to large cosmic-ray showers, showers developing in the roof above the detector, or perhaps muons producing either high energy electrons or photons along the track that can trigger both detectors.

\section{Electronmagnetic component of cosmic-ray showers}\label{sec::electromagnetic_shower}

\begin{wrapfigure}{R}{0.4\textwidth}
\centering
\includegraphics[width=0.4\textwidth]{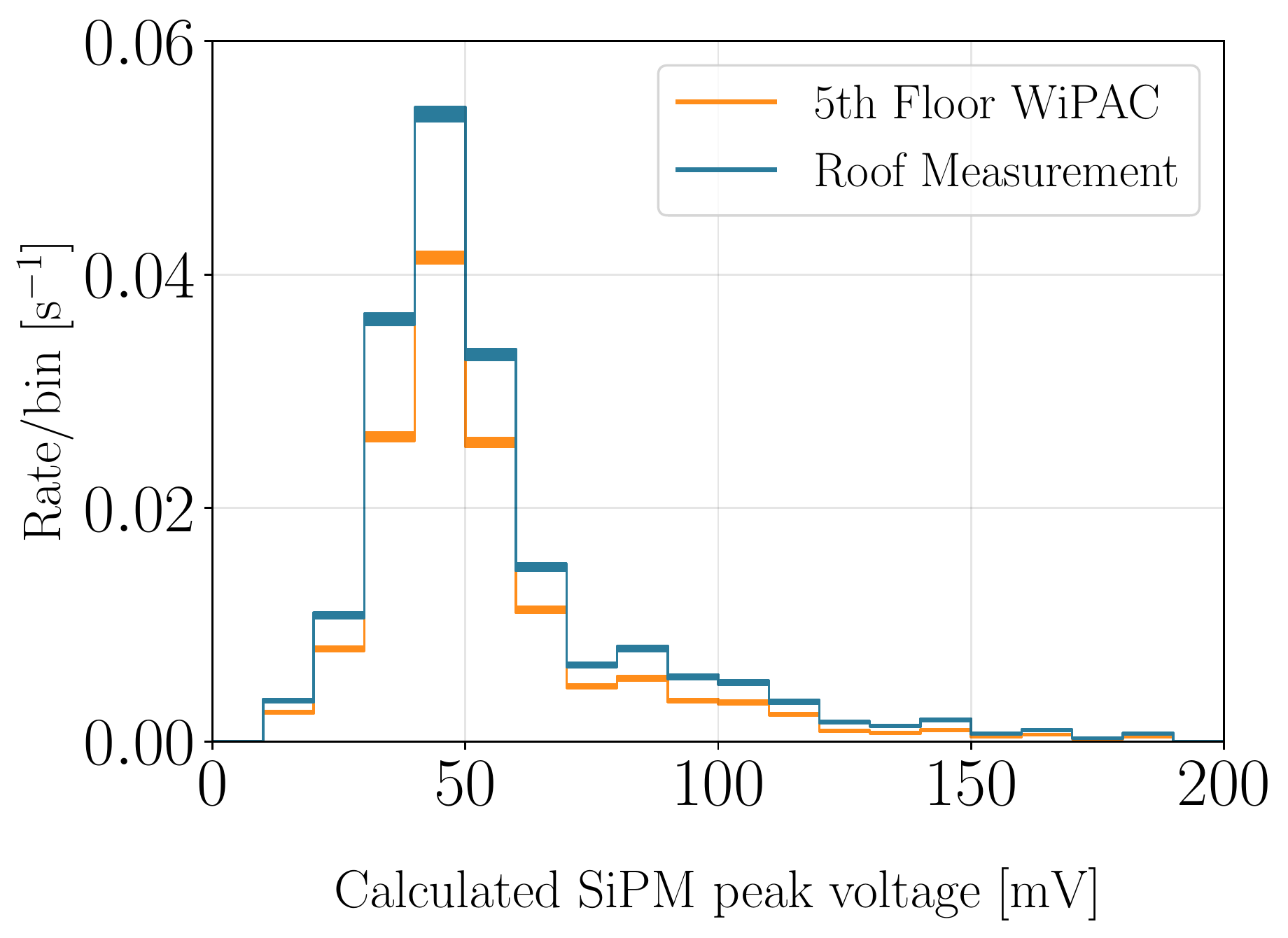}
\caption{The calculated SiPM peak voltage on the roof of the WiPAC building compared to on the 5$^{\mathrm{th}}$ floor.}
\label{fig::roof} 
\end{wrapfigure}
We've discussed in Sec.~\ref{sec::cosmic_rays} that at sea level, there is still have a flux of electrons and positrons (with energies $<~1$GeV) showering down on the Earth. We then discussed in Sec.~\ref{sec::HE_electrons} that this electromagnetic component is not particularly penetrating and can be significantly attenuated with a few tens of cm of concrete (the radiation length in concrete found in Table~\ref{table::radiation} is 10.7~cm). We can test this claim by measuring the coincidence rate on top of a building, then again several floors lower.

We expect a minimal attenuation of the comic ray muons, however a significant attenuation of the electromagnetic component. A typical building may have 15~cm of concrete between each floor. If we measure the vertical muon rate on the top floor, then again 5 floors lower, we expect the energy of the electromagnetic component to be attenuated by 7 radiation lengths (a factor of 1000).

Two detectors were place in coincidence mode one-on-top of the other in configuration (c) of Fig.~\ref{fig::opening_angle}. They were placed on the roof-top of the 10-floor WiPAC building for 24-hours inside a plastic bag to protect them against the weather.  After the measurement, the same two detectors were placed on the 5th floor, and the measurement was repeated.  The results of the measurement are shown in Fig.~\ref{fig::roof}.

We see that the rate on top of the building is 27\% higher than that of the measurement made on the 5$^{\mathrm{th}}$ floor. The attenuation of this component is thought to be mostly due to the elimination of the electromagnetic component. In Sec.~\ref{sec::cosmic_rays}, we explain that approximately 33\% of the total flux from cosmic ionizing radiation at sea level comes from the electromagnetic component.

\section{Correlation between the cosmic-ray muon rate and atmospheric pressure}\label{sec::atm_pressure}

A correlation exists between the atmospheric pressure and the cosmic-ray muon rate. It can be expressed as follows:
\begin{equation}
\frac{\Delta I}{\bar{I}} =  \beta \Delta P,
\end{equation}

where I represents cosmic-ray muon intensity,  $\Delta $P is the measured atmospheric pressure compared to the average pressure, and $\beta$ is the barometric coefficient~\cite{de2013analysis}. This correlation is actually the result of several processes outlined in Sec.~\ref{sec::variations_atm}. The barometric coefficient represents the percent change in detector count rate per hPa change in atmospheric pressure.

\begin{figure}[h]
\begin{center}
\includegraphics[width=1.\columnwidth]{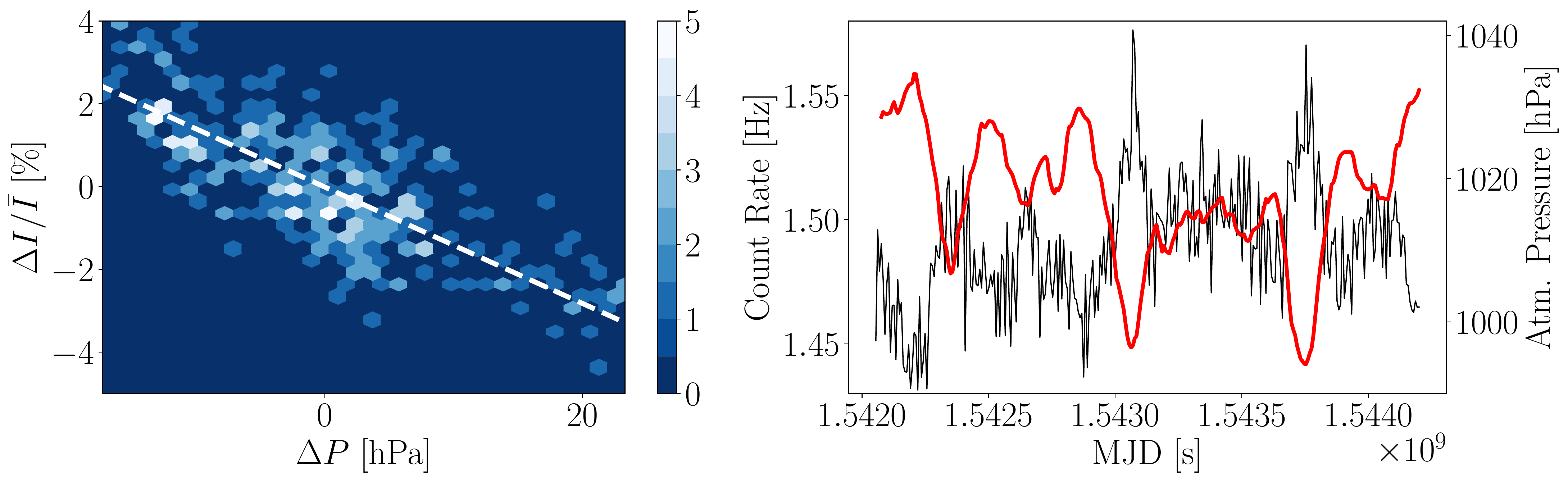}
\caption{Left: The correlation between the atmospheric pressure and the detector count rate. Right: The detector count rate in blue and the atmospheric pressure in red as a function of modified Julian date (MJD). Both plots show the data binned with a bin size of two hours.}
  \label{fig::correlation}
 \end{center}
\end{figure}

For this measurement, we decided to use an array of master-coincident detectors to improve the statistics. Five pairs of detectors were each connected in configuration (c) of Fig.~\ref{fig::opening_angle}, and left to record the cosmic-ray muon rate over the course of 24.8 days (from 13:35~hrs Nov. 12th to 09:00~hrs Dec. 7th 2018.). Data was recorded directly to a microSD card and powered through an 8-way USB hub powered through a wall outlet. Since the Arduino does not keep accurate time over long time-scales, we assume the time drifts linearly with time, and scale the uptime of all detectors such that they are all the same. This is not ideal, and we would recommend recording the data directly to a computer through the import\_data.py script in the future to get an accurate time stamp on all events. The array of detectors was placed in the 4$^{\mathrm{th}}$ floor WiPAC lab in Madison, WI. Atmospheric pressure for Madison was found in Ref.~\cite{weather}. 

Fig.~\ref{fig::correlation} (left) shows the correlation between the detector count rate and the atmospheric pressure. The calculated correlation is shown in the dotted white line. A least squares fit yielded a barometric coefficient of -0.141$\pm$0.007~\%/hPa, in agreement with Ref.~\cite{jourde2016monitoring}.

\section{Reducing the radioactive backgrounds}\label{sec::reducing_backgrounds}

\begin{wrapfigure}{R}{0.45\textwidth}
\centering
\includegraphics[width=0.45\textwidth]{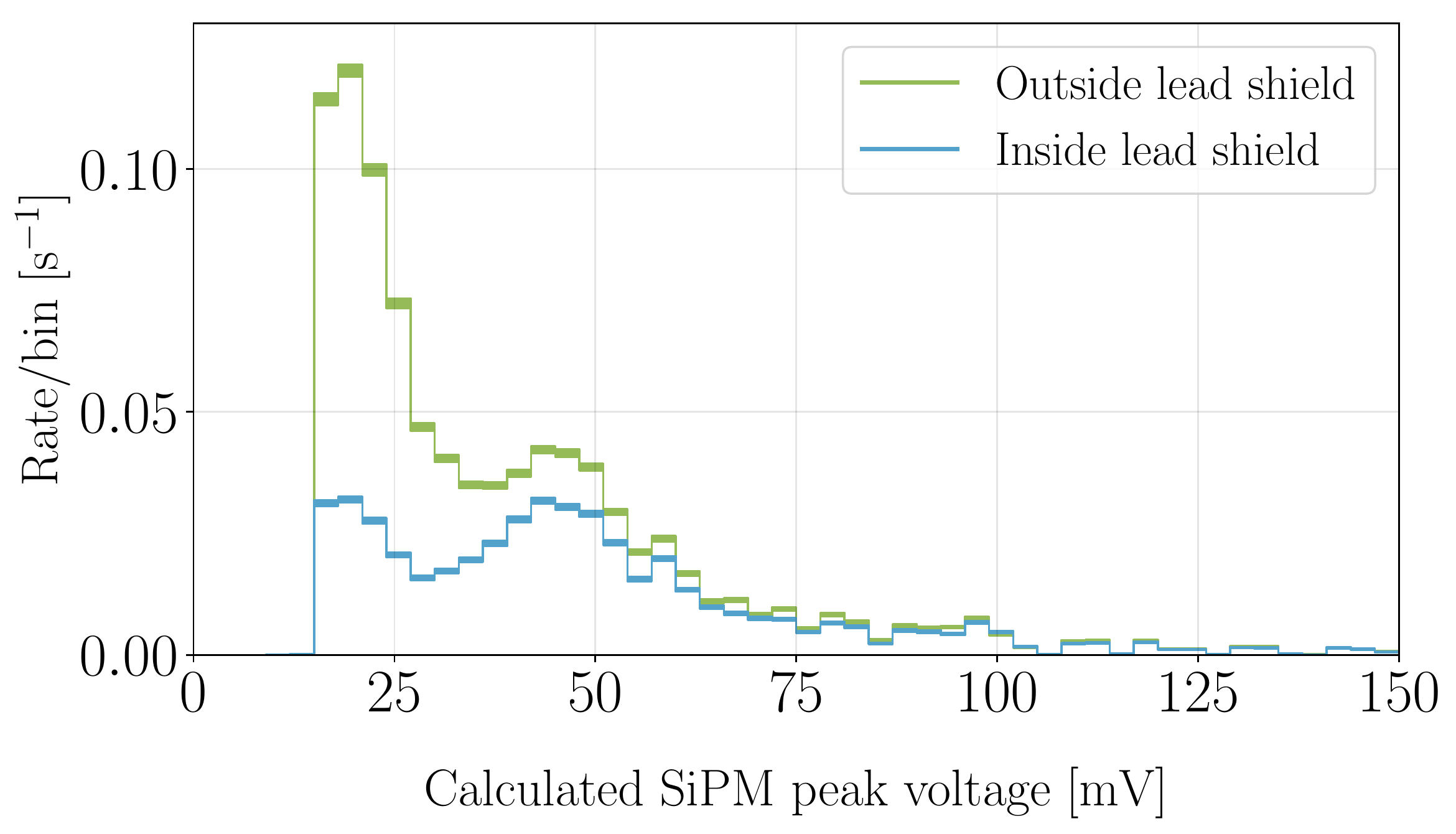}
\caption{The effect of surrounding a single detector in a lead enclosure.}
\label{fig::lead} 
\end{wrapfigure}

It can be difficult to shield against the natural radioactivity found in the environment. Typically at sea level, the largest component of our signal comes from radioactive backgrounds and therefore it would be useful to think about how we can minimize this to improve the purity of the cosmic-ray muon component of the signal (beyond setting detectors into coincidence mode).

Dense material placed around the detector can attenuate the incoming flux from the background radiation. Ideally, we would choose a material that itself is radio-pure, however we can illustrate the effect using bricks of lead. Six lead ingots (each measuring 2"x4"x8") were positioned in such a way to provide 4$\pi$ coverage around a single detector. The detector recorded data directly to the microSD card throughout a full day. It was then placed on a workbench in the same room (far away from the lead), to measure the background spectrum for another full day. The resulting calculated SiPM peak voltage for the two measurements is shown in Fig.~\ref{fig::lead}. As expected, this is shown to significantly reduce the event that contribute to the low SiPM peak voltage region, which is dominated by the radioactive background. 

\begin{figure}[ht]
    \centering
    \begin{minipage}{.55\textwidth}
        \centering
        \includegraphics[width=1.\linewidth]{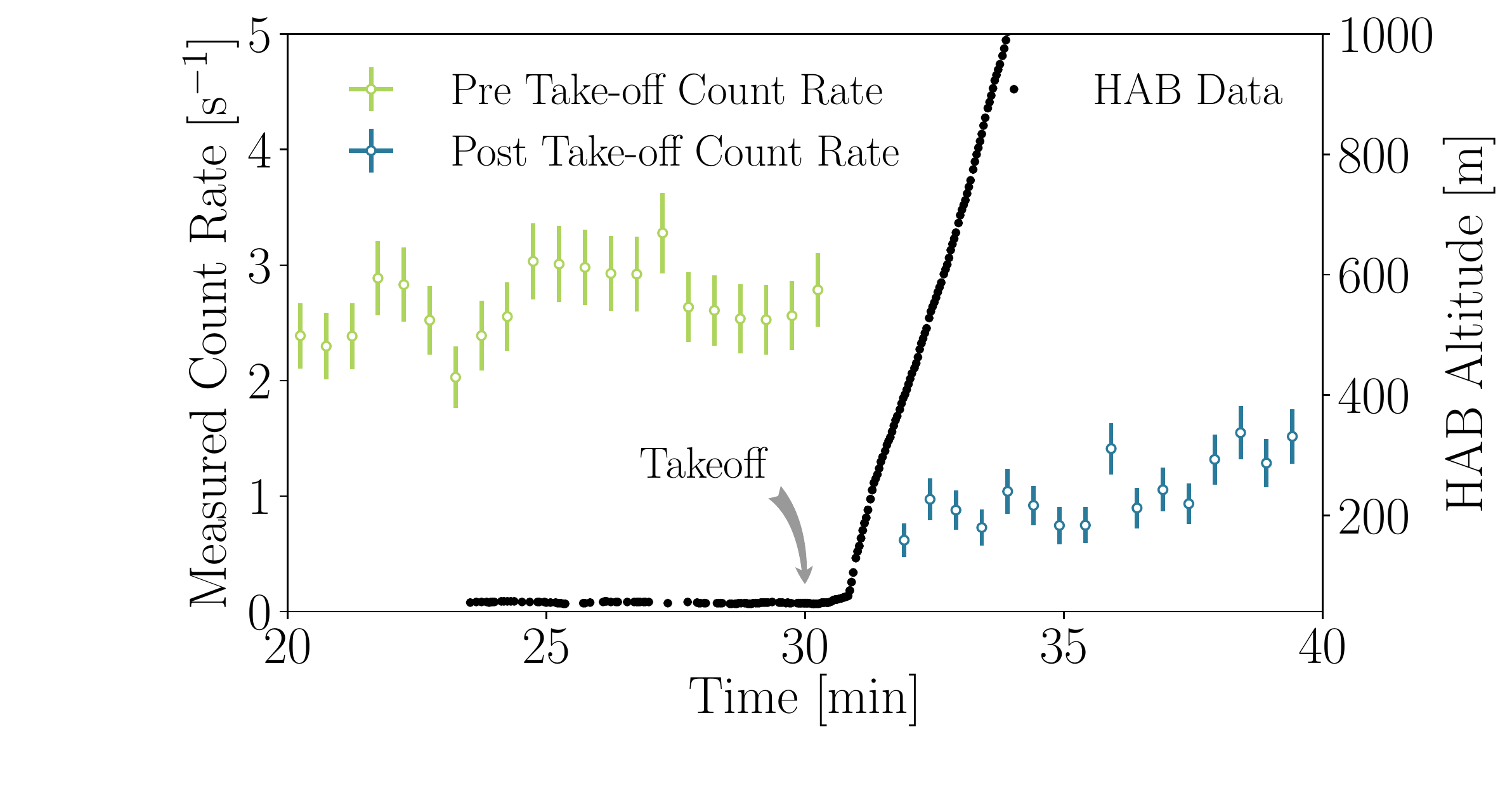}
    \end{minipage}%
    \begin{minipage}{0.35\textwidth}
        \centering
        \includegraphics[width=1\linewidth]{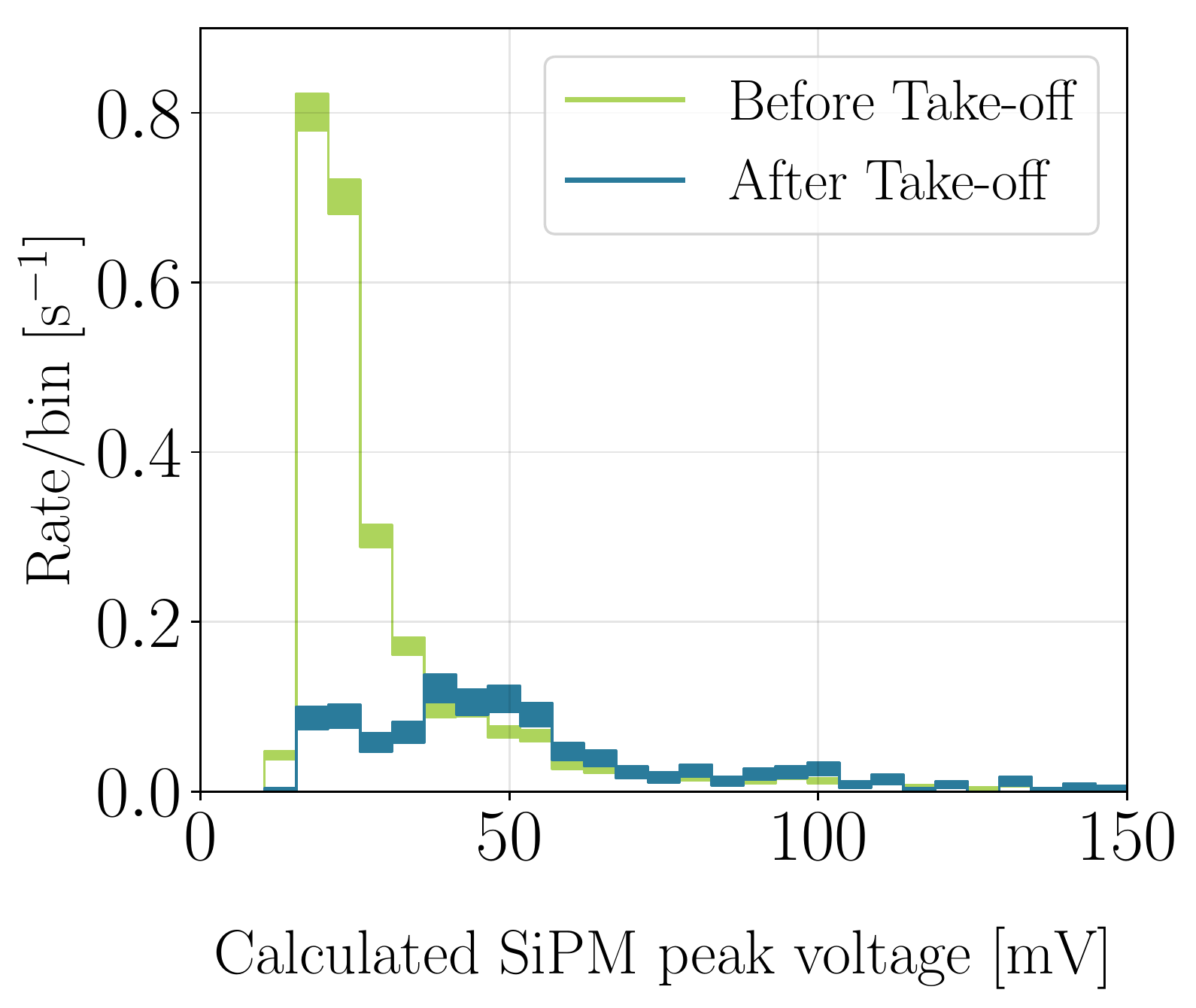}
    \end{minipage}
       \caption{Left: The master detector count rate from the HAB measurement of Sec.~\ref{sec::hab} near take-off.  Right: The calculated SiPM peak voltage for pre and post take off. Pre-take-off shown in green, and post take-off shown in blue.}
       \label{fig::takeoff}
\end{figure}

Another measurement that illustrates the effect of background radiation unexpectedly came from the HAB flight described in Sec.~\ref{sec::hab}. When the balloon pulled the detectors off of the ground, there was a significant decrease in the master detector trigger rate as shown in Fig.~\ref{fig::takeoff} (left).  Fig.~\ref{fig::takeoff} (right) shows the calculated SiPM peak voltage just prior to take-off and after take-off. After take-off, we see the decrease in count rate came from the low SiPM peak voltage region, which we've indicated primarily consists of the background radiation (see Sec.~\ref{sec::coincident}).

\section{Portable trigger system for an accelerator beamline} \label{sec::beamline}
This measurement was previously described in Ref.~\cite{axani2018cosmicwatch} and represents a practical use for the Desktop Muon Detectors.  

\begin{figure}[ht]
\begin{center}
\includegraphics[width=0.6\columnwidth]{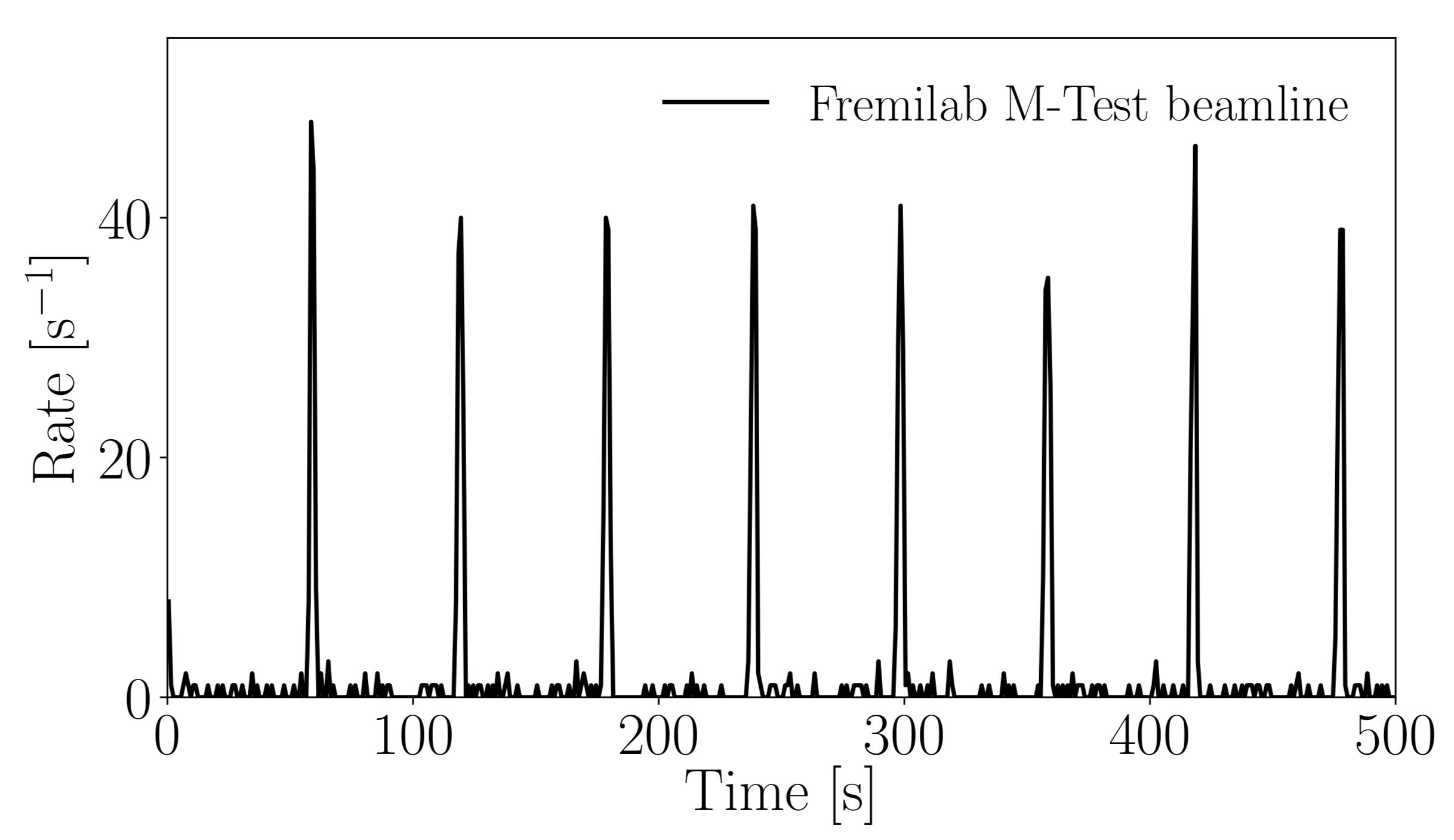}
\caption{The trigger rate as a function of time of a single detector placed in the Fermilab M-Test beamline. Here, the detector is triggering primarily on GeV-scale pions and electrons from the Fermilab Main Injector. }
  \label{fig::beam}
 \end{center}
\end{figure}

A single detector, powered by a 10,000~mAh USB power bank, was placed in the Fermilab M-Test facility to trigger on secondary particles (GeV-scale pions and electrons) from the Main Injector. The purpose of this was to trigger a downstream data acquisition system for another experiment. The BNC output at the back of the detector is the raw SiPM pulse, which has a rise time of a few nanoseconds and a decay time of roughly 0.5~$\mu$s. This signal is useful for experiments that want to use a scintillators but require tens of nanosecond timing. The BNC output was connected to an 80~ft BNC cable to a NIM (Nuclear Instrument Module) rack . The signal passed through a $\times$10 amplifier and into a discriminator. If the amplified signal was above a certain value, a binary signal was sent to a AND gate, where it was compared against another scintillator paddle trigger that was located on the other side of the other experiment. If the AND condition was satisfied (ie. the particle passed through both the scintillator paddle and the Desktop Muon Detector) a binary signal was sent to the data acquisition system that began the recording of data of the downstream experiment. Fig.~\ref{fig::beam} shows the trigger rate of the detector place in the beamline as a function of time. The beam spills occur every minute for two seconds. 

The detector was found to be a useful beamline trigger due to its simplicity. This ability was made possible by including a BNC output connected directly to the SiPM. For this measurement, the approximate 10~ns uncertainty in the trigger was acceptable, however if users would like to use the FAST output of the SiPM, the SiPM PCB could be modified to get down to single ns precision. We plan on investigating this sometime in the future.

\section{A true random number generator}\label{sec::rng}

\begin{figure}[ht]
\begin{center}
\includegraphics[width=1.\columnwidth]{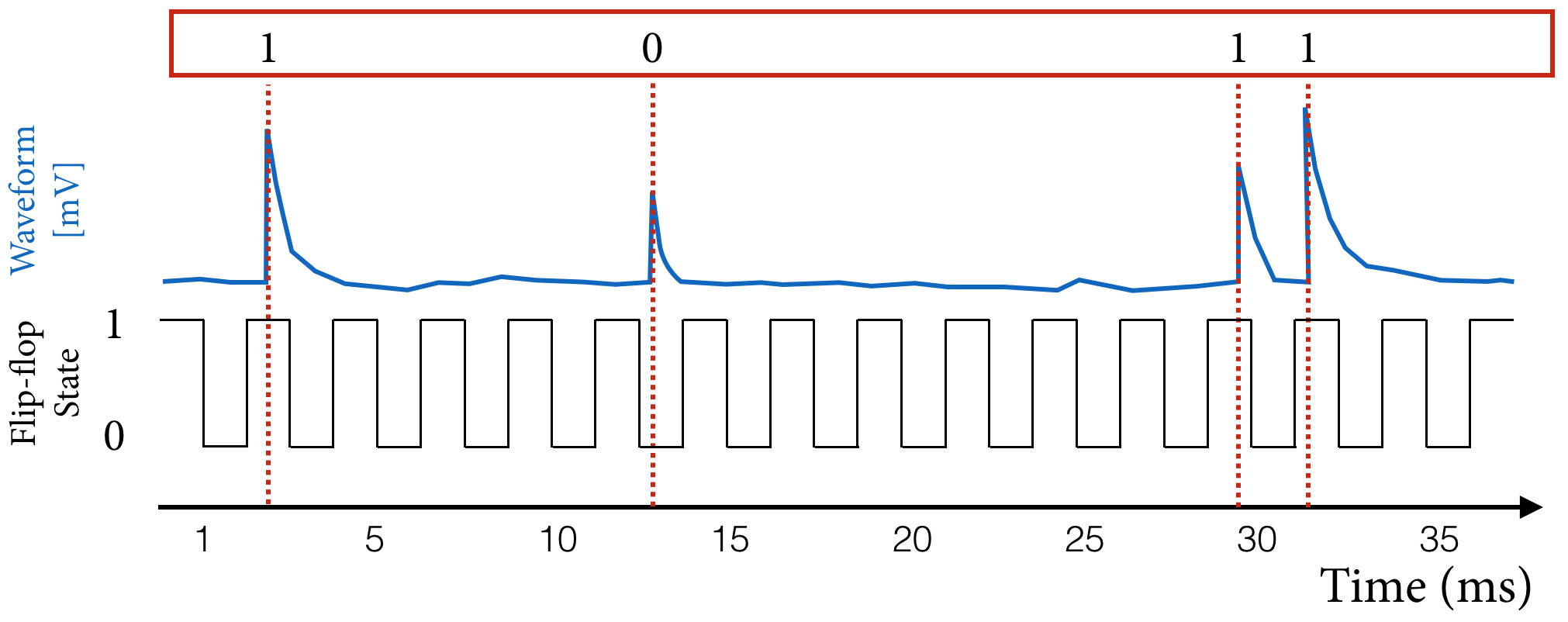}
\caption{An illustrative diagram of the principle behind the random number generator. The x-axis indicates the time measure in milliseconds. Above this, we have the toggle flip-flop, which changes state every millisecond, then an illustrative waveform shown in blue. If the event triggered the detector while in the even state of the flip flop, we assign it a one, otherwise, we assign it a zero. This is shown in the red box at the top of the diagram.}
  \label{fig::rng}
 \end{center}
\end{figure}

Many applications require random numbers. Often, a random number may be generated through some algorithm, but this therefore becomes deterministic since if the user knew the algorithm and starting conditions, they could determine the output. A random number generator would ideally be generated from a truly random process, such as the arrival times of cosmic-ray muons or the radioactive decay of an element. The sum of two random processes will also be random, such as the signal from the radioactive backgrounds in the detector along with the cosmic-ray muon signal. For this measurement, I'm following the description found in Ref.~\cite{grupen2008particle}. 

Any number can be expressed in terms of a sequence of ones and zeros, this is known as binary. An N-length sequence is able to represent a number from zero to 2$^N$-1 (corresponding to 2$^N$ different values). For example, the 4-bit binary sequence "1011", corresponds to $1\times(4^2)+0\times(3^2)+1\times(2^2)+1\times(1^2)$=$8+0+2+1~=~11$. 

We can convert the time stamp of a radioactive decay trigger into a "1" or a "0" using a \textit{toggle flip-flop}. The toggle flip-flop is simply a state that changes from one to zero periodically (we will use a frequency of 1~kHz). If a particle passes through the scintillator during an even time stamp (as measured in milliseconds), we assign it a "1"; if it passes through an odd time stamp, we assign it a "0". After N triggers, we can build an N-bit random number. This is schematically illustrated in Fig.~\ref{fig::rng}.

\begin{figure}[ht]
\begin{center}
\includegraphics[width=1\columnwidth]{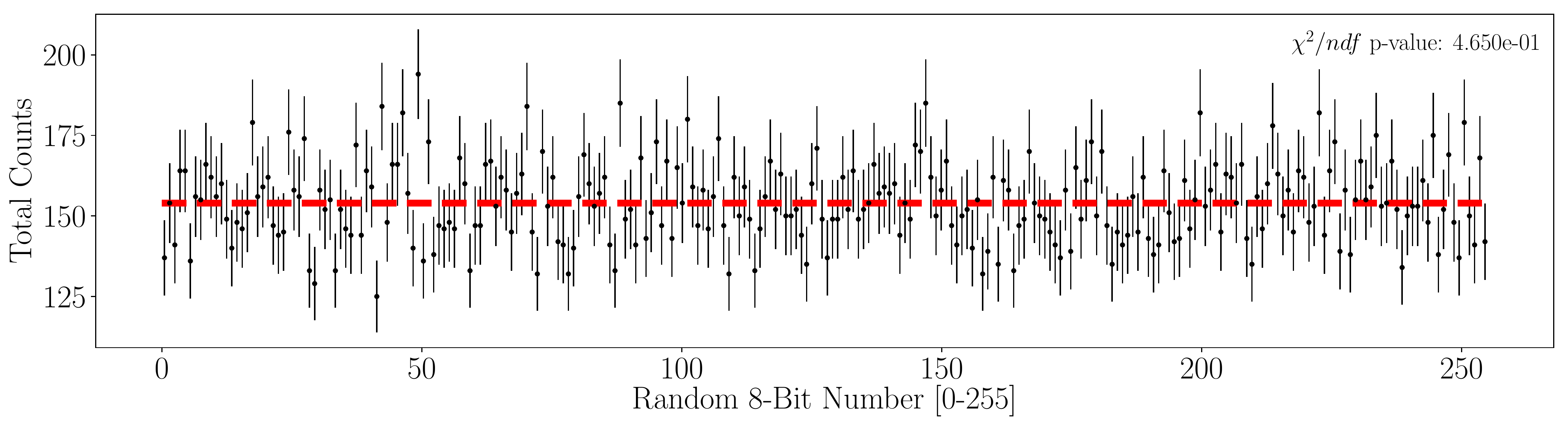}
\caption{The number of occurrences of the generated numbers from 0-255. The reduced $\chi^2$ p-value assuming that the distribution should be randomly distributed about the average number of occurrences is given at the top right of the figure.}
  \label{fig::rng_plot}
 \end{center}
\end{figure}

For this measurement, we use data taken from a 20-day background lab measurement. A single detector was used. After triggering on t total events, we can build t/N N-bit random numbers. To illustrate this, if we chose to generate 8-bit random numbers (from 0 to 255), we can plot the number of each occurrence. This is shown in Fig.~\ref{fig::rng_plot}.

Fig.~\ref{fig::rng_plot} shows the random nature of the triggers. Each number should be equally probable to occur. The reduced $\chi^2$ indicates a p-value for this assumption shown in the top right of the plot. 

There are several ways that the random number generator can become biased. First, let's think about extreme scenarios. Suppose that the trigger rate is roughly 1~Hz, and the flip-flop state only changes every 10 seconds. The first roughly ten triggers would give all ones, then the next roughly triggers would give all zeroes. The first numbers would be biased high, while the later numbers would be biased low. With this extreme example, we see that we need the toggle flip-flop to be changing states at a much higher rate than the trigger rate. Secondly, suppose that the Arduino does not produce equal numbers of even and odd time stamps due to the internal configuration. We found this to be the case when using the microseconds() Arduino function, that is, it only reported the time stamp to the nearest even microsecond. 

\section{The Poissonian nature of radioactive decay}\label{sec::poisson}

\begin{figure}[t]
\begin{center}
\includegraphics[width=1\columnwidth]{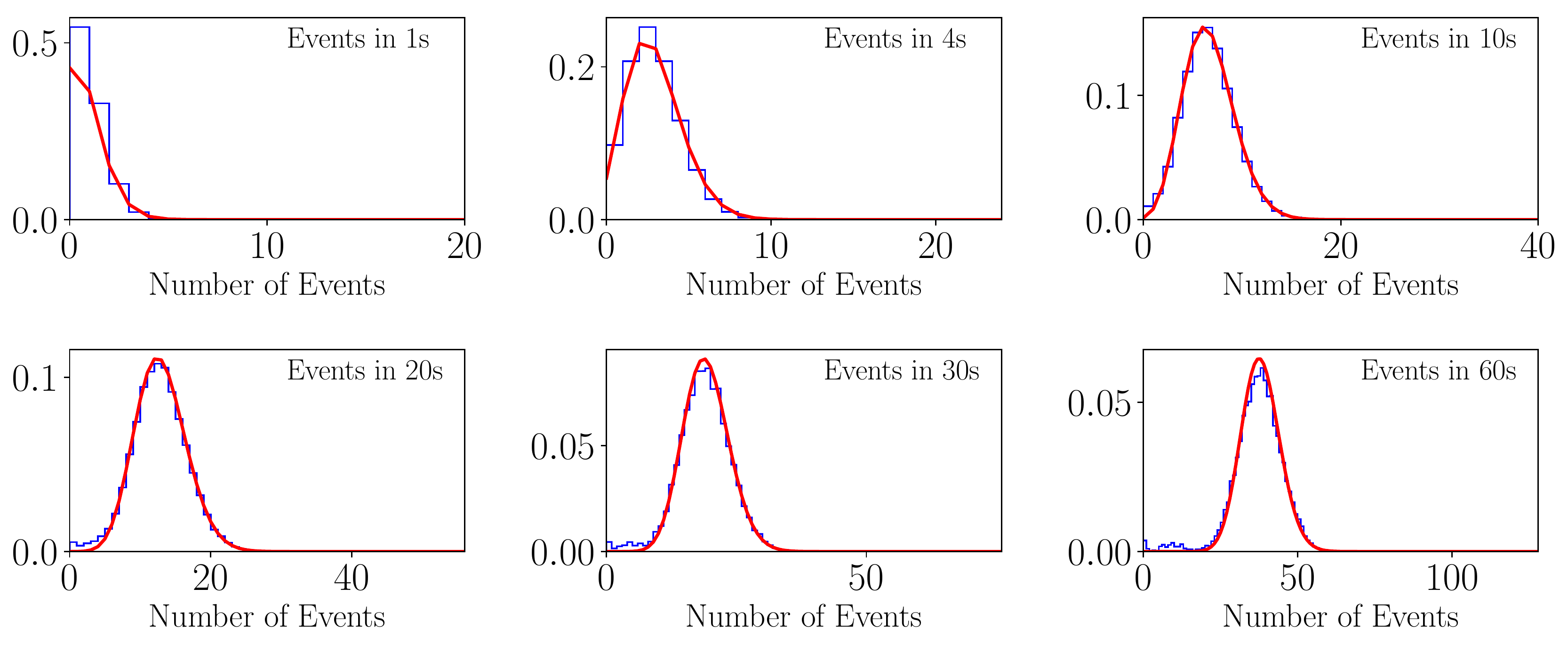}
\caption{The distribution of number of events per unit time. Each cell represents a different amount of time in which the data is binned (labelled on the top right of each cell). The red line is the expected Poissonian distribution from Eq.~\ref{eq::poisson}. }
  \label{fig::poisson}
 \end{center}
\end{figure}

A Poissonian process is one in which an event happens that is completely independent of the occurrence of another event. This is the case for many processes in nature: for example, when atom decays or the arrival times of cosmic rays. The expected probability of observing N events in a time interval t of a Poissonian process, with an average count rate $\mu$ is defined by the \textit{Poisson distribution}, Eq.~\ref{eq::poisson}.

This example uses the same dataset from the previous measurement, however, here we bin the data into bins of width t and then histogram the number of measured events in that time window. This is shown in Fig.~\ref{fig::poisson}.
%
%
%

%% file: Conclusion.tex
\chapter{\textsc{Conclusion}}\label{sec::conclusion}
The CosmicWatch Desktop Muon Detectors are capable of exploring various physical phenomena in nature. This document outlined in detail the physical processes that influence the detector, and how they can be extracted from data. We have investigated various phenomena associated with the geomagnetic field, atmospheric conditions, cosmic-ray shower composition, attenuation of particles in matter, radioactivity, and statistical properties of a Poisson process. Beyond this, the supplementary material contains further information for developing skills in the machine and electrical shop, as well in programming. As the CosmicWatch program grows, we expect more people to contribute to potential measurements and improvements to the detectors. This will benefit the hundreds to thousands of students already building their own CosmicWatch Desktop Muon Detectors and learning about physics.

%
%
%
%